\pdfoutput=1
\documentclass[PAPER,atlasdraft=false,cernpreprint,texlive=2016,UKenglish,texmf]{atlasdoc}
 
\usepackage{atlaspackage}
\usepackage{atlasbiblatex}
 
\usepackage{atlasphysics}
 
\graphicspath{{logos/}{figures/}}
 
\usepackage{ANA-STDM-2018-34-INT1-defs}
\usepackage{PMGRefs-defs}
\usepackage{floatpag}
\usepackage{todonotes}
\usepackage{multirow}
\usepackage{longtable}
\usepackage{pdflscape}
\usepackage{siunitx}
\usepackage{listings}
\usepackage{float}
\usepackage{makecell}
\usepackage{adjustbox}
\usepackage{etoolbox}
 
\providetoggle{paperdraft}
\settoggle{paperdraft}{true}
 
\newcommand{\OnlySupportNote}[1]{\nottoggle{paperdraft}{#1}{}}
\newcommand{\OnlyPaperDraft}[1]{\iftoggle{paperdraft}{#1}{}}

\AtlasTitle{Measurements of $W^+W^-+\ge 1$~jet production cross-sections in $pp$ collisions at $\sqrt{s}=13$~TeV with the ATLAS detector}

\AtlasAbstract{
Fiducial and differential cross-section measurements of $W^+W^-$ production in association with at least one hadronic jet are presented.
These measurements are sensitive to the properties of electroweak-boson self-interactions and provide a test of perturbative quantum chromodynamics and the electroweak theory.
The analysis is performed using proton--proton collision data collected at $\sqrt{s}=13$~TeV with the ATLAS experiment, corresponding to an integrated luminosity of 139~fb$^{-1}$. Events are selected with exactly one oppositely charged electron--muon pair and at least one hadronic jet with a transverse momentum of $p_{\mathrm{T}}>30$~GeV and a pseudorapidity of $|\eta|<4.5$.
After subtracting the background contributions and correcting for detector effects, the jet-inclusive  $W^+W^-+\ge 1$~jet fiducial cross-section and $W^+W^-+$ jets differential cross-sections with respect to several kinematic variables are measured. 
These measurements include leptonic quantities, such as the lepton transverse momenta and the transverse mass of the $W^+W^-$ system, as well as jet-related observables such as the leading jet transverse momentum and the jet multiplicity.
Limits on anomalous triple-gauge-boson couplings are obtained in a phase space where interference between the Standard Model amplitude and the anomalous amplitude is enhanced.
}
 
\author{The ATLAS Collaboration}
 
\AtlasRefCode{STDM-2018-34}

\PreprintIdNumber{CERN-EP-2021-030}

\AtlasJournal{JHEP}
 
\AtlasJournalRef{JHEP 06 (2021) 003}
\AtlasDOI{10.1007/JHEP06(2021)003}

\hypersetup{pdftitle={ATLAS document},pdfauthor={The ATLAS Collaboration}}
 
\begin{document}
 
\maketitle
 
\tableofcontents
 
\section{Introduction}
\label{sec:intro}
 
\newcommand{\AtlasCoordFootnote}{
ATLAS uses a right-handed coordinate system with its origin at the nominal interaction point (IP)
in the centre of the detector and the \(z\)-axis along the beam pipe.
The \(x\)-axis points from the IP to the centre of the LHC ring,
and the \(y\)-axis points upwards.
Cylindrical coordinates \((r,\phi)\) are used in the transverse plane,
\(\phi\) being the azimuthal angle around the \(z\)-axis.
The rapidity is defined as $y=\frac{1}{2}\ln{\frac{E+p_z}{E-p_z}}$ while the pseudorapidity is defined in terms of the polar angle \(\theta\) as \(\eta = -\ln \tan(\theta/2)\).
Angular distance is measured in units of \(\Delta R \equiv \sqrt{(\Delta\eta)^{2} + (\Delta\phi)^{2}}\).
}

The measurement of $W$-boson pair (\WW) production cross-sections is an important test of the Standard Model (SM). \WW production at hadron colliders is sensitive to the properties of electroweak-boson self-interactions and provides a test of perturbative quantum chromodynamics (QCD) and the electroweak (EW) theory.
It also constitutes a large background in the measurement of Higgs boson production as well as in searches for physics beyond the SM. Inclusive and fiducial \WW production cross-sections have been measured in proton--proton ($pp$) collisions at $\sqrt{s}=7\,\TeV$~\cite{STDM-2012-01,CMS-SMP-12-005}, $8\,\TeV$~\cite{STDM-2013-07,CMS-SMP-14-016,STDM-2015-24} and 13\,\TeV~\cite{STDM-2015-20,STDM-2017-24,CMS-SMP-18-004}, as well as in $e^+e^-$ collisions at LEP~\cite{schael:2013ita} and in $p\bar{p}$ collisions at Tevatron~\cite{Abe:1996dw,Aaltonen:2009aa,Abazov:2009ys}.
However, to reduce backgrounds, the measurements of inclusive cross-sections typically require that the WW pair is produced without additional jet activity, or at most with one additional jet. The production of \WWj has therefore not been studied in detail.
 
This article presents results of measurements of fiducial and differential cross-sections for a \WW pair produced in association with one or more jets.
For the first time at the LHC, differential measurements are performed in a jet-inclusive phase space.
This measurement complements previous results as the combination of measurements with and without jets improves the precision of the inclusive \WW cross-section due to an anti-correlation of important systematic uncertainties, for example the jet energy scale uncertainty, as demonstrated in previous measurements from ATLAS~\cite{STDM-2015-24} and CMS~\cite{CMS-SMP-18-004}. 
 
The analysis of one-jet topologies can also improve searches for anomalous triple-gauge-boson couplings (aTGCs), due to the increased interference between the SM amplitude and the anomalous amplitude~\cite{Azatov:2017kzw}. The impact of the $Q_{W}$ aTGC operator, as defined in Ref.~\cite{Grzadkowski:2010es}, increases rapidly with energy, making a measurement at the energies probed by the LHC important. However, at high centre-of-mass energy, the SM amplitude and the anomalous amplitude are dominated by different helicity configurations, so their interference is suppressed, which reduces the impact of the operator. The reduced sensitivity to the interference also poses a problem for the validity of the effective field theory interpretation, as contributions that are quadratic in the dimension-six amplitude, which are expected to be subdominant in the EFT expansion, become large. Requiring hard jets in addition to the diboson pair allows different helicity configurations and, thus, reduces the interference-suppression~\cite{Azatov:2017kzw}.

\begin{figure}
\centering
\includegraphics[width=.3\linewidth]{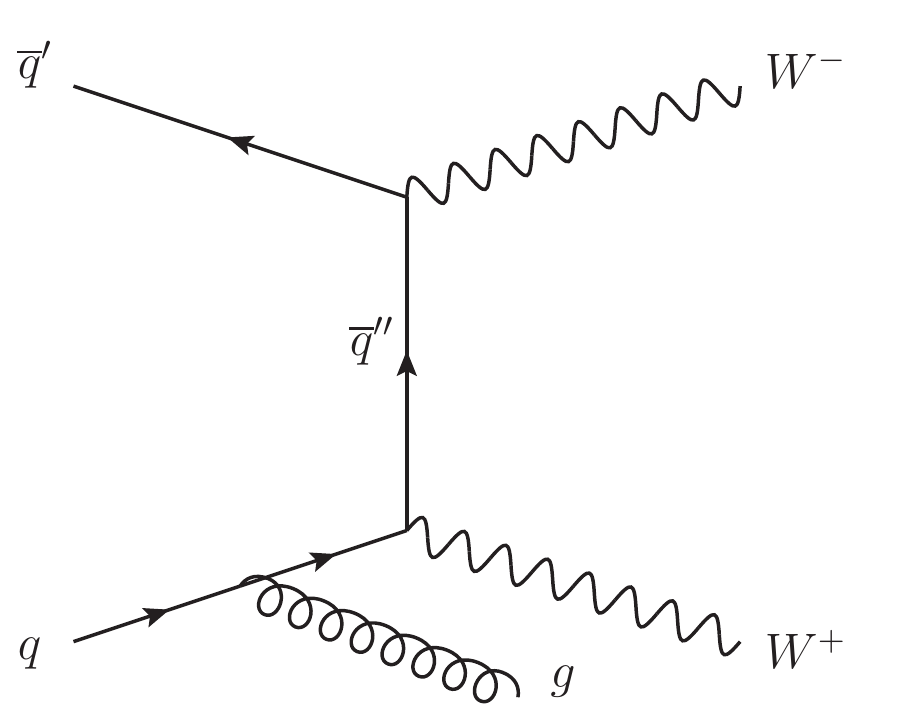}
\includegraphics[width=.3\linewidth]{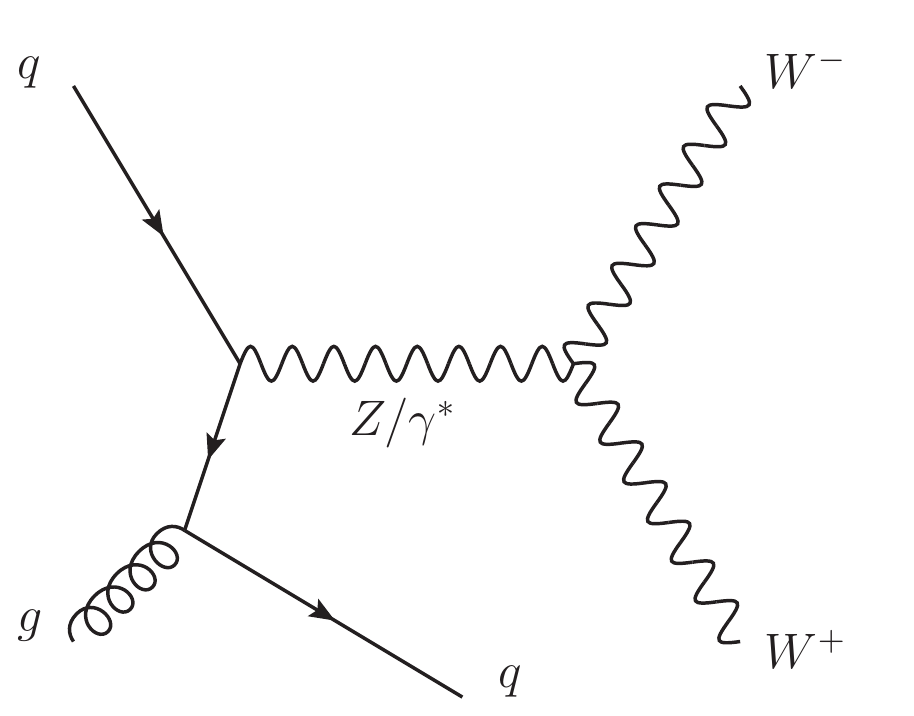}
\includegraphics[width=.3\linewidth]{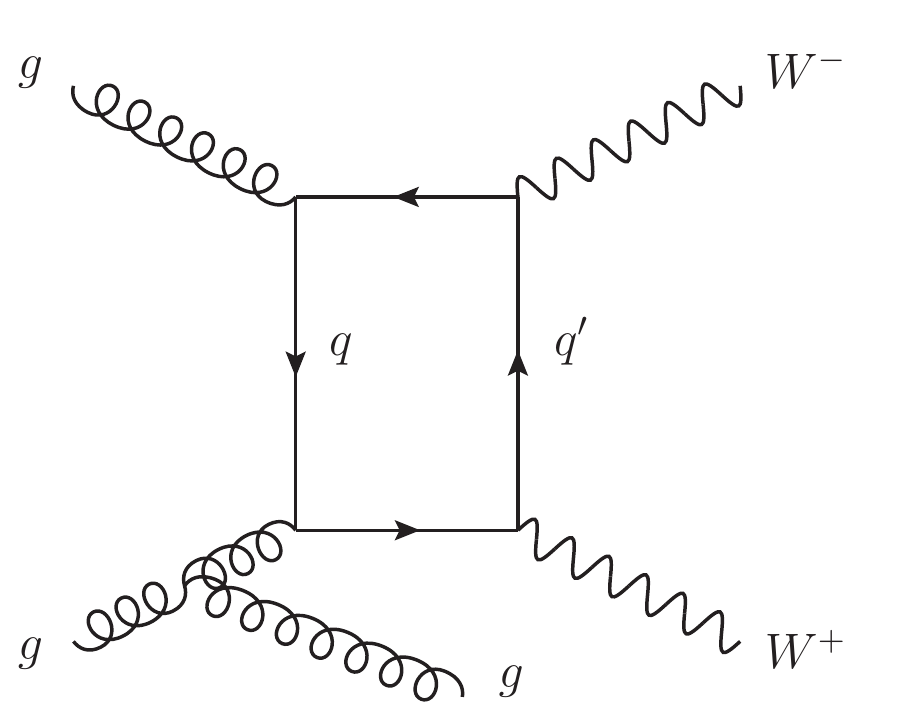}
\caption{Feynman diagrams for the production of a $W^+W^-$ boson pair in association with a jet.}
\label{fig:feyn}
\end{figure}

In $pp$ collisions, two leading processes contribute to $WW$ production: \qqWW in the $t$- and $s$-channel, and loop-induced gluon--gluon fusion processes \ggWW. Beyond leading order in perturbation theory and in particular for \WWj production, additional partonic initial states can contribute to both processes.\footnote{Even though different partonic initial states contribute to both processes, the notation \ggWW is used to identify the loop-induced gluon--gluon fusion channel while \qqWW describes the dominant production mode.} Representative diagrams for $WW\text{+jet}$ production are shown in Figure~\ref{fig:feyn}. In this analysis, the resonant \ggHWW production is included in the signal definition and simulation, although the process is strongly suppressed via kinematic selection requirements.
 
The measurement of $\WW\rightarrow e^\pm \nu \mu^\mp \nu$ production cross-sections at $\sqrt{s}=13\,\TeV$ is performed, using $pp$ collision data recorded by the ATLAS experiment in 2015--2018, corresponding to an integrated luminosity of 139\,\ifb.
The number of events due to top-quark pair production (\ttbar), the largest background for this measurement, is reduced by rejecting events containing jets from $b$-hadron decays (\bjets).
However, the \ttbar background is still sizeable due to the requirement that events contain at least one jet, and a data-driven method is required to reduce its contribution to systematic uncertainties in the measurement. This is achieved by simultaneously measuring the number of \ttbar events and the efficiency of identifying \bjets in these events. The procedure reduces the impact of systematic uncertainties associated with the modelling of \ttbar events and the \btagging efficiency calibration, and provides a precise and accurate estimate of the background up to partonic centre-of-mass energies of the order of 1~\TeV\ and for up to five jets.

\begin{samepage}
The measurement is performed in a fiducial phase space close to the geometric and kinematic acceptance of the experimental analysis.
The cross-section of $\WW$ production is measured differentially as a function of:
\begin{itemize}
\item the transverse momentum\footnote{\AtlasCoordFootnote} of the leading lepton, \ptleadlep,
\item the transverse momentum of the sub-leading lepton, \ptsublep,
\item the transverse momentum of the leading jet, \ptleadjet,
\item the jet multiplicity,
\item the invariant mass of the lepton pair, \mll,
\item the transverse momentum of the lepton pair, \ptll,
\item the scalar sum of all jet transverse momenta, \HT,
\item the scalar sum of all jet and lepton transverse momenta, \ST,
\item the transverse mass of the dilepton system and the missing transverse momentum\footnote{The transverse mass is defined as $\mtll = \sqrt{ (E_{\mathrm{T},e\mu} + \met)^2 - (\vec{p}_{\mathrm{T},e\mu} + \vec{p}_{\mathrm{T}}^{\mathrm{\,miss}} )^2}$, where $E_{\mathrm{T},e\mu} = \sqrt{|\vec{p}_{\mathrm{T},e\mu}|^2 + m_{e\mu}^2}$ and \met is the magnitude of the missing transverse momentum.}, \mtll,
\item the rapidity of the dilepton system, \yll,
\item the azimuthal separation of the two leptons, \dphill, and
\item $\costhetastar = | \tanh(\detall/2)|$\ , which is sensitive to the spin structure of the $W$-boson pair~\cite{Barr:2005dz}.
\end{itemize}
\end{samepage}
 
These observables comprehensively characterize $W$-boson kinematics and jet production in $WW$ events.
To facilitate an anomalous coupling interpretation that is less plagued by the aforementioned interference suppression, the differential cross sections as a function of \mll and \dphill are also measured for  $\ptleadjet > 200\,\GeV$, where the jet \pt threshold is chosen as a compromise between increased interference and good measurement precision.
Additional measurements with $\ptleadlep > 200\,\GeV$ are presented in Appendix~\ref{app:lep200}.
 
\section{The ATLAS detector}
\label{sec:atlas}

The ATLAS experiment~\cite{PERF-2007-01} at the LHC~\cite{Evans:2008zzb} is a multipurpose particle detector
with a forward--backward symmetric cylindrical geometry and a near \(4\pi\) coverage in
solid angle.
It consists of an inner tracking detector surrounded by a thin superconducting solenoid
providing a \SI{2}{\tesla} axial magnetic field, electromagnetic and hadron calorimeters, and a muon spectrometer with three large superconducting toroidal magnets with eight coils each.
 
The inner tracking detector covers the pseudorapidity range \(|\eta| < 2.5\). It consists of a high-granularity silicon pixel detector, including the insertable B-layer installed before Run~2~\cite{ATLAS-TDR-2010-19,PIX-2018-001}, followed by the silicon microstrip tracker. The silicon detectors are complemented by a transition radiation tracking detector, enabling extended track reconstruction within $|\eta| < 2.0$.
 
Lead/liquid-argon (LAr) sampling calorimeters provide electromagnetic (EM) energy measurements
with high granularity.
A steel/scintillator-tile hadron calorimeter covers the central pseudorapidity range (\(|\eta| < 1.7\)).
The endcap and forward regions are instrumented with copper/LAr and tungsten/LAr calorimeters for EM and hadronic energy measurements up to \(|\eta| = 4.9\).
 
The muon spectrometer surrounds the calorimeters and is based on
three large air-core toroidal superconducting magnets with eight coils each.
The field integral of the toroids ranges between \num{2.0} and \num{6.0}\,Tm.
across most of the detector.
The muon spectrometer includes a system of precision tracking chambers and fast detectors for triggering.
 
Events are selected using a two-level trigger system.
The first-level trigger is implemented in hardware and uses a subset of the detector information
to accept events at a rate of about \SI{100}{\kHz}.
The level-1 trigger is followed by a software-based trigger that
reduces the accepted event rate to \SI{1}{\kHz} on average
depending on the data-taking conditions.
 
 
\section{Data and Monte Carlo samples}
\label{sec:data_mc_samples}
 
The analysis uses data collected in proton--proton collisions at a centre-of-mass energy of 13\,\TeV from 2015 to 2018. After applying data quality criteria~\cite{DAPR-2018-01}, the dataset corresponds to 139\,\ifb, with an uncertainty of  1.7\%~\cite{ATLAS-CONF-2019-021}, obtained using the LUCID-2 detector~\cite{LUCID2} for the primary luminosity measurements.
 
Monte Carlo (MC) simulated event samples are used to correct the signal yield for detector effects and to estimate background contributions. All samples were passed through a full simulation of the ATLAS detector~\cite{SOFT-2010-01}, based on $\GEANT4$~\cite{Agostinelli:2002hh}. Table~\ref{tab:nominal-mc-samples} lists the configuration for the nominal MC simulation used in the analysis.

Signal events were modelled using the \SHERPAV{2.2.2}~\cite{Bothmann:2019yzt} generator at next-to-leading order (NLO) accuracy in QCD for up to one additional parton, and leading-order (LO) accuracy for two to three additional parton emissions for $q\bar q$ initial states.
The matrix element calculation of $gg\rightarrow WW$ production, which includes off-shell effects and Higgs boson contributions, incorporates up to one additional parton emission at LO. It was matched
and merged with the \SHERPA parton shower based on Catani--Seymour
dipole~\cite{Gleisberg:2008fv,Schumann:2007mg} using the MEPS@NLO
prescription~\cite{Hoeche:2011fd,Hoeche:2012yf,Catani:2001cc,Hoeche:2009rj}.
The virtual QCD corrections were provided by the
\openloops\ library~\cite{Cascioli:2011va,Denner:2016kdg}. The
NNPDF3.0NNLO set of parton distribution functions~(PDF) was used~\cite{Ball:2014uwa}, along with the
dedicated set of tuned parton-shower parameters developed by the
\SHERPA authors.

To assess the uncertainty in the matrix element calculation and the parton shower modelling, alternative events for \qqWW production were generated using the \POWHEGBOXV{v2}~\cite{Nason:2004rx,Frixione:2007vw,Alioli:2010xd,Nason:2013ydw} generator at NLO accuracy in QCD.  Events were interfaced to \PYTHIAV{8.186}~\cite{Sjostrand:2007gs}
for the modelling of the parton shower, hadronization, and underlying
event, with parameter values set according to the AZNLO
set of tuned parameters~\cite{STDM-2012-23}. The CT10nlo set PDF~\cite{Lai:2010vv} was used
for the hard-scattering processes, whereas the CTEQ6L1 PDF
set~\cite{Pumplin:2002vw} was used for the parton shower. The events were normalized to the next-to-next-to-leading order (NNLO) cross-section~\cite{Gehrmann:2014fva}.
For the \ggWW initial state, which makes up only 5\% of the signal, no alternative simulation is used.
 
The production of \ttbar\ and single-top $Wt$ events was modelled using the
\POWHEGBOXV{v2}~\cite{Frixione:2007nw,Nason:2004rx,Frixione:2007vw,Alioli:2010xd}
generator at NLO with the NNPDF3.0NLO~\cite{Ball:2014uwa} PDF. The events were interfaced
to \PYTHIAV{8.230}~\cite{Sjostrand:2014zea} to model the parton shower,
hadronization, and underlying event, with the A14 set of tuned parameters~\cite{ATL-PHYS-PUB-2014-021} and using the NNPDF2.3LO
set of PDFs~\cite{Ball:2012cx}. For \ttbar event generation, the \hdamp\ parameter\footnote{The
\hdamp\ parameter is a resummation damping factor and one of the
parameters that control the matching of \powheg matrix elements to
the parton shower and thus effectively regulates the
high-\pt\ radiation against which the \ttbar\ system recoils.} was set
to $1.5 \mtop$~\cite{ATL-PHYS-PUB-2016-020}. The diagram-removal scheme~\cite{Frixione:2008yi}
was employed to handle the interference between the $Wt$ and \ttbar\ production processes~\cite{ATL-PHYS-PUB-2016-020}.
Alternative samples were generated to assess the uncertainties in the top-background modelling.
The uncertainty due to initial-state radiation and higher-order QCD effects was estimated by
simultaneous variations of the \hdamp\ parameter and the renormalization and
factorization scales, and by choosing the Var3c up/down variants of the A14 set of tuned parameters
as described in Ref.~\cite{ATL-PHYS-PUB-2017-007}. The impact of
final-state radiation was evaluated with weights that account for
the effect of varying the renormalisation scale for final-state
parton-shower emissions up or down by a factor two.
To assess the dependence on the \ttbar--$Wt$ overlap removal scheme, the diagram-subtraction scheme~\cite{Frixione:2008yi}
was employed as an alternative to the diagram-removal scheme.
The uncertainty due to the parton shower and hadronization model was
evaluated by comparing the nominal sample of events with an event sample
generated by \POWHEGBOXV{v2} and interfaced to
\HERWIGV{7.04}~\cite{Bahr:2008pv,Bellm:2015jjp}, using the H7UE set
of tuned parameters~\cite{Bellm:2015jjp} and the MMHT2014LO PDF set~\cite{Harland-Lang:2014zoa}.
To assess the uncertainty in the matching of NLO matrix elements to the
parton shower, the nominal sample was compared with a sample generated
by \MGMCatNLOV{2.6.2}~\cite{Alwall:2014hca} at NLO in QCD using the five-flavour
scheme and the NNPDF2.3NLO PDF set. The events were
interfaced with \PYTHIAV{8}, as for the nominal sample. The \ttbar\ sample was normalized to the cross-section prediction at NNLO QCD.
in QCD including the resummation of next-to-next-to-leading logarithmic (NNLL) soft-gluon terms calculated using
\textsc{Top++2.0}~\cite{Beneke:2011mq,Cacciari:2011hy,Baernreuther:2012ws,Czakon:2012zr,Czakon:2012pz,Czakon:2013goa,Czakon:2011xx}. The inclusive cross-section for single-top $Wt$ was corrected to the theory prediction
calculated at NLO in QCD with NNLL soft-gluon corrections~\cite{Kidonakis:2010ux,Kidonakis:2013zqa}.
 
The background due to $Z/\gamma^*$+jets production was simulated with the
\SHERPAV{2.2.1} generator using NLO-accurate matrix elements for up to two jets, and LO-accurate matrix elements for three and four jets calculated with the Comix~\cite{Gleisberg:2008fv} and OpenLoops libraries.
They were matched with the \SHERPA parton shower~\cite{Schumann:2007mg} using the MEPS@NLO
prescription~\cite{Hoeche:2011fd,Hoeche:2012yf,Catani:2001cc,Hoeche:2009rj}
and the set of tuned parameters developed by the \SHERPA authors.
The NNPDF3.0NNLO set of PDFs was used, and the samples
were normalised to a NNLO prediction~\cite{Anastasiou:2003ds}.
To assess the uncertainties in modelling the $Z$+jets process, an alternative sample was simulated using LO-accurate matrix elements with up to four final-state partons with \MGMCatNLOV{2.2.2}, with the NNPDF2.3LO set of PDFs. Events were
interfaced to \PYTHIAV{8.186} using the A14 set of tuned parameters. The overlap between matrix-element and parton-shower emissions was removed using the CKKW-L merging procedure~\cite{Lonnblad:2001iq,Lonnblad:2011xx}.
The inclusive cross-section of both the nominal simulation and the alternative simulation was corrected to the theory prediction calculated at NNLO in QCD.
 
The production of $WZ$, $ZZ$, $V\gamma$ (with $V=W,Z$) and triboson ($VVV$, on-shell) final states was simulated with the
\SHERPAV{2.2.2} and \SHERPAV{2.2.8} generators using \openloops\ at NLO QCD accuracy for up to one additional parton and LO accuracy for two to three additional parton emissions, matched and merged with
the \SHERPA parton shower. The $VZ$ simulation includes $V\gamma^*$ contributions for $m(\ell\ell) > 4\,\GeV$. Samples were generated using the  NNPDF3.0NNLO PDF set and normalized to the cross-section calculated by the event generator.
Alternative samples for diboson backgrounds with $WZ$ or $ZZ$ production were generated in the same way as the nominal signal sample: the default \SHERPA simulation was exchanged for \powheg + \PYTHIAV{8}, using NLO-accurate matrix elements. The \powheg diboson cross-section was scaled to NNLO~\cite{Grazzini:2016swo,Grazzini:2017ckn,Grazzini:2015hta,Cascioli:2014yka}, while the cross-section calculated by \SHERPA was found to be in good agreement with the NNLO value.
 
Samples generated with \POWHEGBOX or \MGMCatNLO used the \evtgen~1.2.0 or 1.6.0 program~\cite{EvtGen} to model  the decay of bottom and charm hadrons.
The effect of multiple interactions in the same and neighbouring bunch crossings (pile-up) was modelled by overlaying the hard-scattering event with simulated inelastic $pp$ events generated with \PYTHIAV{8.186} using the NNPDF2.3LO set of PDFs and the A3 set of tuned parameters~\cite{ATL-PHYS-PUB-2016-017}.
 
\OnlySupportNote{
Appendix~\ref{app:mcsets} lists all used datasets.
}
 
\begin{table}
\caption{Summary of the nominal Monte Carlo simulated samples used in the analysis. The $gg\rightarrow WW$ simulation includes Higgs boson contributions. The last two columns give the order in $\alpha_{\textrm{S}}$ of the matrix element calculation and the overall cross-section normalization. All nominal MC samples use the NNPDF3.0 PDF set. The samples generated with \SHERPA use the default set of tuned parton-shower parameters, while for the \POWHEGBOX samples the A14 set of tuned parameters and the NNPDF2.3LO PDF set are used for the parton shower.}
\label{tab:nominal-mc-samples}
\centering
\begin{tabular}{lllll}
\toprule
Process & Generator & Parton shower &  Matrix element $\mathcal{O}(\alpha_\text{S})$& Normalization \\ \midrule
$q\bar{q}\rightarrow WW$ & \SHERPAV{2.2.2}  & \SHERPA & NLO (0--1 jet), LO (2--3 jets)& Generator$^\dagger$ \\
$gg\rightarrow WW$ &  \SHERPAV{2.2.2}  & \SHERPA & LO (0--1 jet) & Generator \\
\ttbar & \POWHEGBOXV{v2} & \PYTHIAV{8} & NLO & NNLO+NNLL \\
$Wt$ & \POWHEGBOXV{v2} & \PYTHIAV{8} & NLO & NLO+NNLL \\
$Z$+jets & \SHERPAV{2.2.1} & \SHERPA & NLO (0--2 jets), LO (3--4 jets) & NNLO\\
$WZ, ZZ$ & \SHERPAV{2.2.2}  & \SHERPA & NLO (0--1 jet), LO (2--3 jets)& Generator$^\dagger$ \\
$W\gamma,Z\gamma$ & \SHERPAV{2.2.8}  & \SHERPA & NLO (0--1 jet), LO (2--3 jets)& Generator$^\dagger$ \\
$VVV$ & \SHERPAV{2.2.2}  & \SHERPA & NLO (0--1 jet), LO (2--3 jets)& Generator$^\dagger$ \\
\bottomrule
\end{tabular}\\
\raggedleft
{\footnotesize$\dagger$: The cross-section calculated by \SHERPA is found to be in good agreement with the NNLO result~\cite{Grazzini:2016swo,Grazzini:2017ckn,Grazzini:2015hta,Cascioli:2014yka,Grazzini:2016ctr}.}
\end{table}
 
 
\section{Event reconstruction and selection}
\label{sec:event-selection}
 
Candidate \WW events are selected by requiring exactly one isolated electron and one isolated muon with opposite charges. Events with two isolated leptons of the same flavour are not considered in the analysis due to the higher background from Drell--Yan events.
 
Events were recorded by either single-electron or single-muon triggers~\cite{TRIG-2018-05,TRIG-2018-01}. The minimum \pt threshold varied during data-taking between 24\,\GeV and 26\,\GeV for electrons, and between 20\,\GeV and 26\,\GeV for muons, both requiring `loose' to `medium' isolation criteria. Triggers with higher \pt thresholds and looser isolation requirements are also used to increase the efficiency. The trigger selection efficiency is more than 99\% for signal events fulfilling all other selection requirements, which are detailed below.
 
Candidate events are required to have at least one vertex having at least two associated tracks with $\pt > 400\,\MeV$. The vertex with the highest $\sum \pt^2$ of the associated tracks is taken as the primary vertex.
 
Electrons are reconstructed from energy deposits in the calorimeter that are matched to tracks~\cite{EGAM-2018-01}. Electron candidates are required to fulfil the `tight' likelihood-based identification criteria as defined in Ref.~\cite{EGAM-2018-01}. Furthermore, they are required to have $E_{\mathrm{T}} > 27\,\GeV$ and $|\eta|<2.47$, excluding the transition region between barrel and endcap regions, $1.37 < |\eta| < 1.52$.
 
Muon candidates are reconstructed by combining a track in the inner detector (ID) with a track in the muon spectrometer~\cite{MUON-2018-03}. Muons are required to have $\pt > 27\,\GeV$ and $|\eta| < 2.5$ and to satisfy the \texttt{Medium} identification selection, as defined in Ref.~\cite{MUON-2018-03}.
 
Leptons are required to be compatible with the primary vertex by imposing requirements on the impact parameters of associated tracks. The transverse impact parameter significance, $d_0 / \sigma_{d_0}$ is required to satisfy $|d_0 / \sigma_{d_0}|  < 5\,(3)$ for electrons (muons). The longitudinal impact parameter $z_0$ must satisfy $|z_0\cdot\sin\theta| < \SI{0.5}{\milli\meter}$, where $\theta$ is the polar angle of the track. Additionally, leptons are required to be isolated using information from the ID tracks and energy clusters in the calorimeters in a cone around the lepton. The \texttt{Gradient} working point is used for electrons~\cite{EGAM-2018-01}, while for muons the \texttt{Tight\_FixedRad} working point is used, which is similar to the \texttt{Tight} selection defined in Ref.~\cite{PERF-2015-10} but with altered criteria at muon $\pt > 50\,\GeV$ in order to increase the background rejection. The electron or muon trigger object is required to match the respective reconstructed lepton.
 
Jets are reconstructed using the anti-$k_t$ algorithm~\cite{Cacciari:2008gp} with a radius parameter of $R=0.4$ using particle-flow objects~\cite{PERF-2015-09}. They are required to have $\pt > 30\,\GeV$ and $|\eta| < 4.5$.  To suppress jets that originate from pile-up, a  jet-vertex tagger~\cite{PERF-2014-03}  is applied to jets with $\pt < 60\,\GeV$ and $|\eta|<2.4$. Jet energy scale and resolution are corrected with $\eta$- and \pt dependent scale factors~\cite{JETM-2018-05}.
Jets with $\pt > 20\,\GeV$ and $|\eta|<2.5$ containing decay products of a $b$-hadron are identified using the DL1r \btagging algorithm~\cite{ATL-PHYS-PUB-2017-013,FTAG-2018-01} at the 85\% efficiency working point.
 
The missing transverse momentum, with magnitude \met, is computed as the negative of the vectorial sum of the transverse momenta of tracks associated with jets and muons, as well as tracks in the ID that are not associated with any other component. The \pt of the electron track is replaced by the calibrated transverse momentum of the reconstructed electron~\cite{PERF-2016-07}.
 
\phantomsection\label{par:overlap-removal}
In order to resolve the overlap between particles reconstructed as multiple physics objects in the detector, non-\btagged jets are removed if they overlap, within $\Delta R < 0.2$, with an electron, or with a muon if the jet has less than three associated tracks with $\pt > 500\,\MeV$ and satisfies $\pt^\mu / \pt^{\mathrm{jet}} > 0.5$, and the ratio of the muon \pt to the sum of the track \pt associated with the jet is greater than 0.7.
Electrons or muons overlapping within $\Delta R < 0.4$  with any jet, including \btagged jets, after the former selection are removed.
 
Events having at least one jet, but no \btagged jets, are selected for the analysis. To reduce the Drell--Yan backgrounds, dominated by $Z$+jets events with $Z\rightarrow\tau^+\tau^-$ decays, the invariant mass of the electron--muon pair is required to be $m_{e\mu} > 85\,\GeV$. This requirement also reduces the contribution of resonant $gg \rightarrow H \rightarrow WW$ production. Events with additional leptons with $\pt > 10\,\GeV$ and satisfying \texttt{Loose} isolation and \texttt{LooseLH} (\texttt{Loose}) identification requirements for electrons (muons), are vetoed to reduce backgrounds due to $WZ$ and $ZZ$ production.
Additionally, the subsets of events with
high leading-jet transverse momentum, $\ptleadjet > 200\,\GeV$, are analysed in detail, to investigate the reduced interference-suppression in the aTGC interpretation.
Table~\ref{tab:selection} gives a summary of the lepton, jet and event selection requirements used to define the signal region.

\OnlySupportNote{
Details on the object and event selection, and their optimisation, are given in Appendices~\ref{app:object-selection} and ~\ref{app:event-selection}.
}
 
\begin{table}
\caption{Summary of the object and event selection criteria.}\label{tab:selection}
\centering
\begin{tabular}{rl}
\toprule
Selection & Criteria \\
\midrule
Lepton $\pt$ & $>27\,\GeV$ \\
Lepton $\eta$ & $|\eta|<2.47$ and not $1.37 < |\eta| < 1.52$ (electron) \\ & $|\eta| < 2.5$ (muon) \\
Lepton identification & \texttt{TightLH} (electron), \texttt{Medium} (muon) \\
Lepton isolation & \texttt{Gradient} (electron), \texttt{Tight\_FixedRad} (muon) \\
Lepton impact parameter & 		$|d_0 / \sigma_{d_0}| < 5, 3$ (electron, muon)  \\
& $|z_0 \cdot \sin\theta| < 0.5$\,\si{\milli\meter} \\
 
Jet selection & $\pt > 30\,\GeV$, $|\eta| < 4.5$ \\
\bjet selection & $\pt > 20\,\GeV, |\eta|<2.5$, DL1r (85\% eff.\ WP)\\
\midrule
Lepton selection & 1 electron and 1 muon of opposite charge,\\ &no additional lepton with $\pt > 10\,\GeV$, \texttt{Loose} isolation,\\& and \texttt{LooseLH} (electron) / \texttt{Loose} (muon) identification \\
Number of jets & $\ge 1$ \\
Number of \bjets& 0\\
$m_{e\mu}$ & $>85\,\GeV$ \\
\midrule
High \ptleadjet selection & $ \ptleadjet > 200\,\GeV$ \\
\bottomrule
\end{tabular}
\end{table}
 
 
 
\section{Background estimate}
\label{sec:background}
 
The top-quark background, from either \ttbar or single-top $Wt$ production, comprises about 60\% of the events passing the event selection and about 90\% of the total background. Additional backgrounds considered are $Z$+jets production, events with non-prompt or misidentified leptons, diboson production ($WZ$, $W\gamma$, $ZZ$, and $Z\gamma$), and triboson production.
 
\subsection{Top-quark background}
An estimate of the \ttbar background is obtained using a data-driven technique, while the single-top $Wt$ background is estimated using simulation and is found to contribute about 16\% of the top-quark background.
Following the procedure used in a measurement of the \ttbar cross-section \iftoggle{paperdraft}{\cite{TOPQ-2018-17}}{\cite{TOPQ-2018-17,TOPQ-2018-17-int}}, two control regions requiring exactly one and exactly two \btagged jets are defined. All other selection criteria are the same as in the signal region.
These regions are dominated by \ttbar events and can be used to infer the number of \ttbar events in the signal region with minimal dependence on the selection and \btagging efficiencies.
The contribution of non-\ttbar events in the 1-\bjet and 2-\bjet control regions is 13\% and 4\% of the expected events, respectively, of which 90\% can be attributed to single-top $Wt$ production.
 
The numbers of \ttbar events in the two control regions, as well as in the signal region, are given by
\begin{align}
N_{1b}^{\ttbar} &= N_{1b} - N_{1b}^{\mathrm{others}} = \mathcal{L} \sigma_{\ttbar} \varepsilon_{e\mu} \cdot 2 \varepsilon_{b} \left(1 -  C_b \varepsilon_{b} \right) \ ,\label{eq:1b}\\
N_{2b}^{\ttbar} &= N_{2b} - N_{2b}^{\mathrm{others}} = \mathcal{L} \sigma_{\ttbar} \varepsilon_{e\mu} \cdot C_b \varepsilon_{b}^2 \ , \label{eq:2b}\\
N_{0b}^{\ttbar} &= \mathcal{L} \sigma_{\ttbar} \varepsilon_{e\mu} \cdot \left(1 - 2 \varepsilon_{b} +  C_b \varepsilon_{b}^2 \right) \ , \label{eq:0b}
\end{align}
where $N_{ib}$ and $N_{ib}^{\mathrm{others}}$ are, respectively, the number of selected events in data and the number of non-\ttbar events, estimated using simulation, with exactly $i$ \btagged jets. The term $\mathcal{L} \sigma_{\ttbar} \varepsilon_{e\mu}$ is the product of the integrated luminosity, the \ttbar cross-section, and the general selection efficiency, and $\varepsilon_{b}$ is the efficiency of selecting a \bjet in a \ttbar event. The correction factor $C_b=\varepsilon_{bb}/\varepsilon_{b}^2$ accounts for correlation effects between selecting one and two \bjets. It is determined from \ttbar simulation as $ C_b = 4 \cdot N_{\mathrm{MC}}^{\ttbar} N_{2b,\mathrm{MC}}^{\ttbar} / \left(N_{1b,\mathrm{MC}}^{\ttbar} + 2\cdot N_{2b,\mathrm{MC}}^{\ttbar}\right)^2$, and typically has values close to unity. The \bjet selection efficiency, $\varepsilon_{b}$, accounts for the efficiency of the \btagging algorithm and also for the acceptance of \bjets.
Using Eqs.~\eqref{eq:1b}--\eqref{eq:0b}, the number of \ttbar events in the signal region can be expressed as
\begin{equation*}
N_{0b}^{\ttbar} = \frac{C_b}{4} \frac{\left(N_{1b}^{\ttbar} + 2 N_{2b}^{\ttbar}\right)^2}{N_{2b}^{\ttbar}} - N_{1b}^{\ttbar} - N_{2b}^{\ttbar}\ ,
\end{equation*}
which depends only on $N_{ib}$ and $N_{ib}^{\mathrm{others}}$ ($i=1,2$), as well as $C_b$.
The \ttbar background estimate is performed in each analysis bin, i.e.\ for the fiducial selection as well as in each individual bin for the differential measurements. Because \btagged jets are selected with a lower \pt threshold than regular jets, this method also works for events with exactly one regular jet.
 
As the \ttbar background estimate is largely based on observed yields in data control regions and the only input from \ttbar simulation is the correlation factor $C_b$, this method strongly reduces experimental and theoretical uncertainties in the \ttbar background, and, thus, lowers the total uncertainty in the background by a factor of approximately five.
In regions of phase space where, for a large fraction of events, one or both \bjets are outside the detector acceptance, the reliance on \ttbar simulation for the extrapolation into the signal region increases. In such cases, the \ttbar estimate remains valid because modelling uncertainties cover rate and shape differences between data and simulation for \bjet kinematic distributions in the control region.
Uncertainties in the single-top $Wt$ production rate that are independent of the \bjet multiplicity, such as the cross-section uncertainty, partially cancel out because single-top $Wt$ is the dominant background to \ttbar in the \ttbar control regions. A variation leading to a larger $Wt$ prediction in the control regions reduces the \ttbar estimate, so if the same variation also leads to a larger $Wt$ prediction in the signal region, the overall effect on the combined top background is reduced.
The total uncertainty in the top background in the signal region is $2.8\%$.
 
The top background estimate is validated in a top-enriched subset of the signal region which requires $m_{\ell j} < 140\,\GeV$ and $\dphill < \pi/2$ in addition to the normal event selection. Here $m_{\ell j}$ is the invariant mass of the leading jet and the closest lepton. This region is approximately 70\% pure in top events and shows good agreement between the data and the combined signal and background prediction, which uses the data-driven top background estimate. The level of agreement of the prediction with the observed events in the control regions and the top-enriched selection is summarized in Table~\ref{tab:bkg-summary}.
Figure~\ref{fig:vr-datamc} shows the distributions of the \ptleadlep and the jet multiplicity, confirming the accurate modelling of lepton and jet-related properties in events without \bjets. 
 
\subsection{Drell--Yan background}
The Drell--Yan $Z$+jets background is estimated using MC simulation. The $\mll>85~\GeV$ requirement strongly suppresses this background by a factor of about nine. The contribution of this background to the selected events in the signal region is about 3\%, almost entirely due to $Z/\gamma^*\rightarrow\tau^+\tau^-$+jets events.
 
The $Z$+jets estimate is checked in a validation region requiring a dilepton invariant mass between 45\,\GeV and 80\,\GeV and either $\ptll< 30\,\GeV$ or $\met < 20\,\GeV$, in addition to the \bjet veto and the requirement of at least one jet with $\pt>30\,\GeV$ and $|\eta|<4.5$.
The $Z$+jets purity of this region is 75\% and good modelling of the data is observed, as shown in Table~\ref{tab:bkg-summary}.
Figure~\ref{fig:vr-datamc} shows the distribution of the dilepton invariant mass \mll in the validation region, which features the resonant $Z\rightarrow\tau\tau$ distribution over a rising background of top events.
 
In addition to the theoretical uncertainty in the $Z$+jets cross-section of  5\%~\cite{STDM-2015-03}, uncertainties are estimated by comparing the nominal MC simulation with events simulated by \mgamc.
This uncertainty estimate was found to bracket the effect of scale uncertainties.
In the signal region, the total uncertainty in the $Z$+jets background is about 30\%.
\OnlySupportNote{
Appendix~\ref{app:dy-bkg}  gives details to the $Z$+jets background validation.
}

\subsection{Backgrounds with non-prompt or misidentified leptons}
 
Reducible backgrounds from events with non-prompt or misidentified leptons are called fake-lepton backgrounds or `fakes'. Fake leptons correspond to leptons from heavy-flavour hadron decays and jets misidentified as electrons.
Fake-lepton events stem mainly from $W$+jets production and contribute about 3\% of the selected events. Top backgrounds with one prompt lepton contribute about 10\% of the fake-lepton backgrounds.

Fake-lepton backgrounds are estimated using a data-driven technique. A control region is defined, where one of the two lepton candidates fails the nominal selection with respect to the impact parameters and isolation criteria, but instead fulfils a looser set of requirements designed to increase the contribution of fake leptons. The fake-lepton background in the signal region is, then, obtained by scaling the number of data events in this control region by an extrapolation factor, after subtracting processes with two prompt leptons using simulation.
The extrapolation factor is determined in a data sample that is dominated by fake leptons, and it depends on the \pt, $|\eta|$, and flavour of the lepton. The data sample is selected by requiring events with a dijet-like topology with one lepton candidate recoiling against a jet, with $|\Delta \phi (\ell,j)|>2.8$. To suppress contamination from $W$+jets events in this sample, the sum of \met and the transverse mass of the lepton and \met system is required to be smaller than 50~\GeV.
The approach used closely follows the one applied in Ref.~\cite{STDM-2017-06}.
 
Systematic uncertainties in the composition of the different sources of fake leptons are estimated by varying the selection of the data sample in which the extrapolation factors are determined. The variations include selecting events with a \bjet recoiling against the lepton candidate, as well as changing the \met requirements to increase the fake-lepton contributions.
The normalization of the prompt-lepton background in the control region used for the extrapolation factor determination is varied by 10\%, which covers the largest discrepancies between simulation and data observed in a dedicated validation region.
An additional 25\% uncertainty in the fake-lepton background normalization covers a potential mismodelling of the identification efficiency of prompt leptons that fail the tight, but fulfil the looser, lepton identification requirements.
The uncertainty in the signal contamination in the control region, which is subtracted using simulation, is determined from the typical size of the largest deviations between the measured and predicted differential cross-sections, which is 20\%.
The total relative uncertainty in the fake-lepton background is about 40\%.

In order to validate the estimate of the fake-lepton backgrounds, the opposite-charge requirement of the signal region selection is inverted, and events with an electron--muon pair of the same charge are selected. As many processes leading to fake leptons are charge symmetric, while most Standard Model processes are not, this selection increases the contribution of $W$+jets events to about $25\%$.
The modelling of the fake-lepton backgrounds can be validated despite the relatively low purity
since the dominant diboson background in this region is known with a precision of about 10\%.
Reasonable agreement of the prediction with the data is observed, as is shown in Figure~\ref{fig:vr-datamc} in the \ptsublep distribution, and in Table~\ref{tab:bkg-summary} comparing the numbers of observed and predicted events.
\OnlySupportNote{
Appendix~\ref{app:fakefactor} gives the details of the non-prompt background estimate.
}
 
\subsection{Other backgrounds}
 
Backgrounds from $WZ$, $ZZ$, $W\gamma$ and $Z\gamma$ production are estimated from simulation, and are found to contribute about 3\% of the total selected events, dominated by $WZ$ events, which are observed to be well described by the nominal \SHERPA simulation in Ref.~\cite{STDM-2018-03}.
Uncertainties are derived by comparing the nominal simulation with events simulated by \powheg + \PYTHIAV{8}.
The difference in generator predictions was found to be larger than the impact of scale uncertainties in the Sherpa simulation, and thus the assigned uncertainty is the conservative option. Additionally, the uncertainty in the diboson cross-section of 10\% \cite{ATL-PHYS-PUB-2016-002,Grazzini:2015nwa} is included.
 
The $VZ$ ($WZ$ and $ZZ$) prediction is validated in events containing a third lepton having $\pt \ge 10$\,\GeV that must satisfy loosened identification criteria. The invariant mass of the resulting same-flavour opposite-charge pair of leptons is required to be between 80\,\GeV and 100\,\GeV, close to the $Z$ boson mass. These selections gives a very pure sample of diboson events, and the prediction is in good agreement with the data, as seen in Figure~\ref{fig:vr-datamc} and Table~\ref{tab:bkg-summary}. In Figure~\ref{fig:vr-datamc} the \met distribution in the $VZ$ validation region shows separation between $ZZ$ and $WZ$ events.
 
$V\gamma$ ($W\gamma$ and $Z\gamma$) events enter the signal region as backgrounds when the photon is reconstructed and selected as an electron candidate. To validate estimates of these backgrounds, the electron identification requirements are changed such that contributions from photon conversions increase. As the electron candidates reconstructed from photon conversion are charge symmetric, both opposite-charge and same-charge candidates are selected with respect to the selected muon.
For the $V\gamma$ validation region the \pt distribution of the electron candidates is shown in Figure~\ref{fig:vr-datamc}. It is dominated by electrons from photon conversion. Good agreement with the observed data in the validation regions is found.
 
Based on MC simulations, it is estimated that the triboson background contributes less than 0.1\% of the inclusive selected events and at most 0.5\% of the selected events in a single bin and is thus neglected in the analysis.
 
\OnlySupportNote{
Appendix~\ref{app:vvbkg} gives details on the diboson background validation, Appendix~\ref{app:vy} to the $V\gamma$ background, and Appendix~\ref{app:triboson} shows the fraction of triboson events in each analysis bin.
}
 
\subsection{Selected \WW candidate events}
 
\begin{table}[tb]
\caption{Summary of the observed and predicted events in the background control regions (CR) and validation regions (VR), and in the top-background enriched selection. The uncertainty in the prediction includes statistical and systematic effects, excluding theory uncertainties on the signal. The purity column gives the purity of the target process, relative to the total prediction. The \ttbar prediction in the two \ttbar control regions is from simulation, while in the top-enriched region the data-driven estimate is used.}
\label{tab:bkg-summary}
\centering
\begin{tabular}{lrr@{$\,\pm\,$}rr}
\toprule
Region         &   Observed &   Predicted &   Error~~ & Purity   \\
\midrule
$t \bar{t}$ CR 1b    & 260\,971 & 268\,000 & 19\,000 & 87\%      \\
$t \bar{t}$ CR 2b    & 257\,777 & 267\,000 & 21\,000 & 96\%      \\
Top enriched         & 7167 & 7000 & 1000 & 72\%      \\
Same-sign VR        & 5095 & 5000 & 600 & 25\%      \\
Drell--Yan VR   & 11\,824 & 13\,000 & 1600 & 74\%      \\
$V Z$ VR     & 14\,770 & 14\,000 & 1900 & 94\%      \\
$V \gamma$ VR (OS) & 2720 & 2670 & 240 & 63\%      \\
$V \gamma$ VR (SS) & 2401 & 2250 & 240 & 76\%      \\
\bottomrule
\end{tabular}
\end{table}
 
\begin{figure}
\centering
\includegraphics[width=.45\linewidth]{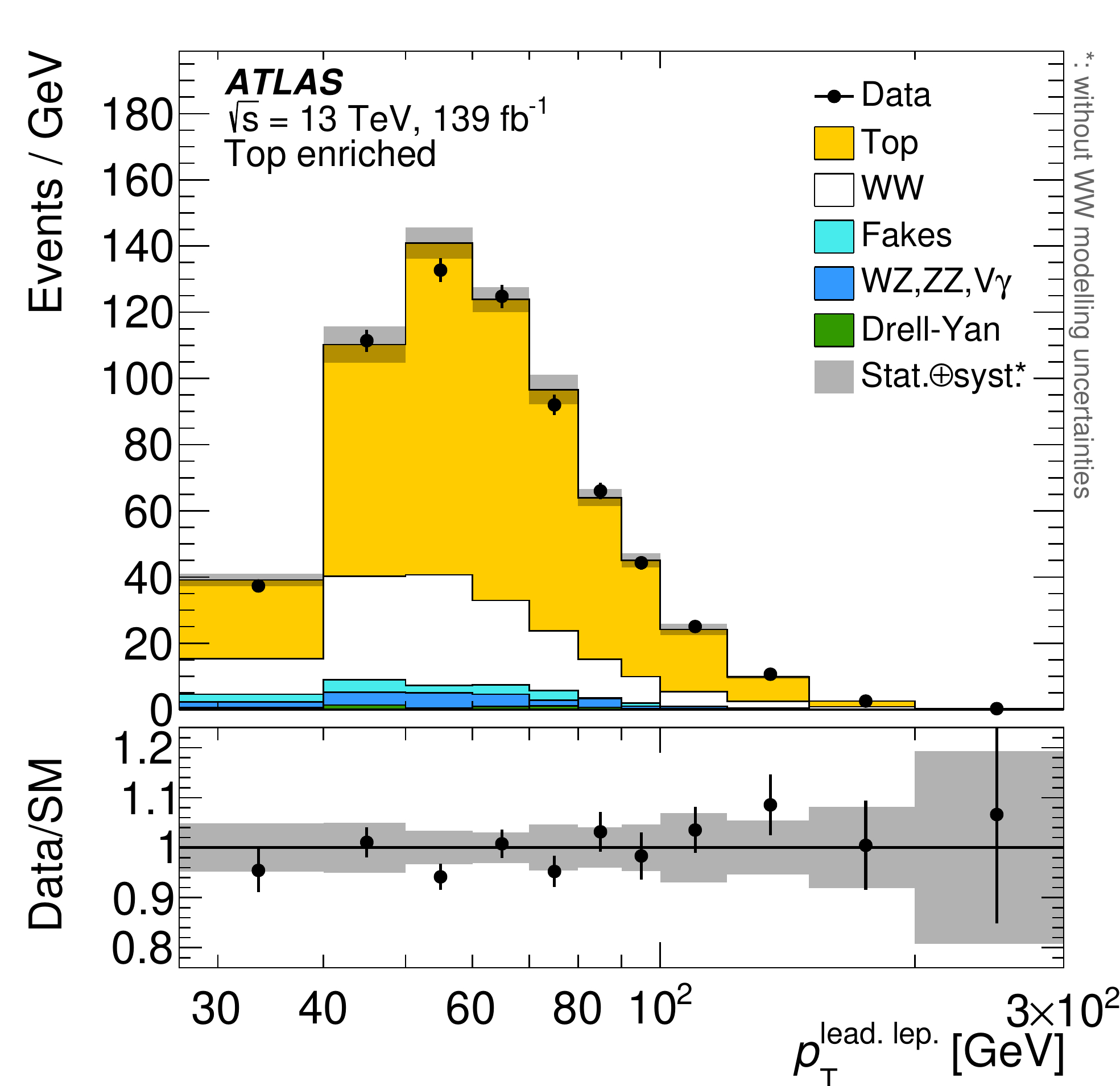}
\includegraphics[width=.45\linewidth]{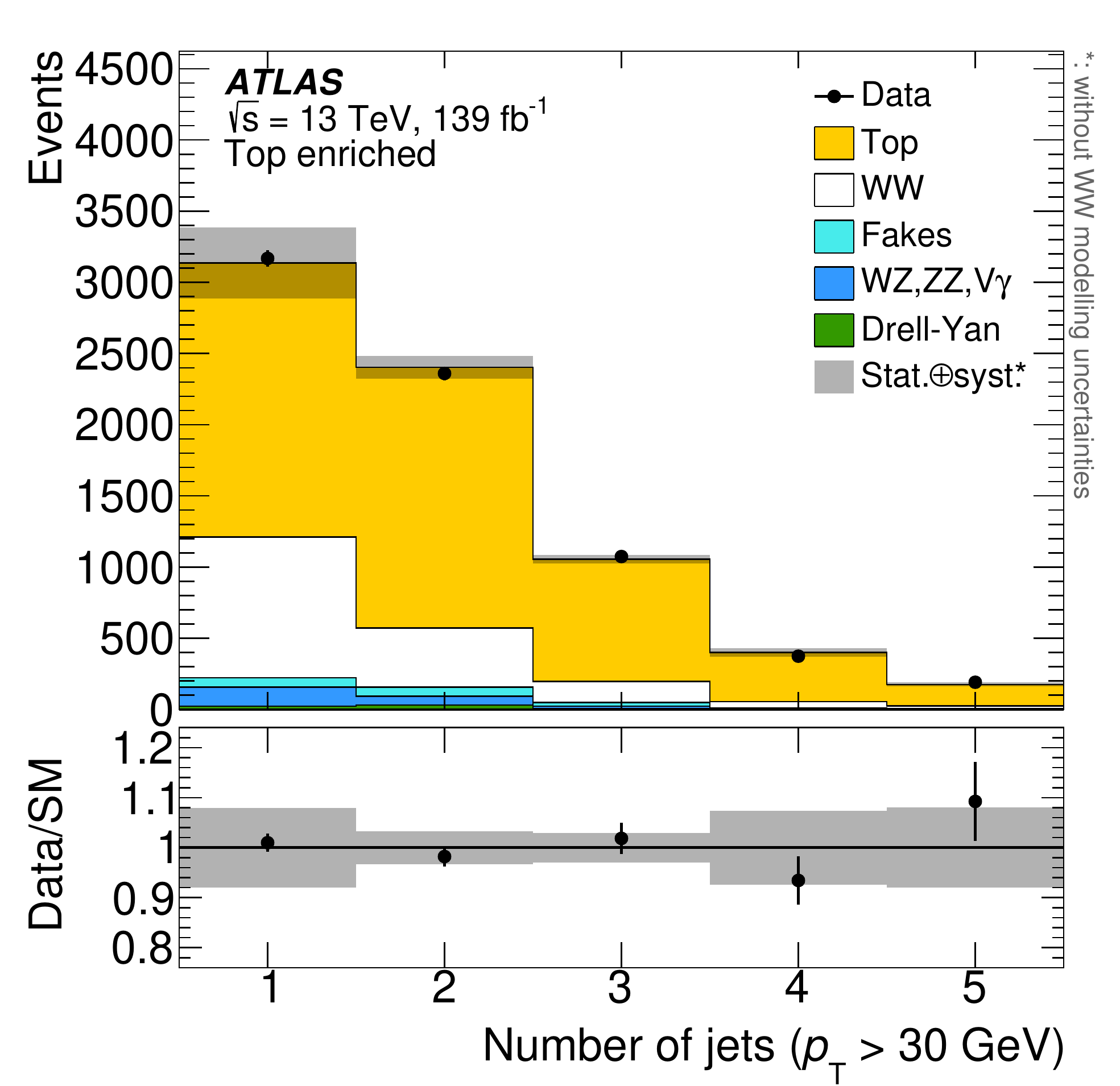}
\includegraphics[width=.45\linewidth]{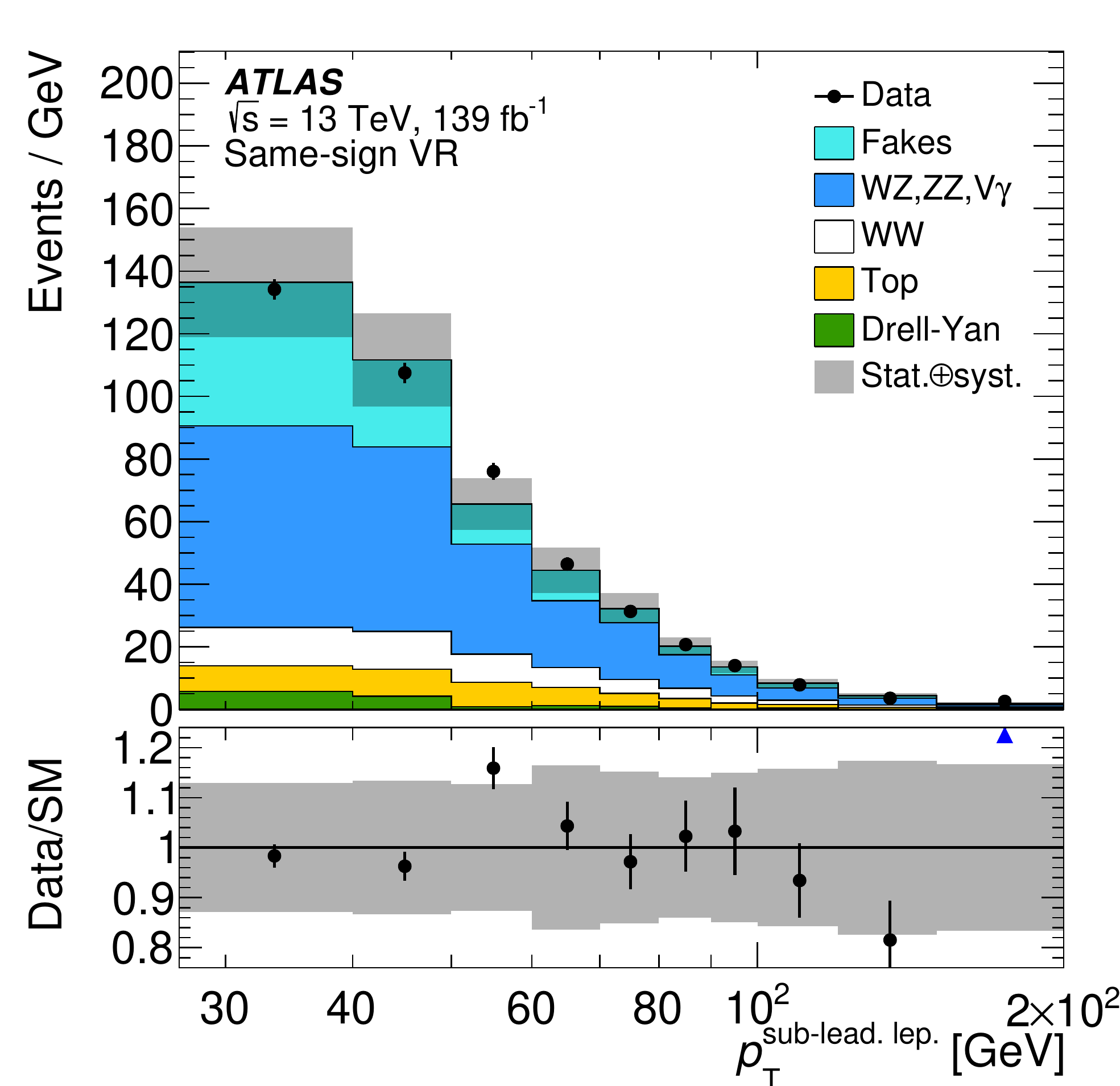}
\includegraphics[width=.45\linewidth]{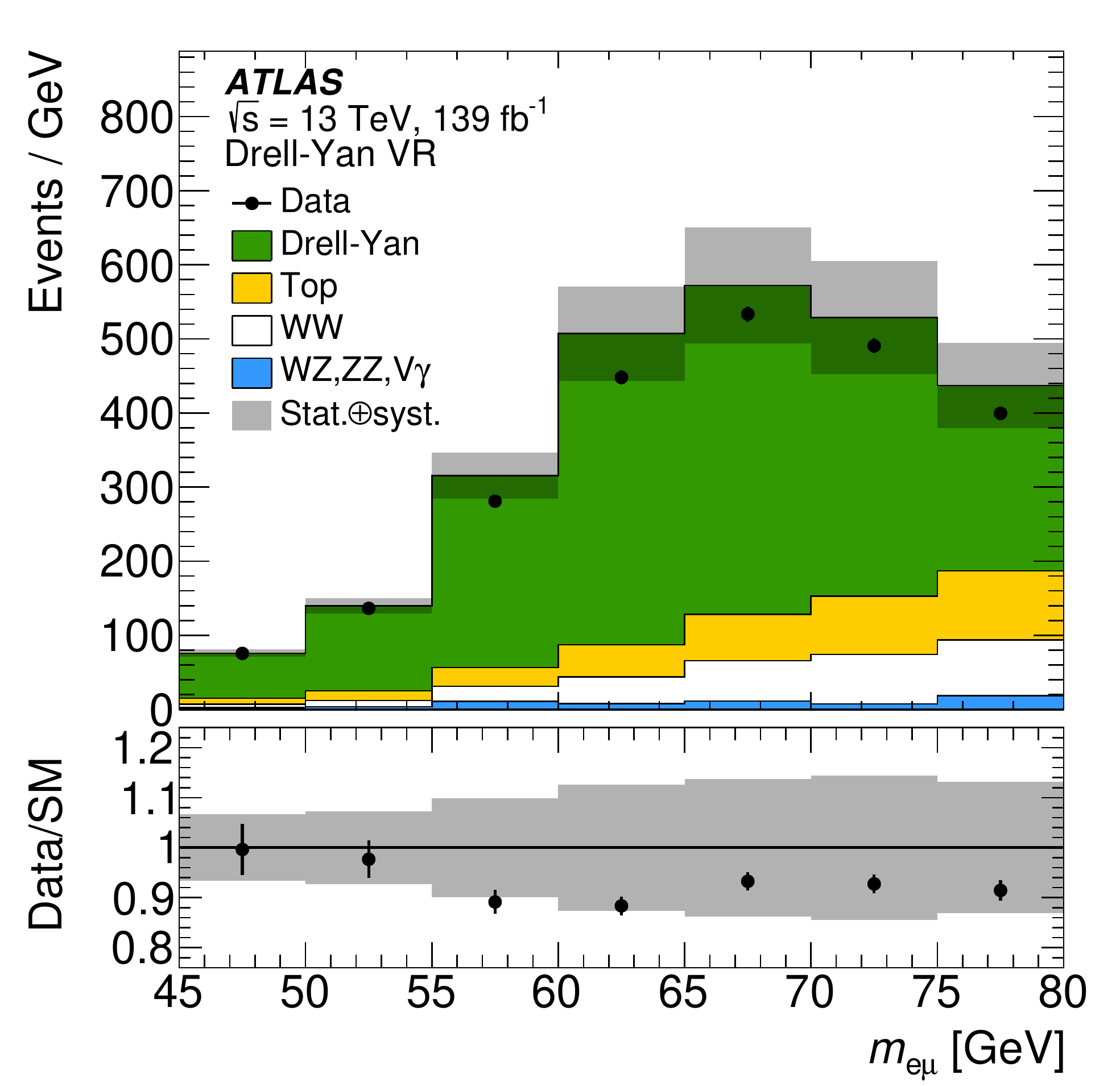}
\includegraphics[width=.45\linewidth]{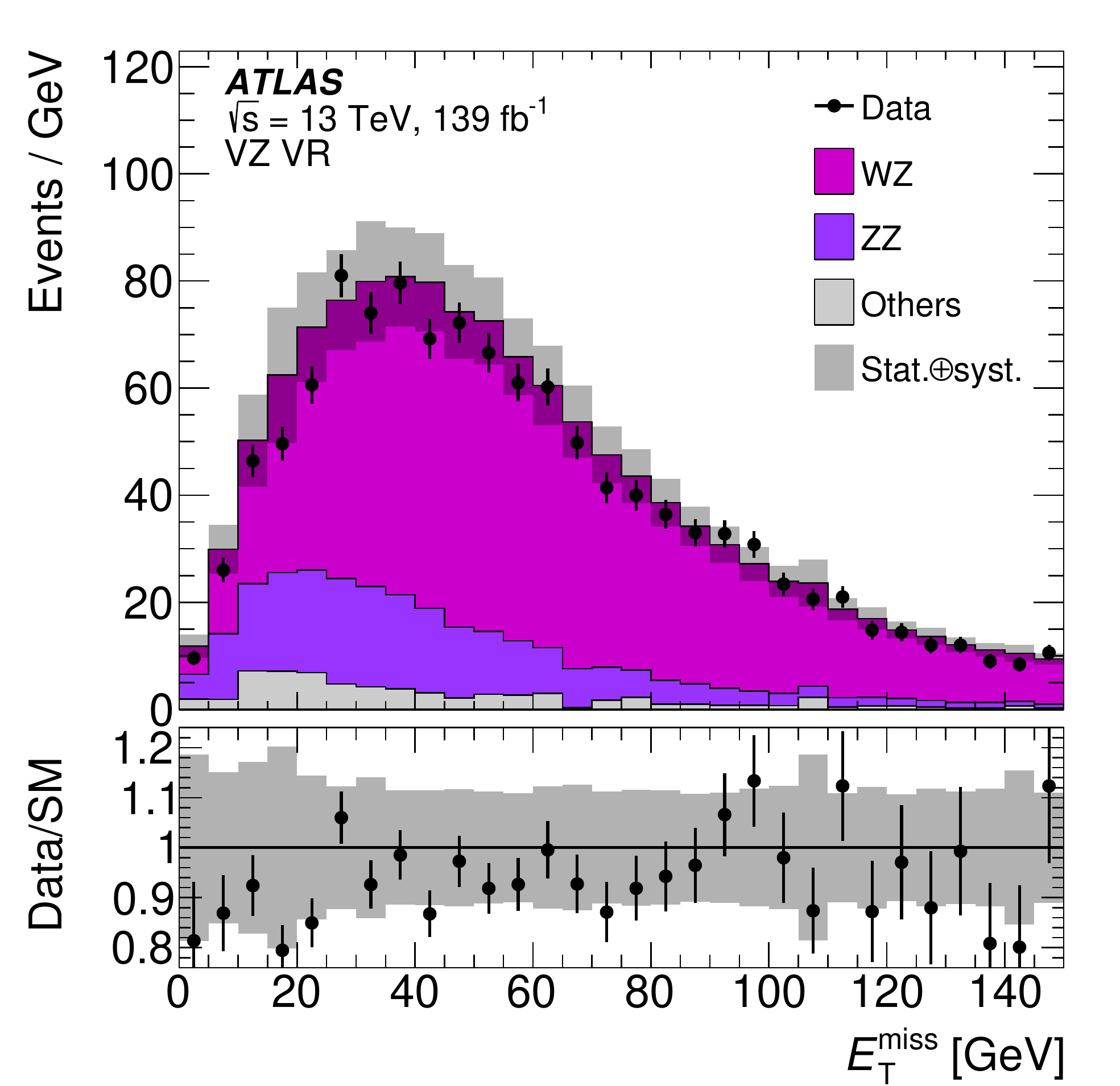}
\includegraphics[width=.45\linewidth]{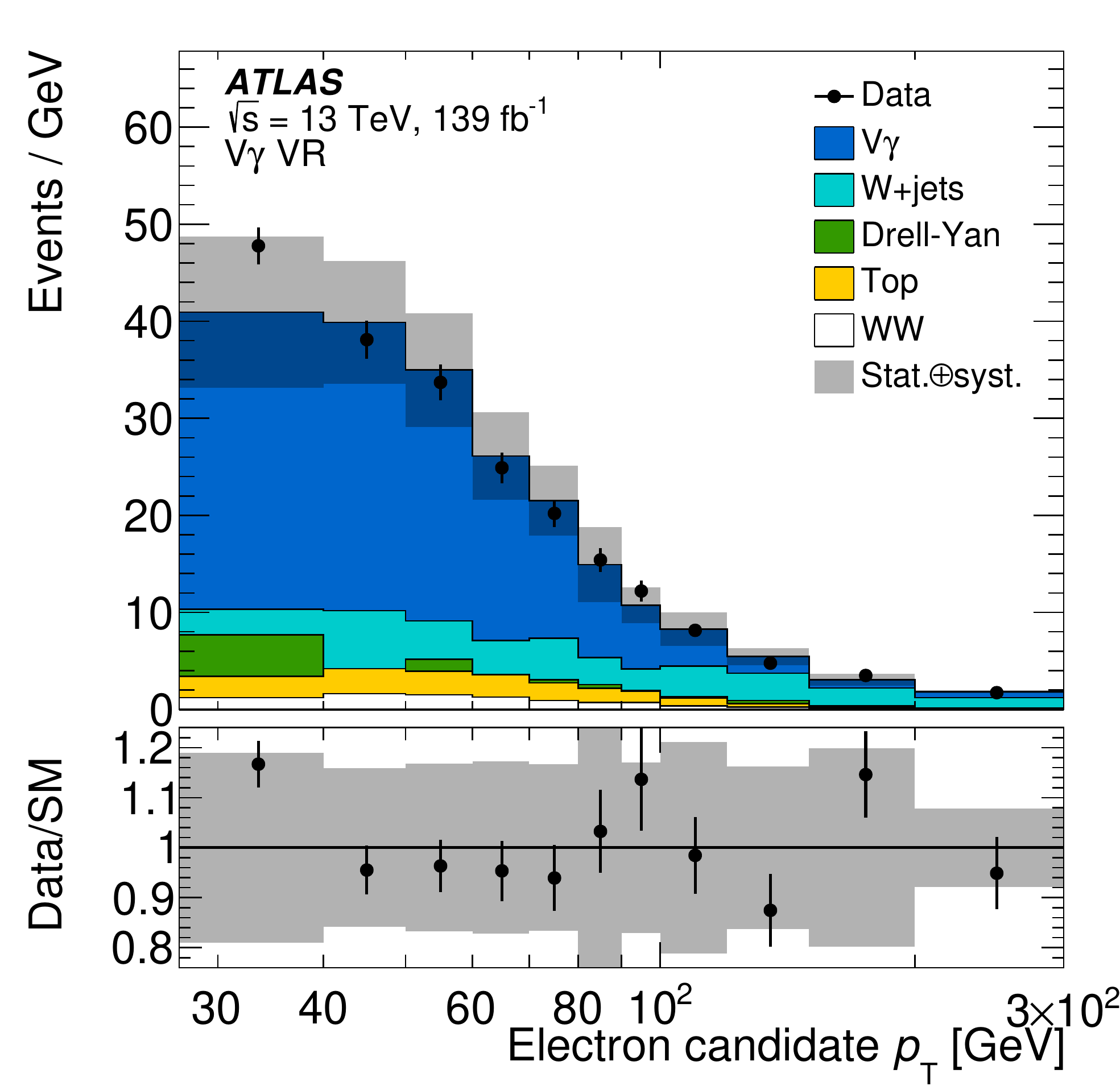}
\hspace{.45\linewidth}
\caption{Detector-level distributions of the \ptleadlep (top left) and the jet multiplicity (top right) in the top-enriched region, the \ptsublep in the same-sign validation region (VR) (middle left), the \mll in the Drell--Yan VR (middle right), the \met in the $VZ$ VR (bottom left), and the electron candidate \pt in the $V\gamma$ VR with opposite-sign leptons (bottom right). The last bin contains overflow events.
Data are shown as black markers, together with the predictions for the signal and background production processes. The top background in the top-enriched region is estimated using the data-driven method explained in the text; in all other regions the nominal MC prediction is used.
The lower panels show the ratio of the data to the total prediction.
The uncertainty bands shown include statistical and systematic uncertainties, excluding theory uncertainties on the signal.
}
\label{fig:vr-datamc}
\thisfloatpagestyle{empty}
\end{figure}

Table~\ref{tab:sr-yields} lists the number of selected \WW candidate events, as well as the breakdown of the background predictions. Details of the systematic uncertainties are given in Section~\ref{sec:systematics}.
Figure~\ref{fig:reco-level-dist} shows selected distributions at detector level in the final analysis binning and compares the observed data with the signal prediction and the background estimate.
Reasonable agreement between data and expectations is observed for both the event yields and the shapes of the distributions. For the nominal signal model, small excesses are seen in the predictions at low \ptleadlep, as well as at low \mll in the high-\ptleadjet selection~\OnlyPaperDraft{(both in Figure~\ref{fig:reco-level-dist})}. These are, however, covered by the theory uncertainties of the signal, which are not included in the error bands in this figure.

\begin{table}
\caption{Selected \WW candidate events, together with the signal prediction and the background estimates. The uncertainties include statistical and systematic contributions, excluding theory uncertainties on the signal. The fractions in percent give the relative contribution to the total SM prediction. The individual uncertainties are correlated, and do not add up in quadrature to the total uncertainty. }
\label{tab:sr-yields}
\centering
\begin{tabular}{lr@{$\,\pm\,$}rrrr@{$\,\pm\,$}rr}
\toprule	 & \multicolumn{3}{c}{Signal region} & \multicolumn{4}{r}{$\ptleadjet > 200\,\GeV$} \\
\midrule
Data & \multicolumn{2}{l}{89\,239} &  & \rule{0.5em}{0pt} & \multicolumn{2}{l}{5825} & \\
Total SM & 91\,600 & 2500 &  & & 5980 & 150 & \\
\midrule
\WW & 28\,100 & 1200 & 31\% & & 2480 & 60 & 42\%\\
Total bkg. & 63\,500 & 1800 & 69\% & & 3500 & 140 & 58\%\\
\midrule
Top & 55\,800 & 1500 & 61\% & & 3030 & 110 & 51\%\\
Drell--Yan & 2200 & 700 & 2\% & & 66 & 9 & 1\%\\
Fake leptons & 2700 & 1100 & 3\% & & 140 & 70 & 2\%\\
$WZ,ZZ,V\gamma$ & 2800 & 500 & 3\% & & 270 & 70 & 4\%\\
\bottomrule
\end{tabular}
\end{table}

\begin{figure}
\centering
\includegraphics[width=.45\linewidth]{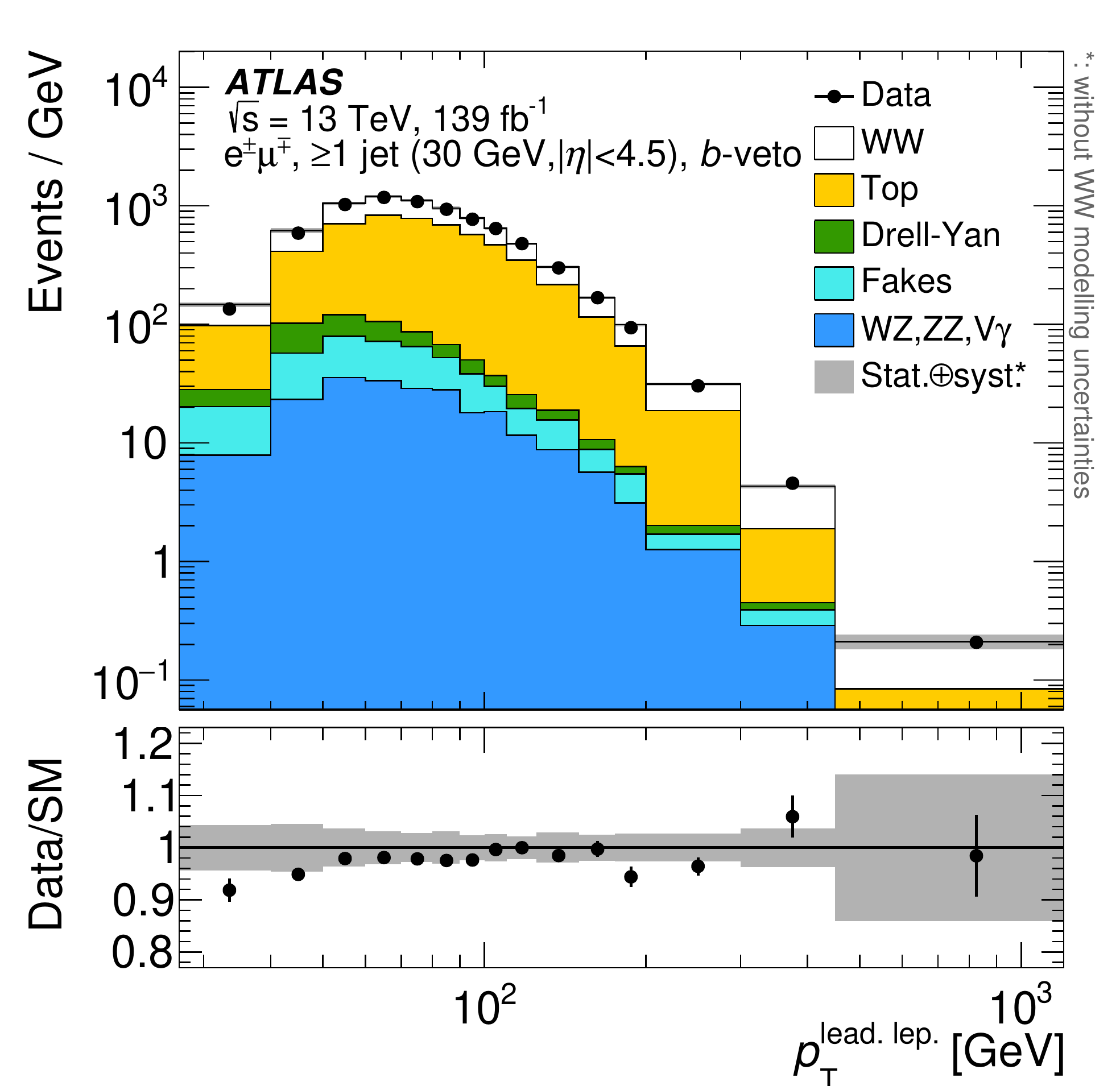}
\includegraphics[width=.45\linewidth]{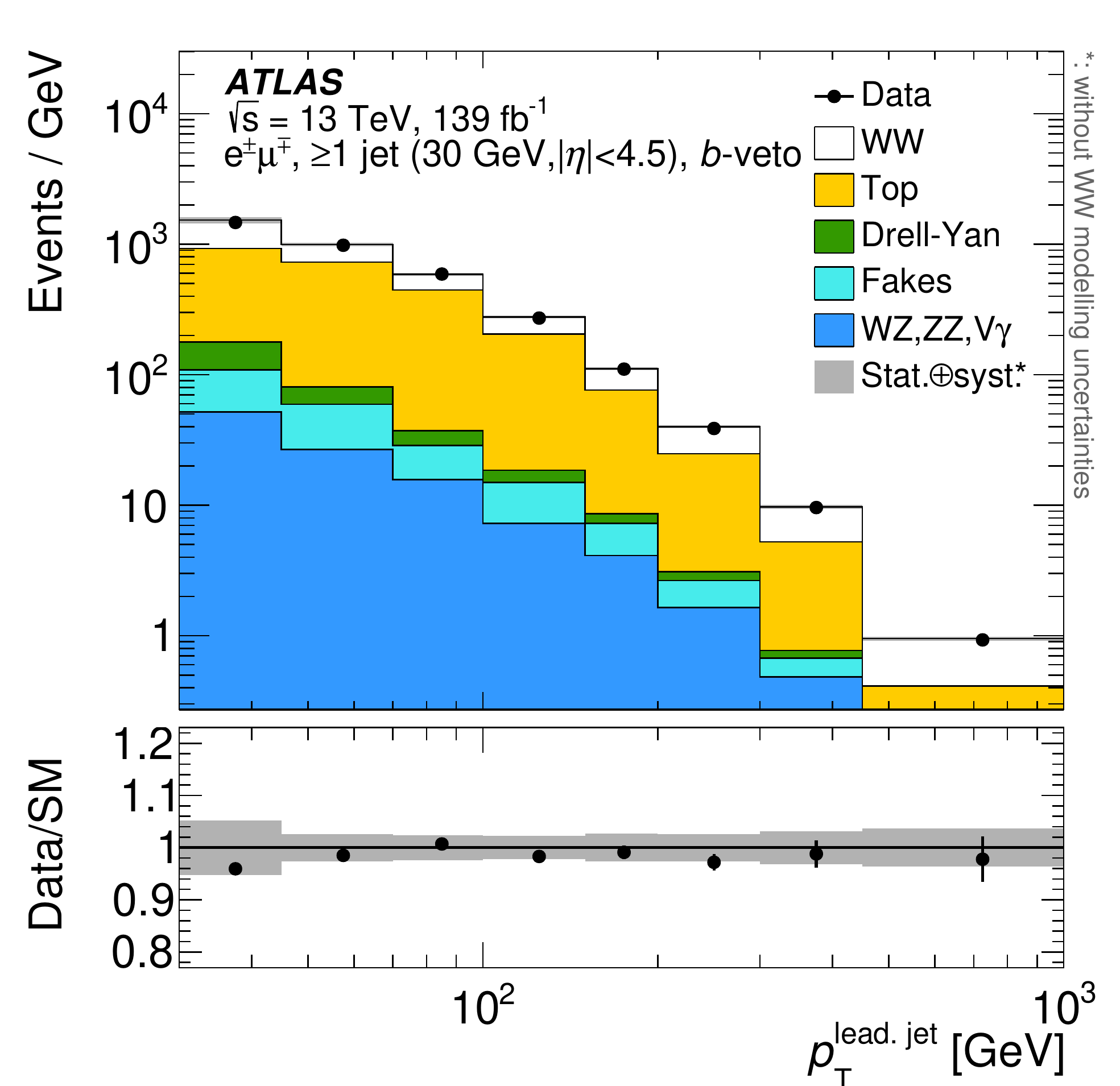}
 
\includegraphics[width=.45\linewidth]{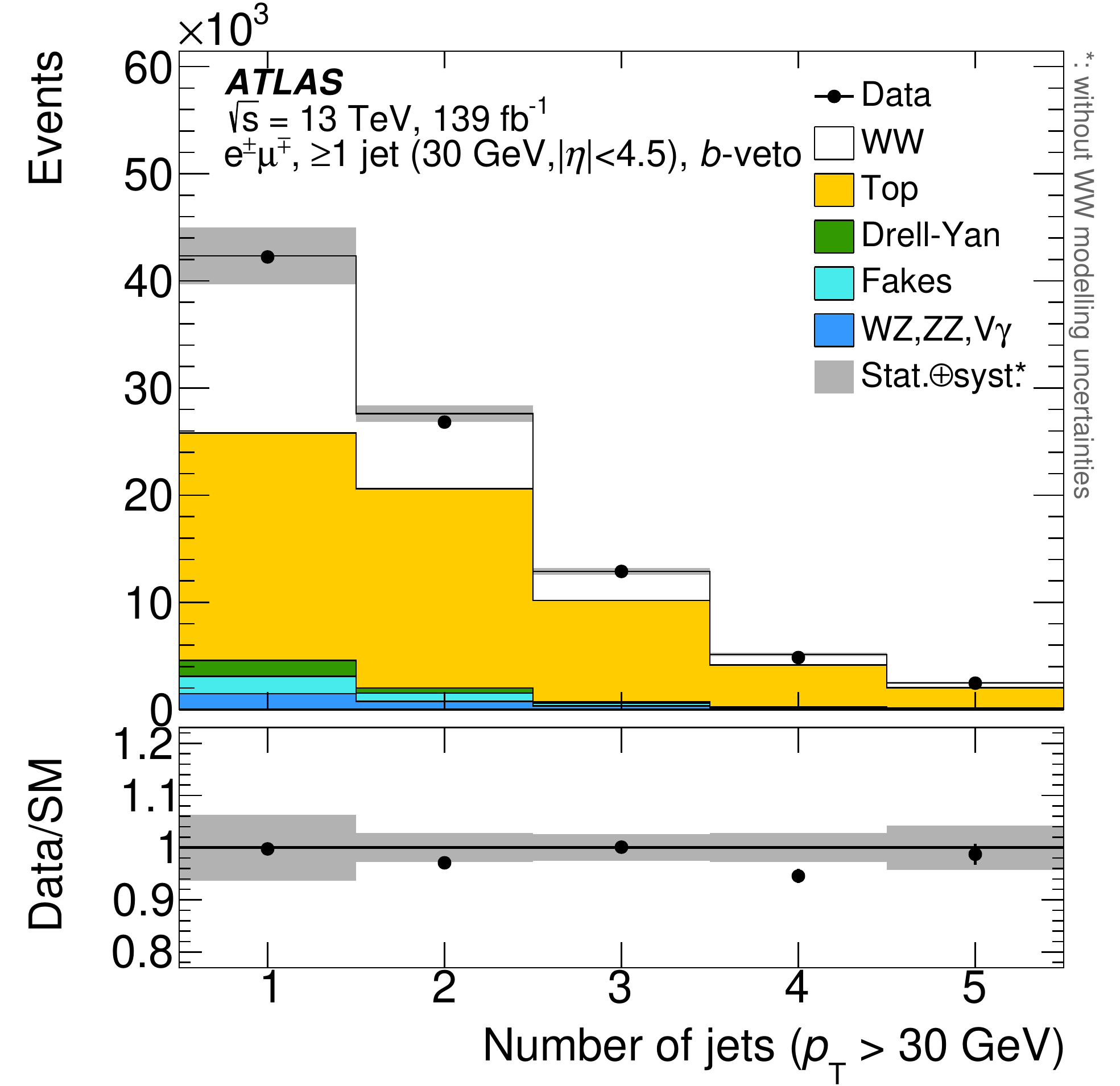}
\includegraphics[width=.45\linewidth]{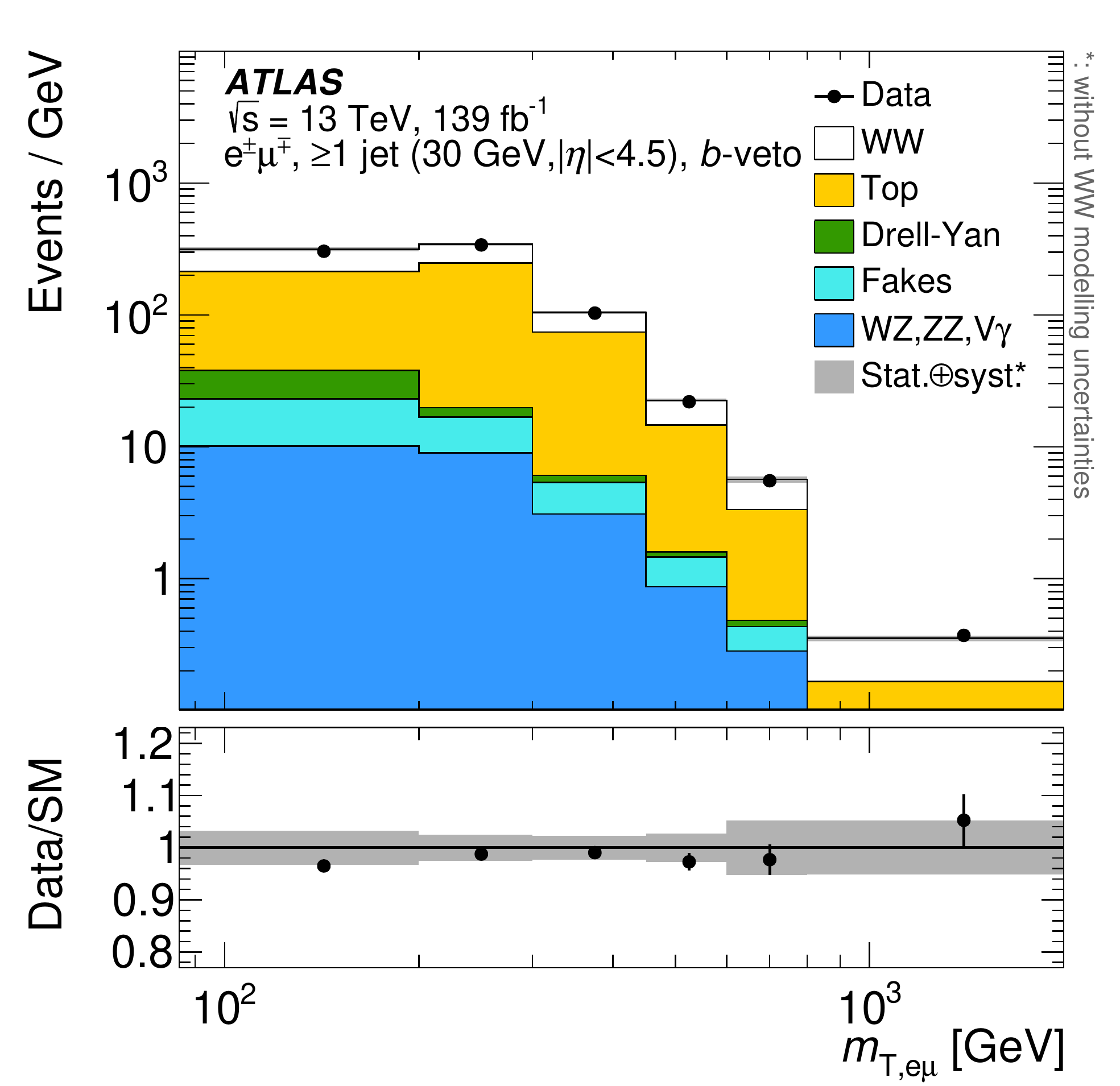}
 
\includegraphics[width=.45\linewidth]{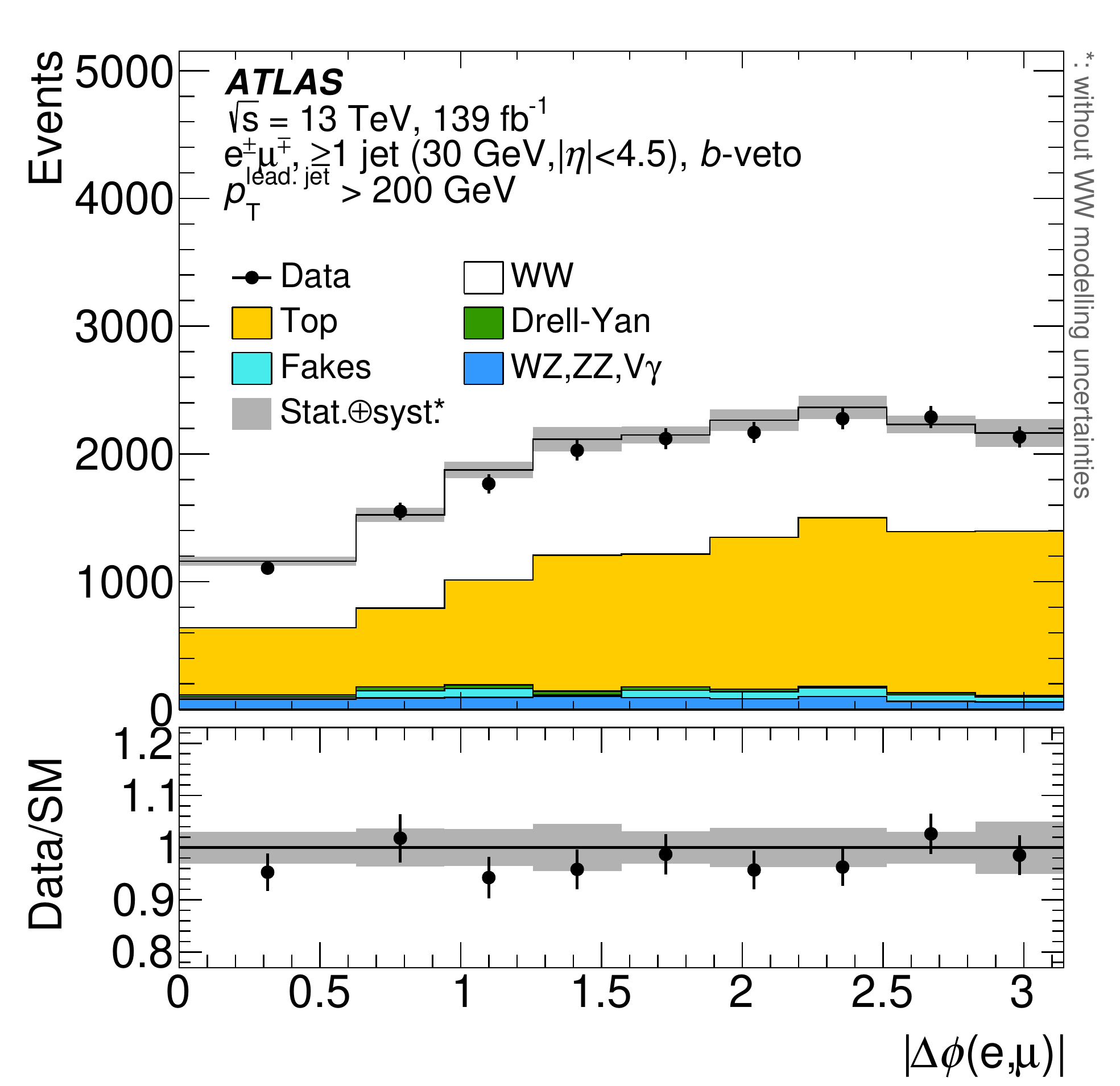}
\includegraphics[width=.45\linewidth]{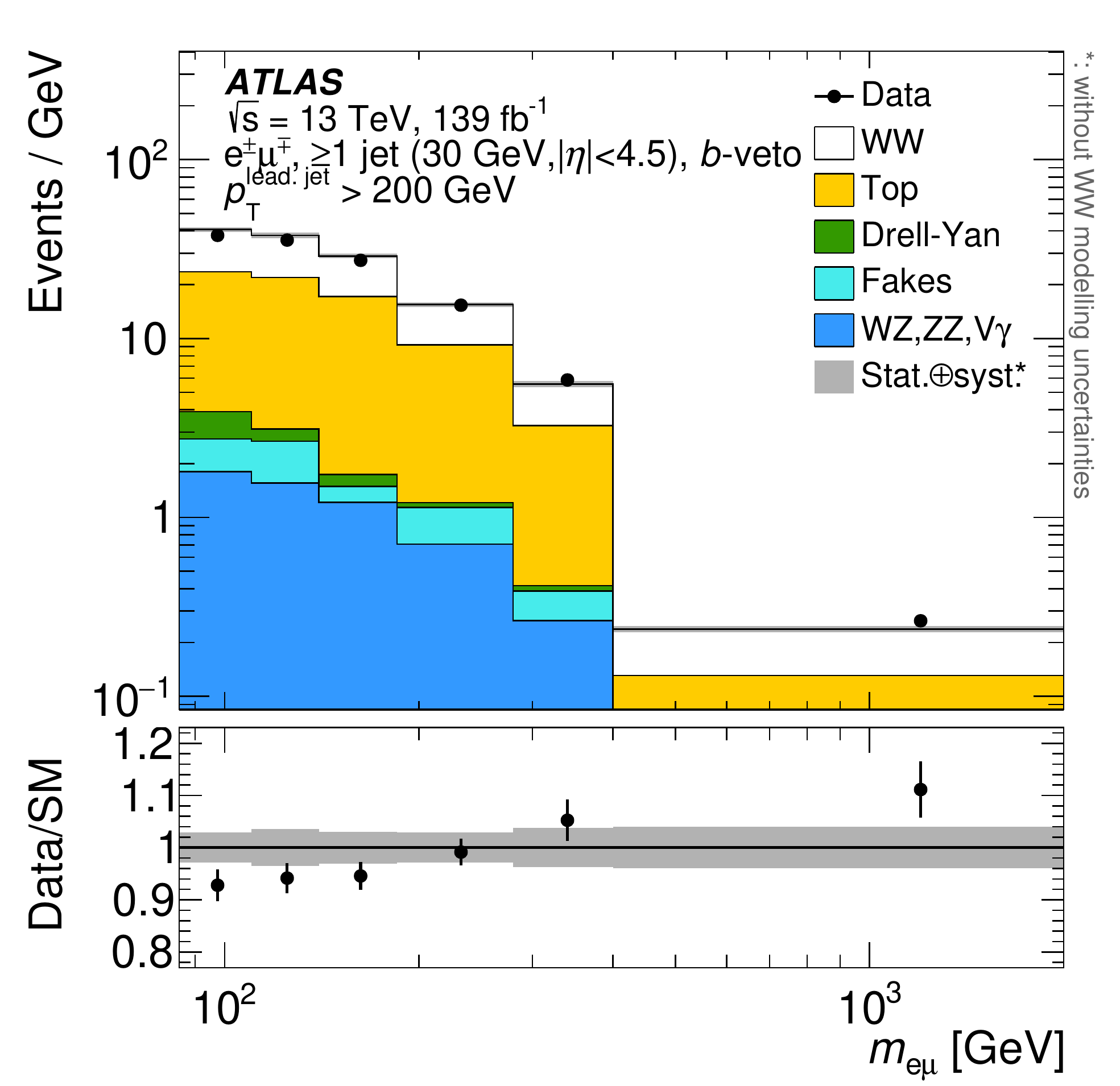}
 
\caption{Signal region detector-level distributions of the \ptleadlep (top left), the \ptleadjet (top right), the jet multiplicity (middle left), the \mtll (middle right), and, for events with $\ptleadjet > 200\,\GeV$, the \dphill (bottom left) and \mll (bottom right). Data are shown as black markers together with the predictions for the signal and background production processes. The last bin contains overflow events.
The lower panels show the ratio of the data to the total prediction.
The uncertainty bands shown include statistical and systematic uncertainties, excluding theory uncertainties on the signal.
}
\label{fig:reco-level-dist}
\thisfloatpagestyle{empty}
\end{figure}
 
 
\section{Fiducial and differential cross-section determination}
\label{sec:unfolding}
 
The \WWj cross-section is evaluated in the fiducial phase space of the \WWemu decay channel as defined in Table~\ref{tab:fiducialtruthlevelcuts}.
In simulated events, electrons and muons are required to originate directly from the hard interaction and not from $\tau$-lepton or hadron decays.
The momenta of photons emitted in a cone of size $\Delta R = $~0.1 around the lepton direction that do not originate from hadron decays are added to the lepton momentum to form `dressed' leptons.
Stable final-state particles,\footnote{Particles are considered stable if their decay length $c\tau$ is greater than 1 cm.} excluding prompt leptons and the associated photons, are clustered into particle-level jets using the anti-$k_t$ algorithm with radius parameter $R$ = 0.4.
The missing transverse momentum is defined at particle level as the transverse component of the vectorial sum of the neutrino momenta.
The nominal definition of the particle-level fiducial phase space does not include a veto on \bjets\OnlySupportNote{, to simplify the comparison to theoretical predictions}.
Alternative results that include a veto on particle-level \bjets\footnote{At particle level, \bjets are defined by ghost-association~\cite{Cacciari:2007fd}, wherein $b$-hadrons are included in the jet clustering as infinitely soft particles (ghosts). Jets with $b$-hadron ghosts among their constituents are \bjets.} with $\pt>20$ GeV are provided in HEPData\footnote{\url{https://www.hepdata.net/record/100511}}.

The fiducial cross-section is obtained as follows:
\begin{linenomath*}
\begin{equation*}
\sigma^{\mathrm{fid}}
= \frac{N_\mathrm{ obs} - N_\mathrm{ bkg}}{C\, \times {\cal L} },
\label{eq:xSectFid}
\end{equation*}
\end{linenomath*}
where ${\cal L}$ is the integrated luminosity, $N_\mathrm{obs}$ is the observed number of events, $N_\mathrm{bkg}$ is the estimated number of background events and $C$ accounts for detector inefficiencies, resolution effects, and contributions from $\tau$-lepton decays.
$C$ is calculated as the number of simulated signal events passing the reconstruction-level event selection divided by the events in the fiducial phase space. Its numerical value is $C=0.747\pm 0.061$ and its uncertainty is dominated by uncertainties in jet energy scale, jet energy resolution, and pile-up modelling. The fraction of \WW events passing the event selection but containing at least one lepton from $\tau$-lepton decays is 9\%.
 
The differential cross-sections are determined using an iterative Bayesian unfolding method~\cite{BayesUnfold1,BayesUnfold2}.
The unfolding procedure corrects for migrations between bins in the distributions during the reconstruction of the events, and applies fiducial as well as reconstruction efficiency corrections. The fiducial corrections take into account events that are reconstructed in the signal region, but originate from outside the fiducial region; the reconstruction efficiency corrects for events inside the fiducial region that are not reconstructed in the signal region due to detector inefficiencies. Tests with MC simulation demonstrate that the method is successful in retrieving the true distribution in the fiducial region from the reconstructed distribution in the signal region.
To reduce bias due to the assumed true distribution, the method can be applied iteratively, at the cost of an increased statistical uncertainty. Two iterations are used to unfold the \HT, \ST, and \ptleadjet distributions and the exclusive jet multiplicity, which are subject to large modelling uncertainties. For the remaining distributions, either the result is independent of the number of iterations, or the modelling uncertainty is not reduced and the statistical uncertainties increase. For these cases, only one unfolding iteration is performed.
\OnlySupportNote{
More details on the unfolding procedure can be found in Appendix~\ref{app:unfolding}.
More details on the binning of distributions can be found in Appendix~\ref{app:binning}.
}
 
\begin{table}[t!]
\caption{Definition of the $WW\rightarrow e\mu$+jets fiducial phase space, where $\pt^{\ell}$ ($\eta^{\ell}$) refers to the transverse momentum  (pseudorapidity) of charged leptons and $\pt^{j}$ ($y^{j}$) to the transverse momentum (rapidity) of jets.}
\begin{center}
\begin{tabular}{rp{0.1cm}l}
\toprule
\multicolumn{3}{c}{Fiducial selection requirements}   \\
\midrule
$\pt^{\ell}$ & > & $27$~\GeV  \\
$|\eta^{\ell}|$ & < & $2.5$         \\
$\mll$ & > & $85$~\GeV               \\
$\pt^{j}$ & > & $30$ \GeV\\
$|y^j|$ & < & 4.5 \\
\bottomrule
\end{tabular}
\end{center}
\label{tab:fiducialtruthlevelcuts}
\end{table}
 
 
\section{Uncertainties}
\label{sec:systematics}
Systematic uncertainties in the \WWj cross-section measurements arise from experimental sources, the background determination, the procedures used to correct for detector effects, and theoretical uncertainties in the signal modelling.
 
The dominant experimental systematic uncertainties arise in the calibration of the jet energy scale and resolution and the calibration of the \btagging efficiency and mis-tag rates.
Experimental uncertainties also encompass uncertainties in the calibration of lepton trigger, reconstruction, identification and isolation efficiencies, the calibration of the lepton momentum or energy scale and resolution, and the modelling of pile-up.
All experimental uncertainties are evaluated by varying the respective calibrations, and propagating their effects through the analysis, affecting both the background estimates and the unfolding of detector effects.
 
Systematic uncertainties in the estimate of fake leptons are derived by changing the selection used to estimate the weights, in order to change the composition of the sources of fake leptons. Additionally, the subtraction of the prompt-lepton sources in the control region is varied, and the statistical uncertainties of the weights are propagated. More details on the uncertainties affecting the fake-lepton estimate can be found in Section~\ref{sec:background}.
 
The estimate of the top background is affected by the statistical uncertainty of the number of events in the control region, and by uncertainties in the modelling of \ttbar and single-top $Wt$ events, such as the uncertainty in the matrix element calculation, the parton shower modelling, the QCD scale choices, the initial- and final-state radiation and the interference between \ttbar and single-top $Wt$ events.
These  are e\-val\-uat\-ed by using the alternative simulations described in Section~\ref{sec:data_mc_samples} and propagating the results through the top background estimate.
The effect of the PDF uncertainty on the top background was evaluated, but found to be negligible.
 
The uncertainty in minor backgrounds is estimated by varying their total cross-section within its uncertainty and by using alternative simulations, as described in Section~\ref{sec:background}. The difference between nominal and alternative simulations covers PDF uncertainties, missing higher-order QCD corrections, and the parton shower model.
 
The bias introduced by using distributions generated by the nominal signal simulation as a prior in the unfolding is estimated by reweighting these distributions at generator level with a smooth function such that, after including simulated detector effects, they closely resemble the background-subtracted data. This reweighted detector-level prediction is unfolded using the nominal unfolding set-up. The unfolding procedure is able to very accurately recover the generator-level distribution, so this uncertainty source is negligible.
Uncertainties in the unfolding procedure due to the theoretical modelling of the signal are evaluated by repeating the unfolding procedure with alternative signal simulations. The uncertainty due to missing higher-order QCD corrections is evaluated by varying the renormalization and factorization scales. The uncertainty due to the choice of generator for the hard interaction, the parton shower model and the underlying-event modelling is estimated using the alternative simulation of $\qqWW$ production, from \POWHEGBOXV{v2}, interfaced to \PYTHIAV{8.186}.
For the uncertainty estimation, the alternative model is first reweighted to the nominal model, so that uncertainties due to disagreement in the predicted shape of distributions can be ignored, and only the difference in the prediction of the migration matrix and fiducial and efficiency corrections are taken into account.
Statistical uncertainties are evaluated by creating pseudo data samples that are obtained by varying the data within their Poisson uncertainties in each bin and then propagating these varied samples through the unfolding.
The statistical uncertainties of the background estimates, which include statistical uncertainties in MC predictions and due to the control regions used in estimating the top and fake-lepton backgrounds, are evaluated using the same method.
If not stated otherwise, `statistical uncertainties' refers to the combined statistical uncertainties from signal and control regions.
 
Table~\ref{tab:syst-breakdown} gives a breakdown of the uncertainties in the fiducial cross-section measurements, and Figure~\ref{fig:syst-breakdown} displays the uncertainties as a function of the unfolded \ptleadlep and \ptleadjet distributions.
Jet-related uncertainties generally decrease with \ptleadjet and with correlated quantities such as \ptleadlep, while statistical uncertainties increase at high energy. This leads to a minimum of the total uncertainty for intermediate values of \ptleadjet and \ptleadlep.
 
\OnlySupportNote{
 
Additional details on systematic uncertainties are given in Appendix~\ref{app:systematics}. The appendices detailing the background estimates also discuss the respective systematic uncertainties.}

\begin{table}
\caption{Breakdown of the uncertainties in the measured fiducial cross-section. ``Jet calibration'' uncertainties encompass jet energy scale and resolution uncertainties, ``Top modelling'' and ``Signal modelling'' are uncertainties in the theoretical modelling of the respective processes, ``Fake-lepton background'' is the uncertainty in the fake-lepton estimate while ``Other background'' is the uncertainty due to minor prompt-lepton backgrounds, ``Flavour tagging'' is all uncertainties in flavour tagging efficiency and mis-tag rate, and ``Luminosity'' is the uncertainty in the measurement of the integrated luminosity. All systematic uncertainties belonging to none of the above categories are included in ``Other systematic uncertainties''. Statistical uncertainties arise in both the signal region and control region used for the data-driven top and fake-lepton estimates and also from backgrounds that are estimated using MC simulations.
}
\label{tab:syst-breakdown}
\centering
\begin{tabular}{lr}
\toprule
Uncertainty source & Relative effect \\
\midrule
Total uncertainty & 10\%\\
\midrule
Signal region statistical uncertainty & 1.1\%\\
Data-driven background and MC statistics & 1.2\%\\
\midrule
Jet calibration & 6.3\%\\
Top modelling & 4.5\%\\
Fake-lepton background & 4.3\%\\
Signal modelling & 2.7\%\\
Other background & 2.3\%\\
Flavour tagging & 2.3\%\\
Luminosity & 1.9\%\\
Other systematic uncertainties & 0.6\%\\
\bottomrule
\end{tabular}
\end{table}
 
\begin{figure}
\centering
\includegraphics[width=.49\linewidth]{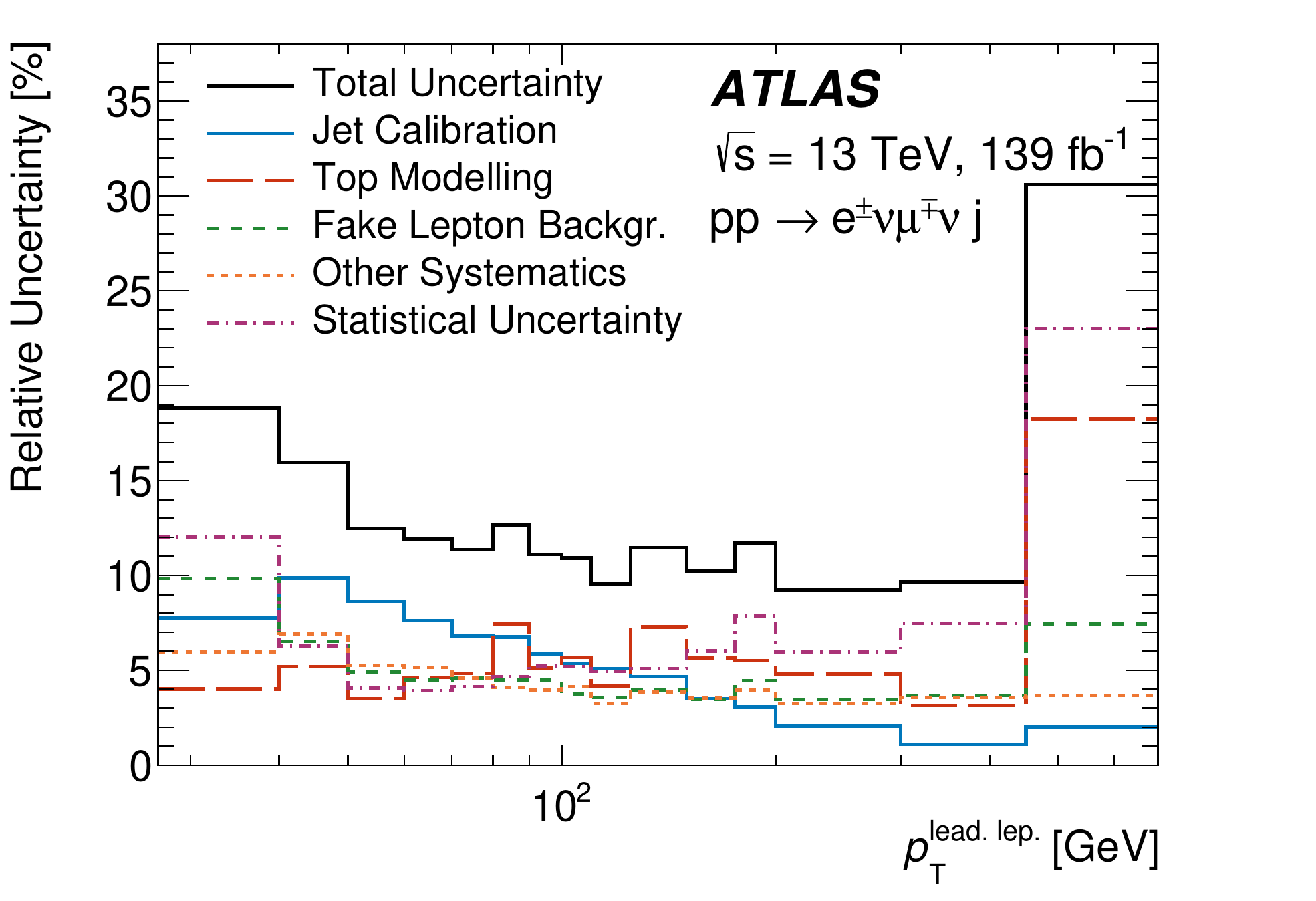}
\includegraphics[width=.49\linewidth]{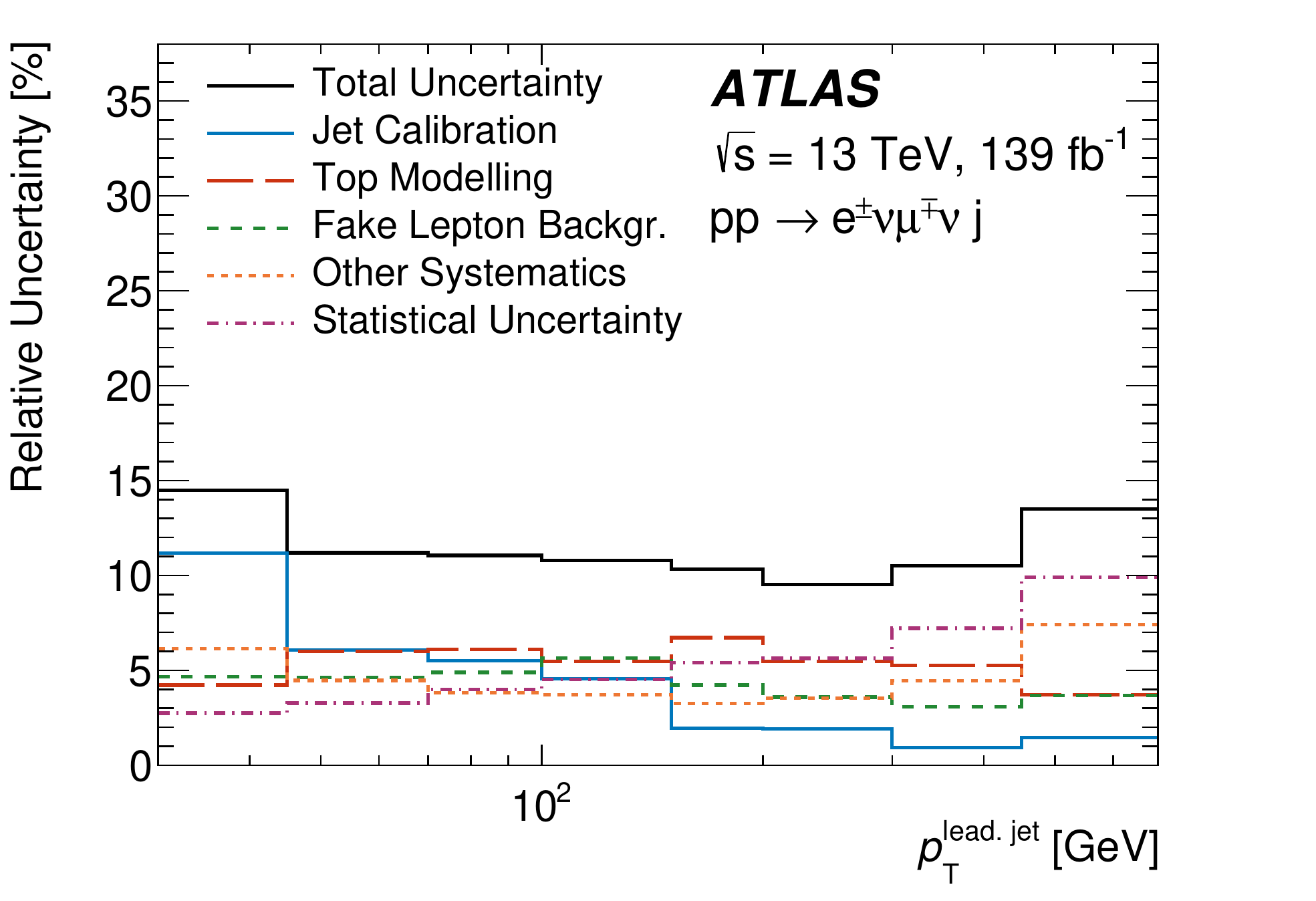}
\caption{Relative size of uncertainties for the unfolded \ptleadlep and \ptleadjet distributions.  ``Jet Calibration'' uncertainties encompass jet energy scale and resolution uncertainties while ``Top Modelling'' encompasses all \ttbar and single-top modelling uncertainties. ``Fake Lepton Backgr.'' is the uncertainty in the non-prompt-lepton estimate from the fake-factor method.  ``Other Systematics'' includes modelling and total cross-section uncertainties in the remaining backgrounds, lepton-related uncertainties as well as uncertainties due to pile-up reweighting and the signal modelling in the unfolding, while ``Statistical Uncertainty'' is the combined statistical uncertainty in the signal region, from control regions, and from MC simulations.}
\label{fig:syst-breakdown}
\end{figure}
 
\OnlySupportNote{
Appendix~\ref{app:systematics} gives a summary of the systematics pruning strategy, as well as a complete breakdown of the detector level uncertainties and their effect for all observables.
}
 
\newcommand{\ResFigCaption}{The measured cross-section values are shown as points with error bars giving the statistical uncertainty and solid bands indicating the size of the total uncertainty.
The results are compared with the NNLO prediction with extra NLO EW corrections and NLO corrections for $gg\rightarrow \WW$ production (denoted MATRIX $\otimes$ NLO EW) as well as NLO+PS predictions from \SHERPAV{2.2.2}, \mgamc+ \PYTHIAV{8} with FxFx merging, and \minlo + \PYTHIAV{8}  for \qqbar initial states, combined with \SHERPA{} + \openloops\ (LO+PS) for the $gg$ initial state. The \SHERPAV{2.2.2} + \openloops\ prediction is normalized to the total cross-section calculated at NLO in QCD. Theoretical predictions are indicated as markers with vertical lines denoting PDF and scale uncertainties.}
 
\section{Results}
\label{sec:results}
 
The measured fiducial cross-section for \WWj production, with \WWemu,  at $\sqrt{s}=13\,\tev$, for the phase space defined in Table~\ref{tab:fiducialtruthlevelcuts} is
\[
\sigma_{\mathrm{fid}}=258 \pm 4~\mathrm{(stat.)} \pm 25~\mathrm{(syst.)} \,\si{\femto\barn},
\]
 
with a total uncertainty of 10\%.
In Figure~\ref{fig:TotResult}, the measured result is compared with various predictions for \WWj production, and good agreement is found.
Differential fiducial cross-sections are presented in Figures~\ref{fig:DiffResults-1} to~\ref{fig:DiffResults-3}. Figure~\ref{fig:DiffResults-4} displays distributions in a phase space that additionally requires a jet with a transverse momentum of at least 200\,\GeV.
 
\begin{figure}
\centering
\includegraphics[width=0.6\textwidth]{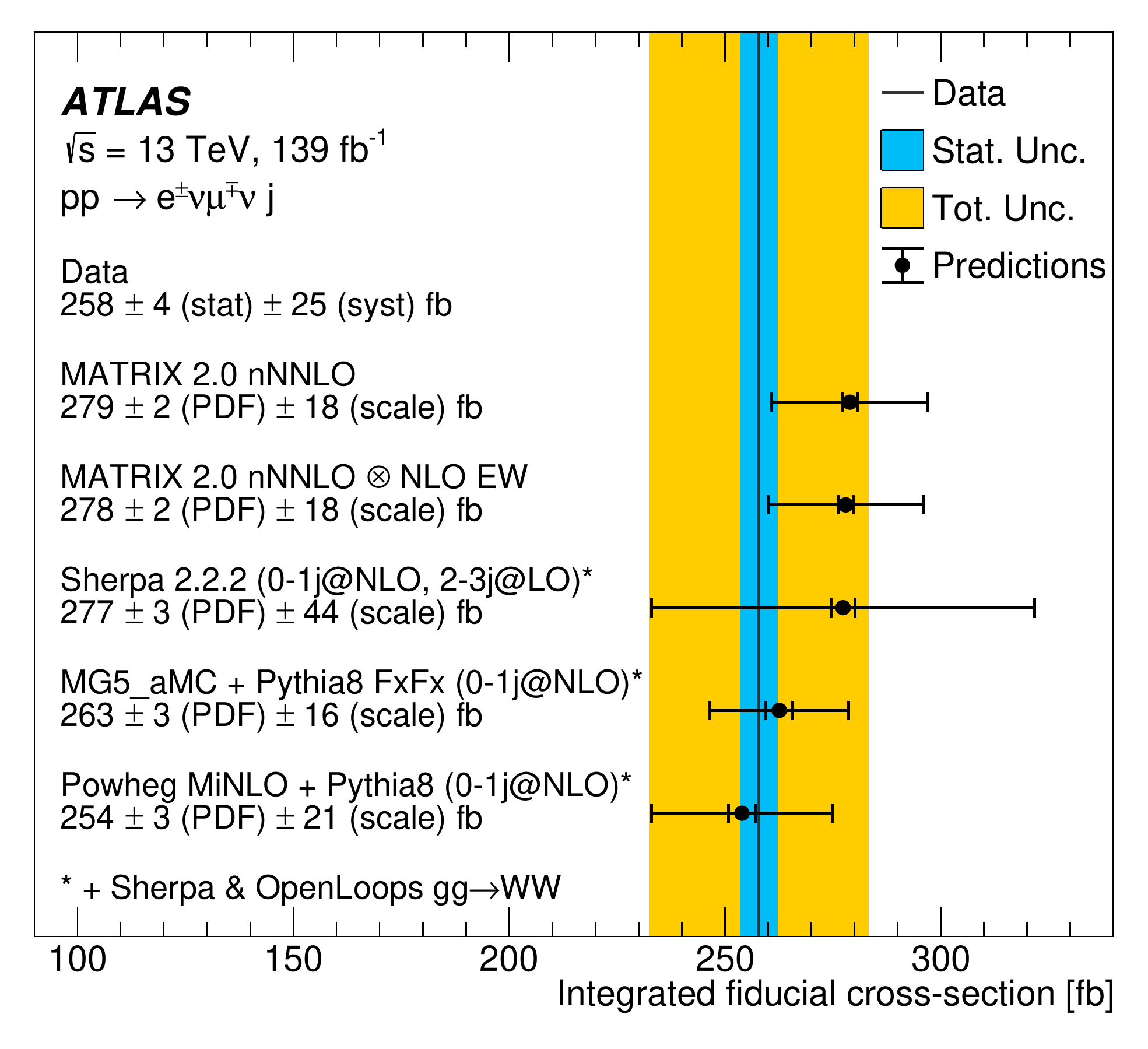}
\caption{
Comparison of the measured fiducial \WWj cross-section with various theoretical predictions.
Theoretical predictions are indicated as points with inner (outer) error bars denoting PDF (PDF+scale) uncertainties.
The central value of the measured cross-section is indicated by a vertical line with the narrow band showing the statistical uncertainty and the wider band the total uncertainty including statistical and systematic uncertainties.
The result is compared with a fixed-order parton-level prediction from MATRIX~2.0 that is accurate to NNLO (NLO) for \qqWW (\ggWW) production, and a prediction that additionally accounts for EW corrections to \WW + jet production, which have been calculated with \SHERPAV{2.2.2} + \openloops.
It is also compared with predictions from  \SHERPAV{2.2.2}, \mgamc+~\PYTHIAV{8} with FxFx merging, and  \minlo~+~\PYTHIAV{8}, which are all supplemented by a \SHERPAV{2.2.2}~+~\openloops \ggWW LO+PS prediction.}
\label{fig:TotResult}
\end{figure}
 
\subsection{Comparison with theoretical predictions}
\label{sub:results-comp}
The measurement is compared to the theory predictions listed in Table~\ref{tab:theory-predictions}.
The measured fiducial cross-section is compatible with the prediction of $279 \pm 2~\text{(pdf)}~^{+20}_{-16}~\text{(scale)}$~fb from MATRIX~\cite{Catani:2007vq,Catani:2012qa,Gehrmann:2014fva,Gehrmann:2015ora,Denner:2016kdg,Grazzini:2016ctr,Grazzini:2017mhc,Cascioli:2011va,Grazzini:2020stb}, which is accurate to NNLO (NLO) for \qqWW (\ggWW) production, denoted nNNLO, but only NLO (LO) accurate for \qqWW (\ggWW) production in association with a jet.
For this prediction, the NNPDF3.1NNLO parton distribution function is used, while renormalization and factorization scales are set to $m_W$.
In Figure~\ref{fig:TotResult}, the measured integrated fiducial cross-section is also compared with a prediction that combines the QCD corrections from MATRIX with NLO EW corrections to \WWj production that were generated with \SHERPAV{2.2.2} + \openloops~\cite{Kallweit:2014xda,Schonherr:2017qcj,Bothmann:2019yzt,Buccioni:2019sur}. Photon-induced contributions are included as an additive correction, while the EW correction to \qqWW is taken into account multiplicatively. The latter correction decreases the cross-section by 4\%, while the former leads to an increase of 4\%. The importance of both corrections increases with energy. The difference between an additive and multiplicative combination scheme for QCD and EW corrections is typically of the order 1\% but can be as large as 10\% in the highest \HT and \ST bins. 
 
Also displayed in Figure~\ref{fig:TotResult} are the nominal \qqWW \SHERPAV{2.2.2} prediction, a prediction from \MADGRAPHV{2.3.3}  using \fxfx~merging~\cite{Frederix:2012ps} and interfaced to \PYTHIAV{8.212}, and a \minlo~\cite{Hamilton:2016bfu} prediction interfaced to \PYTHIAV{8.244}. All three calculations are NLO-accurate for \WW production with one jet and use the NNPDF3.0 PDF set. The effects of scale uncertainties for all predictions are estimated by varying the factorization and renormalization scales of the hard process.
The effects of scale uncertainties on the \SHERPAV{2.2.2} prediction are large in comparison with the other generators as this calculation includes leading-order matrix elements with up to three jets, which are strongly affected by scale variations.
Both predictions are supplemented by the \SHERPAV{2.2.2} + \openloops simulation of \ggWW, which is normalized to the total NLO QCD cross-section~\cite{Grazzini:2020stb}.

\begin{table}
\caption{Summary of the theoretical predictions that are compared with the measured cross-sections. The samples generated with \SHERPA use the default parton-shower tune, while for those using \PYTHIAV{8} the A14 tune and the NNPDF2.3LO PDF set are used for the parton shower.}
\label{tab:theory-predictions}
\centering
\begin{tabular}{@{}lllll@{}}
\toprule
Process & Generator & Parton shower & PDF & Matrix element $\mathcal{O}(\alpha_\text{S})$\\
\midrule
$q\bar{q}\rightarrow WW$ & MATRIX 2.0  & -- & NNPDF3.1  & NNLO \\
$gg\rightarrow WW$ & MATRIX 2.0 & -- & NNPDF3.1 &  NLO \\
 
$q\bar{q}\rightarrow WW$ & \SHERPAV{2.2.2}  & \SHERPA & NNPDF3.0  & NLO (0--1 jet), LO (2--3 jets) \\
$q\bar{q}\rightarrow WW$ & \minlo  & \PYTHIAV{8} & NNPDF3.0 & NLO (0--1 jet) \\
$q\bar{q}\rightarrow WW$ & \MADGRAPHV{2.3.3}  & \PYTHIAV{8}  & NNPDF3.0 & NLO (0--1 jet) \\
$gg\rightarrow WW$ &  \SHERPAV{2.2.2} + \openloops  & \SHERPA & NNPDF3.0 & LO (0--1 jet) \\
 
\bottomrule
\end{tabular}\\
\raggedright
 
\end{table}

The measured distributions in Figures~\ref{fig:DiffResults-1}--\ref{fig:DiffResults-4} are also compared with the predictions described above.
Within uncertainties, all predictions give an excellent description of the observed data.
For the nominal \SHERPAV{2.2.2} prediction, values of $\chi^2$ divided by the number of degrees of freedom are below one, except for the \mll distribution measured for $\ptleadjet > 200\,\GeV$, for which the value is 1.4.
Comparisons of the remaining predictions with the data yield similar $\chi^2$ values, except for the jet multiplicity, \HT, and \ST distributions, where,  for the highest multiplicities and energies, small discrepancies exist between data and predictions.

\OnlySupportNote{
More details on theoretical predictions can be found in Appendix~\ref{app:theory}.
In Appendix~\ref{app:powheg} the measured differential cross-sections are further compared to predictions by \POWHEG that are only LO for \WW + jet production. Additional results for $\ptleadlep > 200\,\GeV$ are shown in Appendix~\ref{app:lep200}.
}
 
\begin{figure}
\centering
\includegraphics[width=0.49\textwidth]{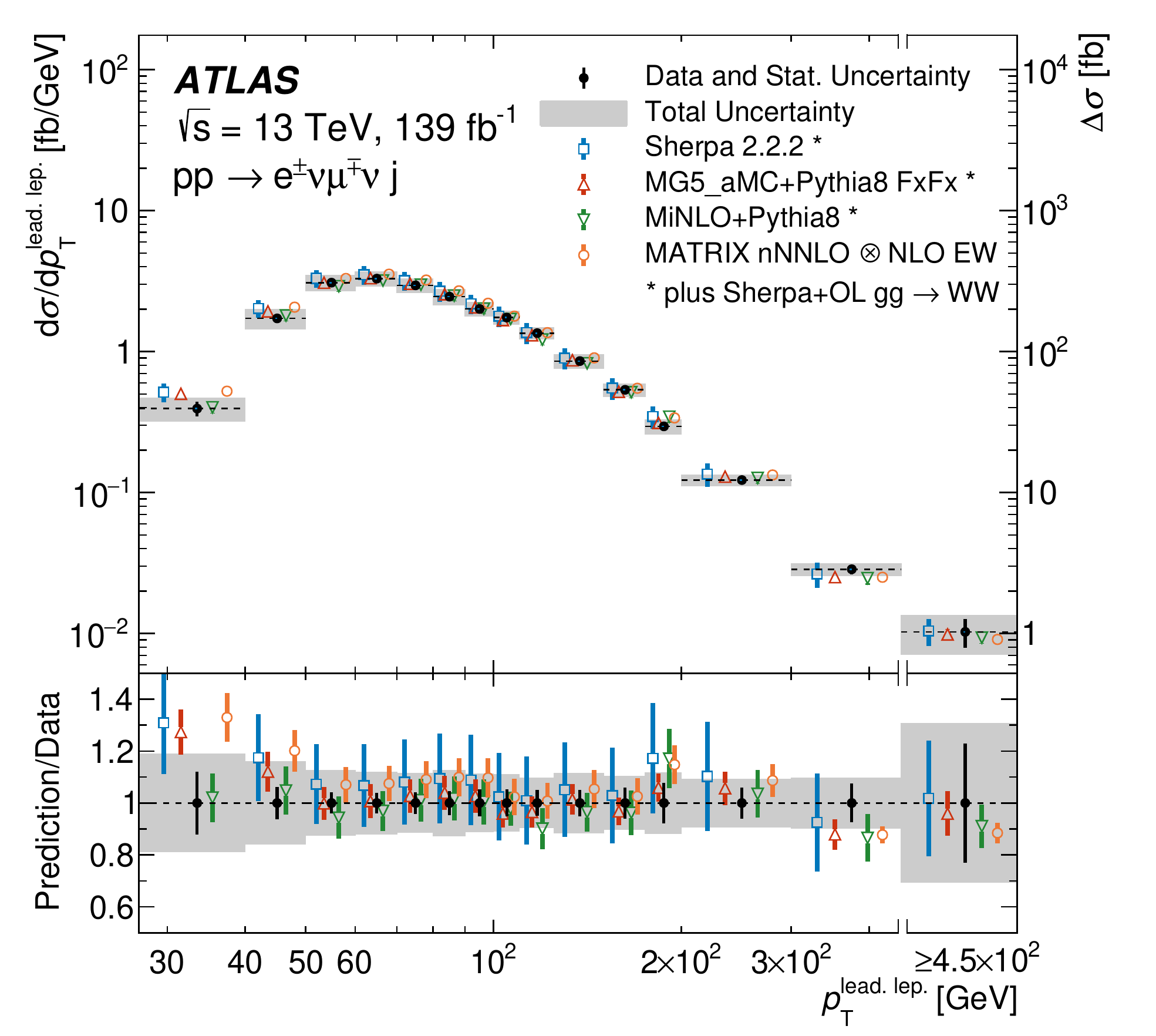}
\includegraphics[width=0.49\textwidth]{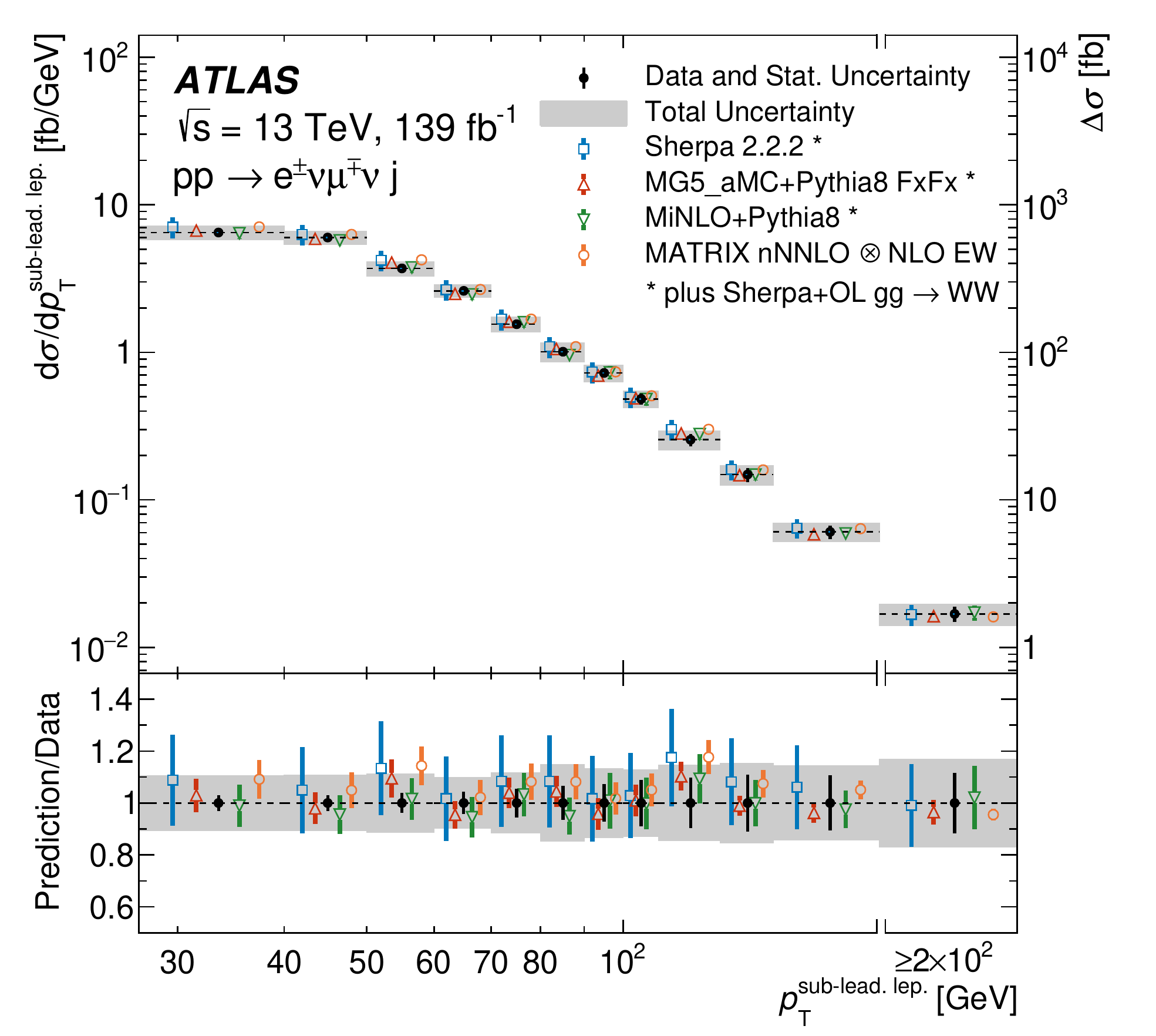}
\includegraphics[width=0.49\textwidth]{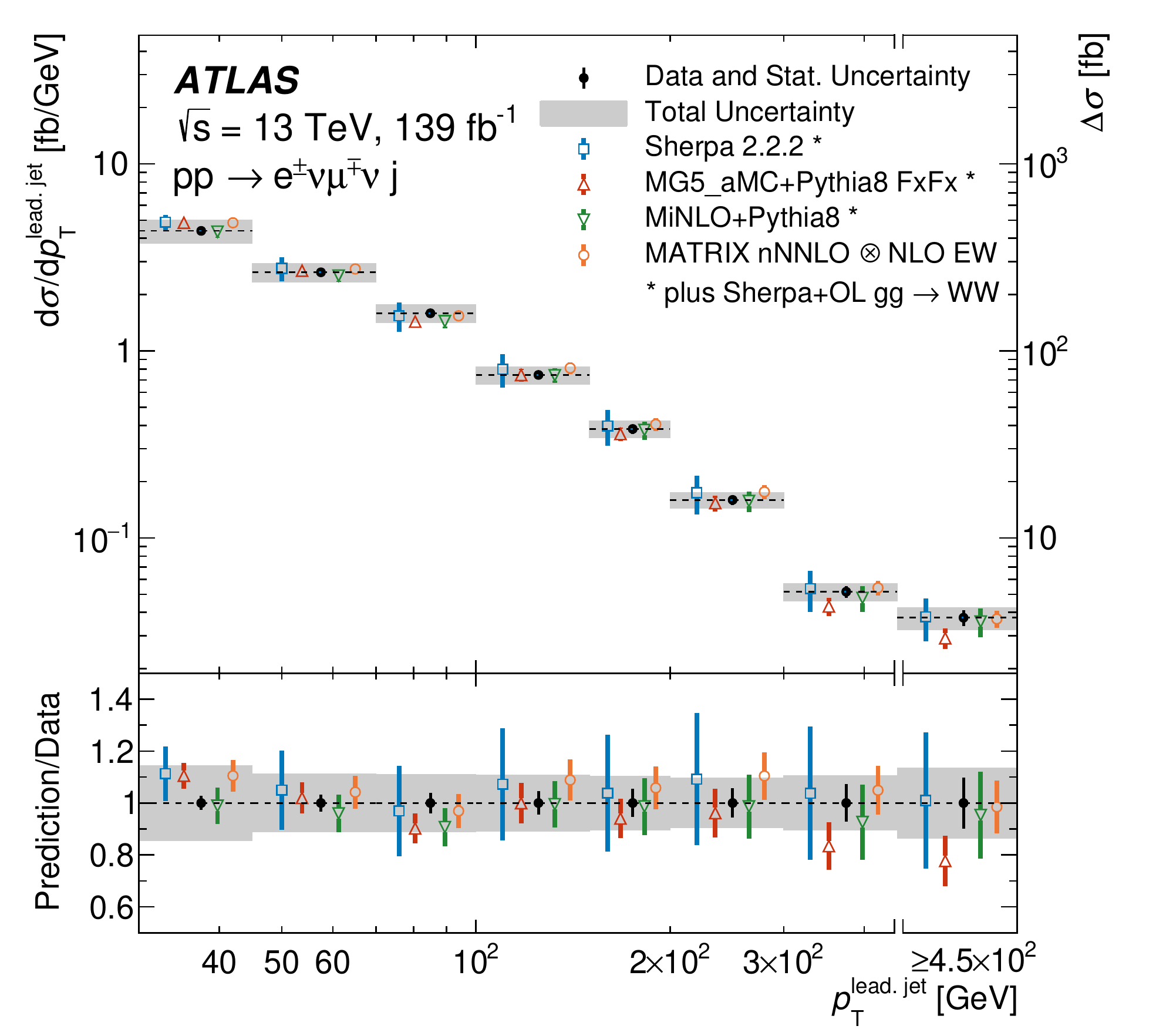}
\includegraphics[width=0.49\textwidth]{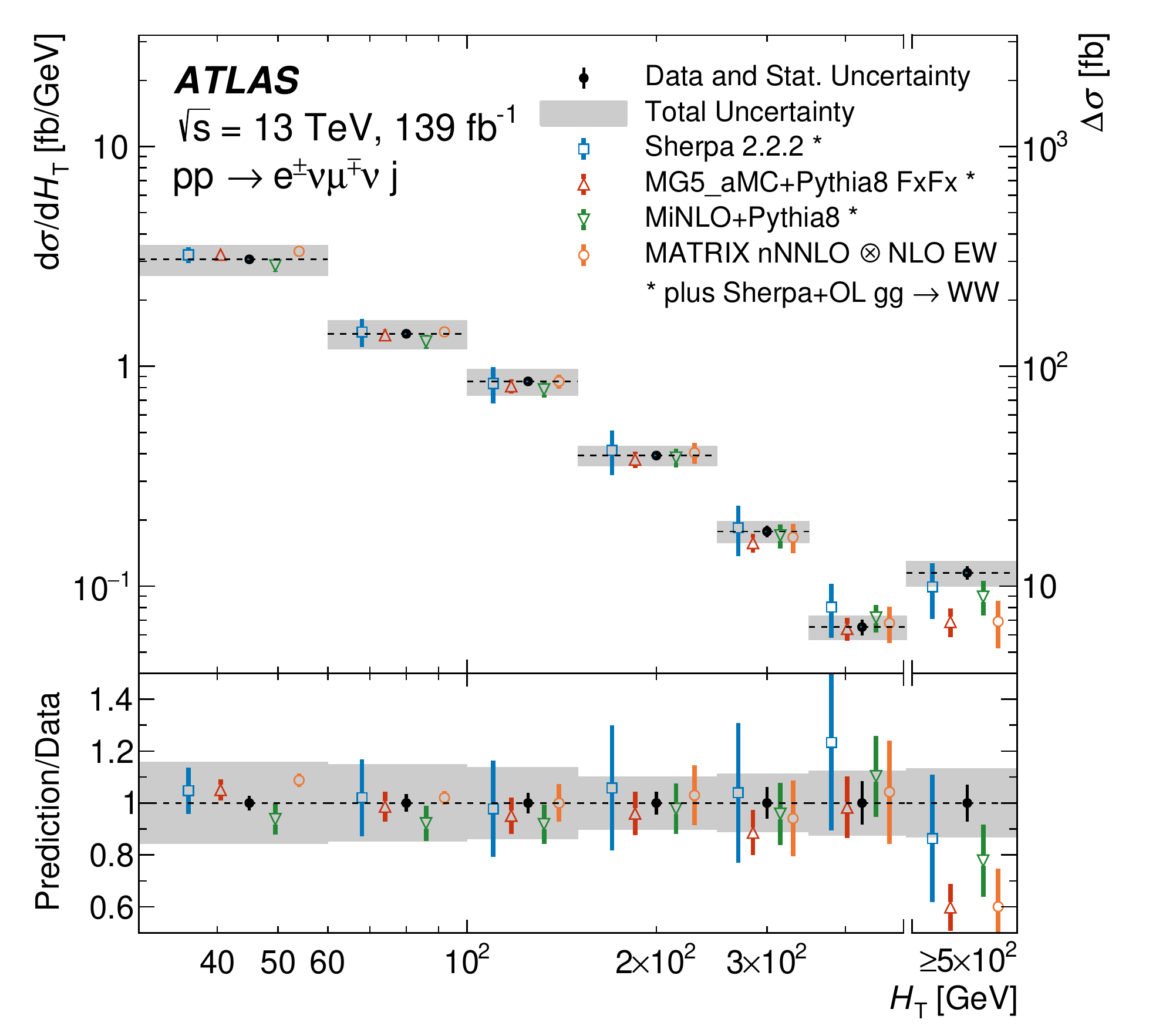}
 
\caption{Measured fiducial cross-sections of \WWj production for (from left to right and top to bottom): \ptleadlep, \ptsublep, \ptleadjet, and \HT.
The last bin of each distribution is inclusive in the measured observable and the corresponding integrated cross-section is indicated by the right-hand-side axis.
\ResFigCaption}
\label{fig:DiffResults-1}
\end{figure}

\begin{figure}
\centering
\includegraphics[width=0.49\textwidth]{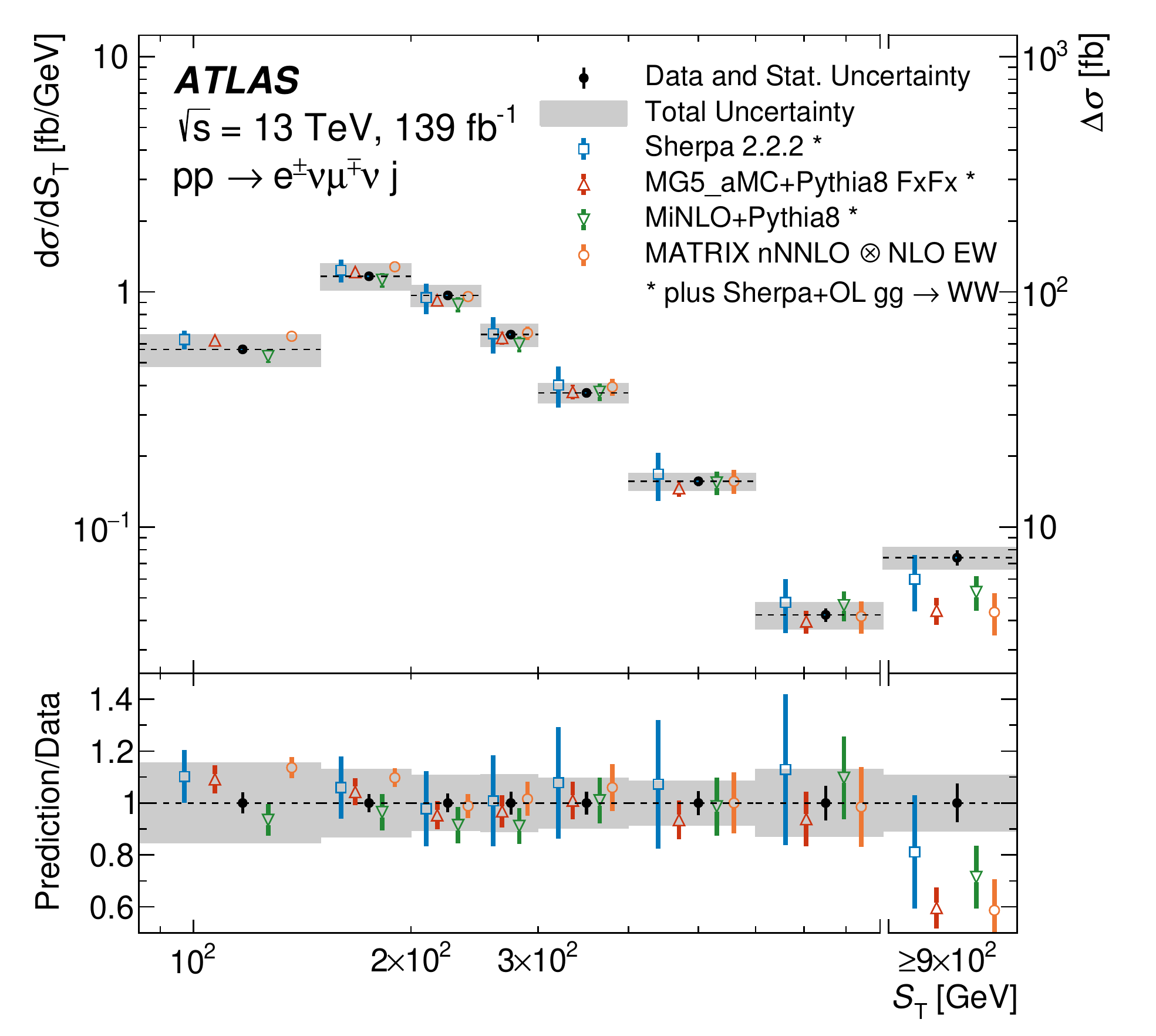}
\includegraphics[width=0.49\textwidth]{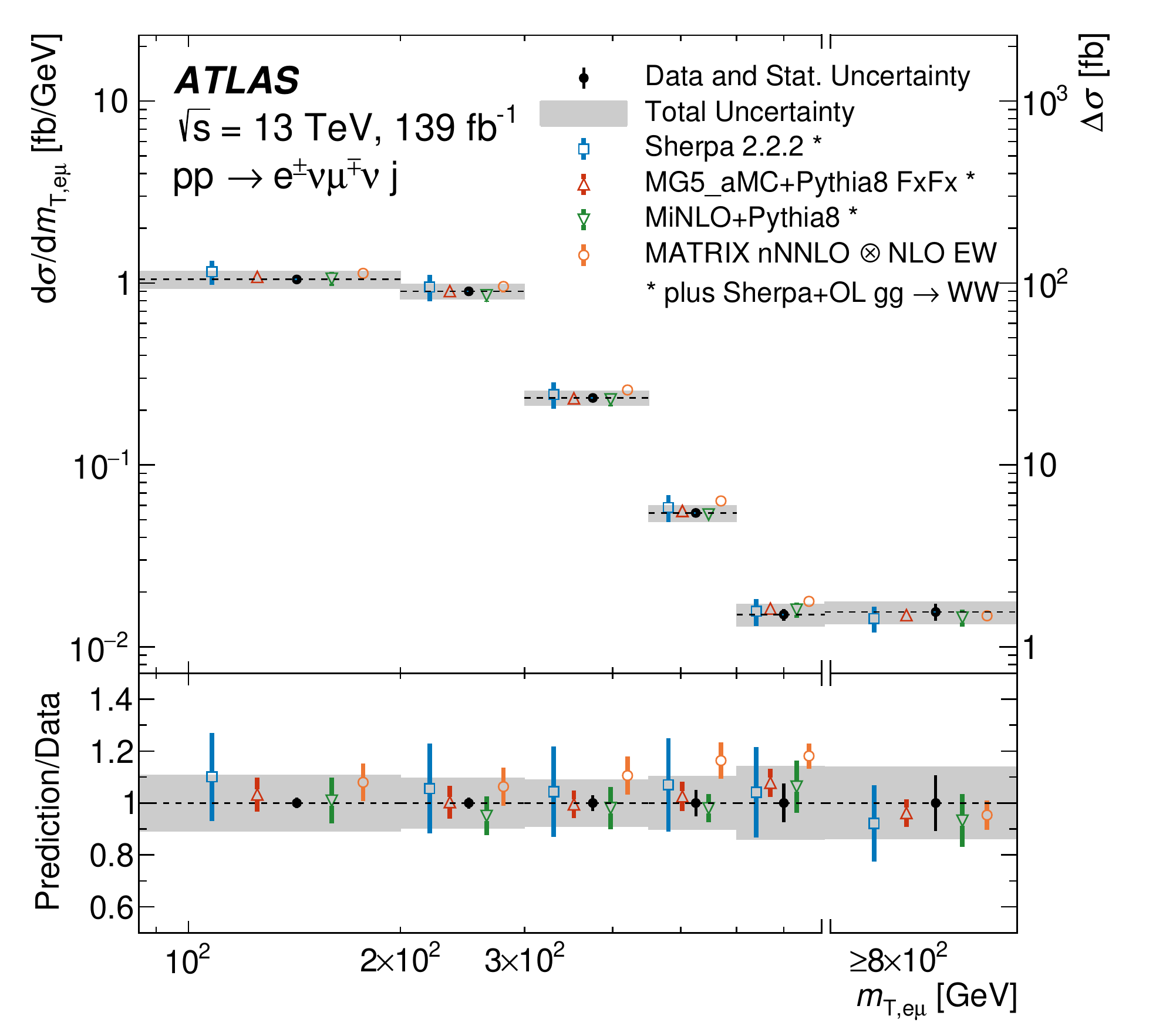}
\includegraphics[width=0.49\textwidth]{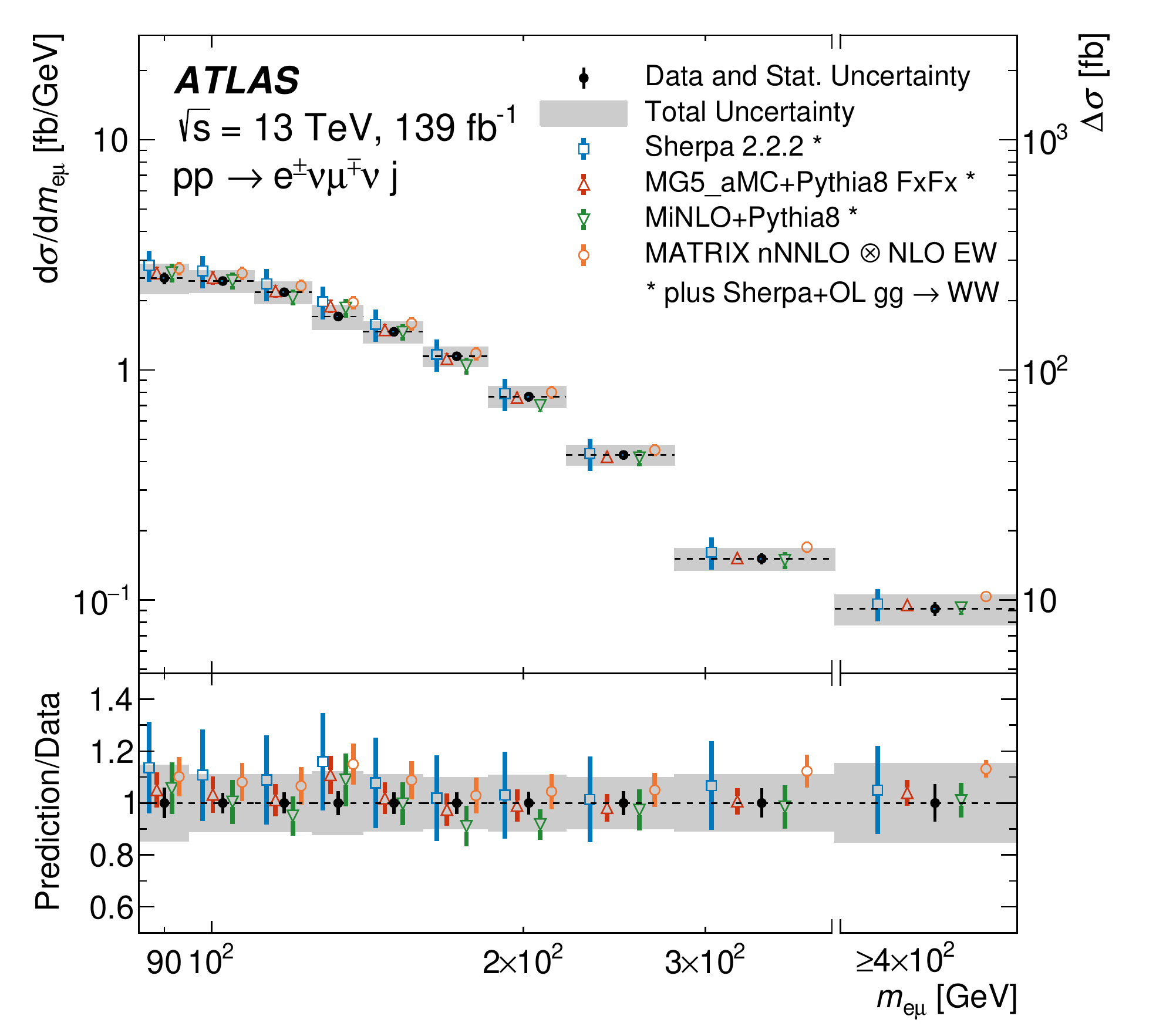}
\includegraphics[width=0.49\textwidth]{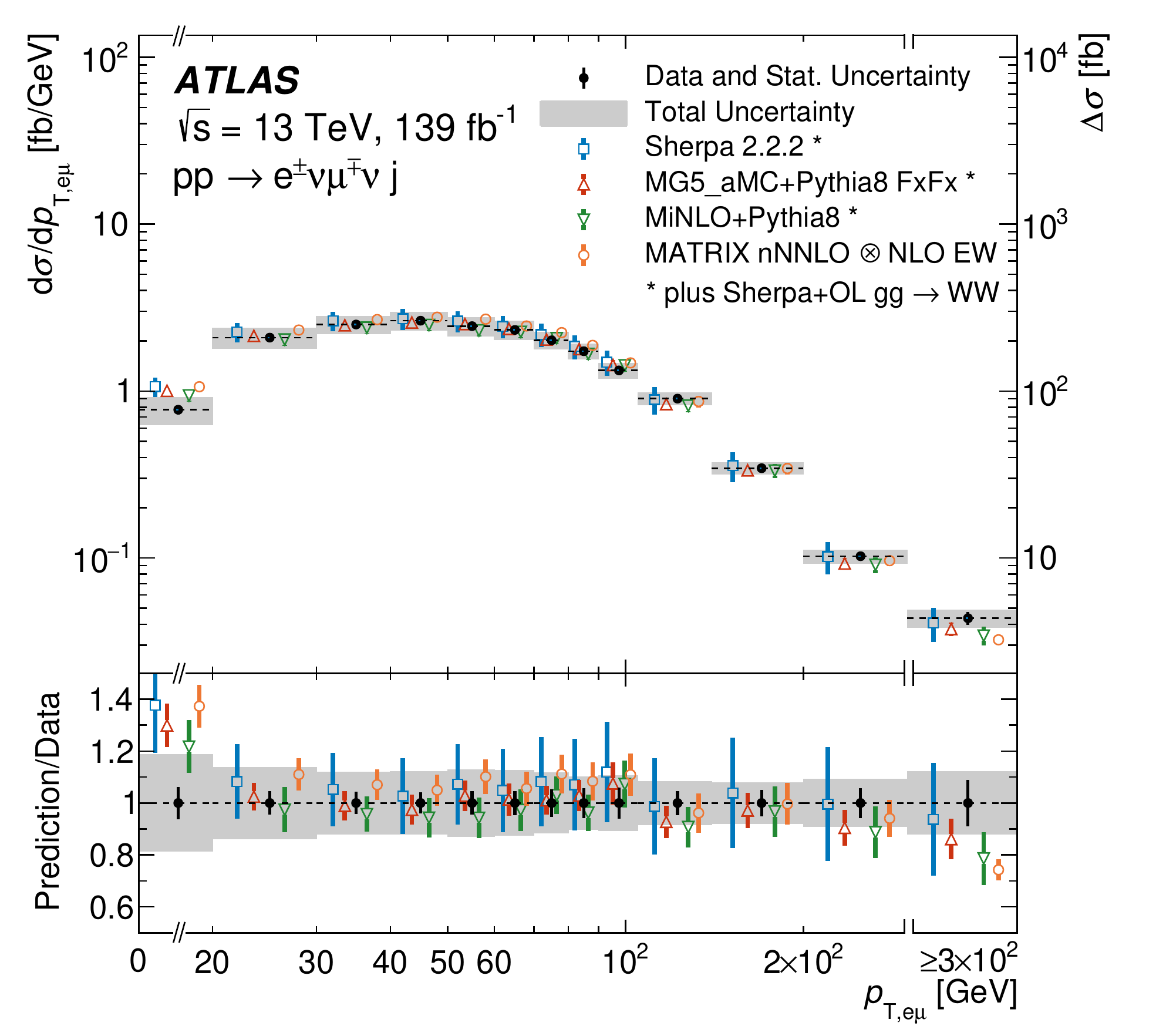}
\caption{
Measured fiducial cross-sections of \WWj production for (from left to right and top to bottom): \ST, \mtll, \mll, and \ptll.
The last bin of each distribution is inclusive in the measured observable and the corresponding integrated cross-section is indicated by the right-hand-side axis.
\ResFigCaption
}
\label{fig:DiffResults-2}
\end{figure}
 
\begin{figure}
\centering
\includegraphics[width=0.49\textwidth]{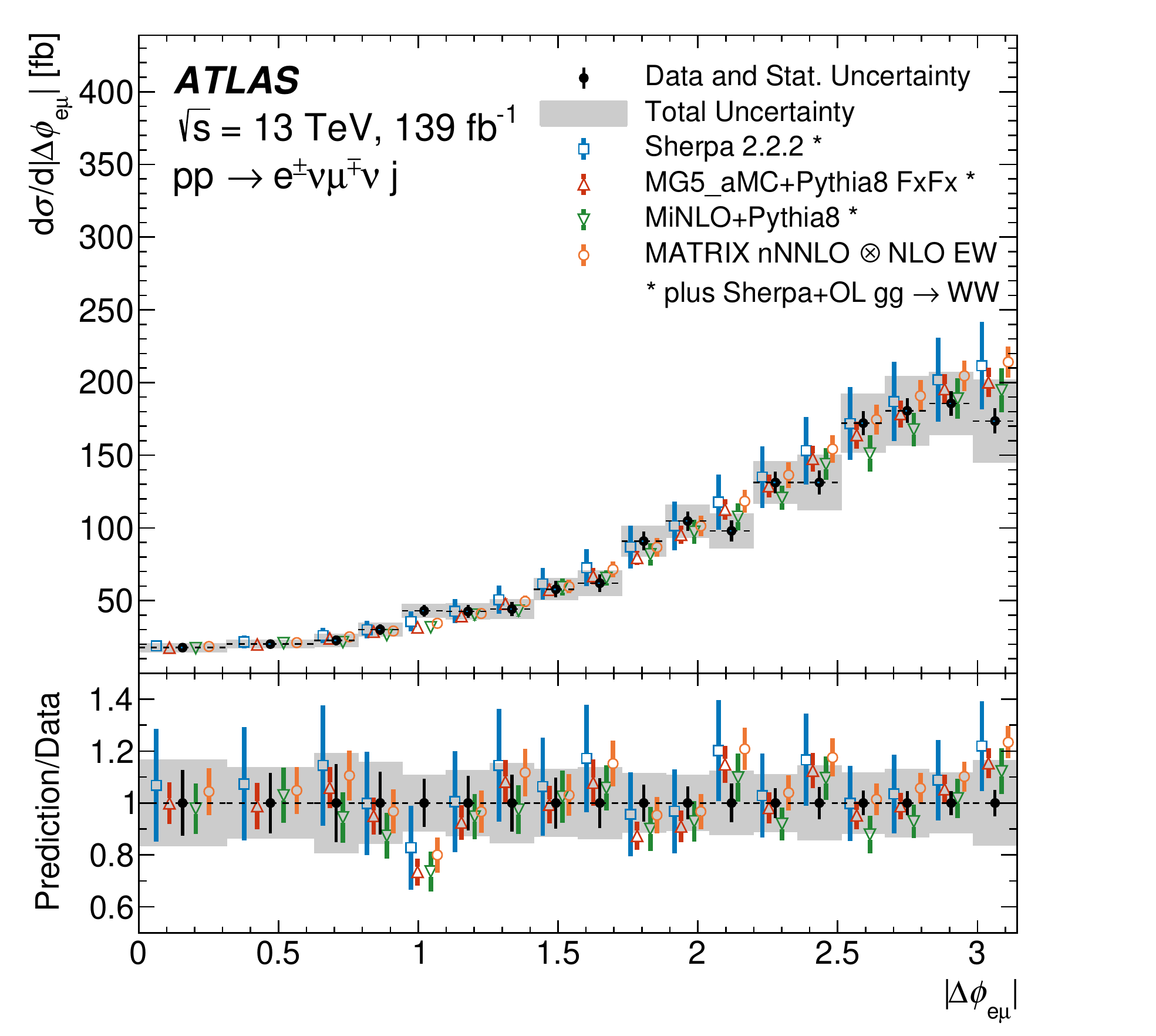}
\includegraphics[width=0.49\textwidth]{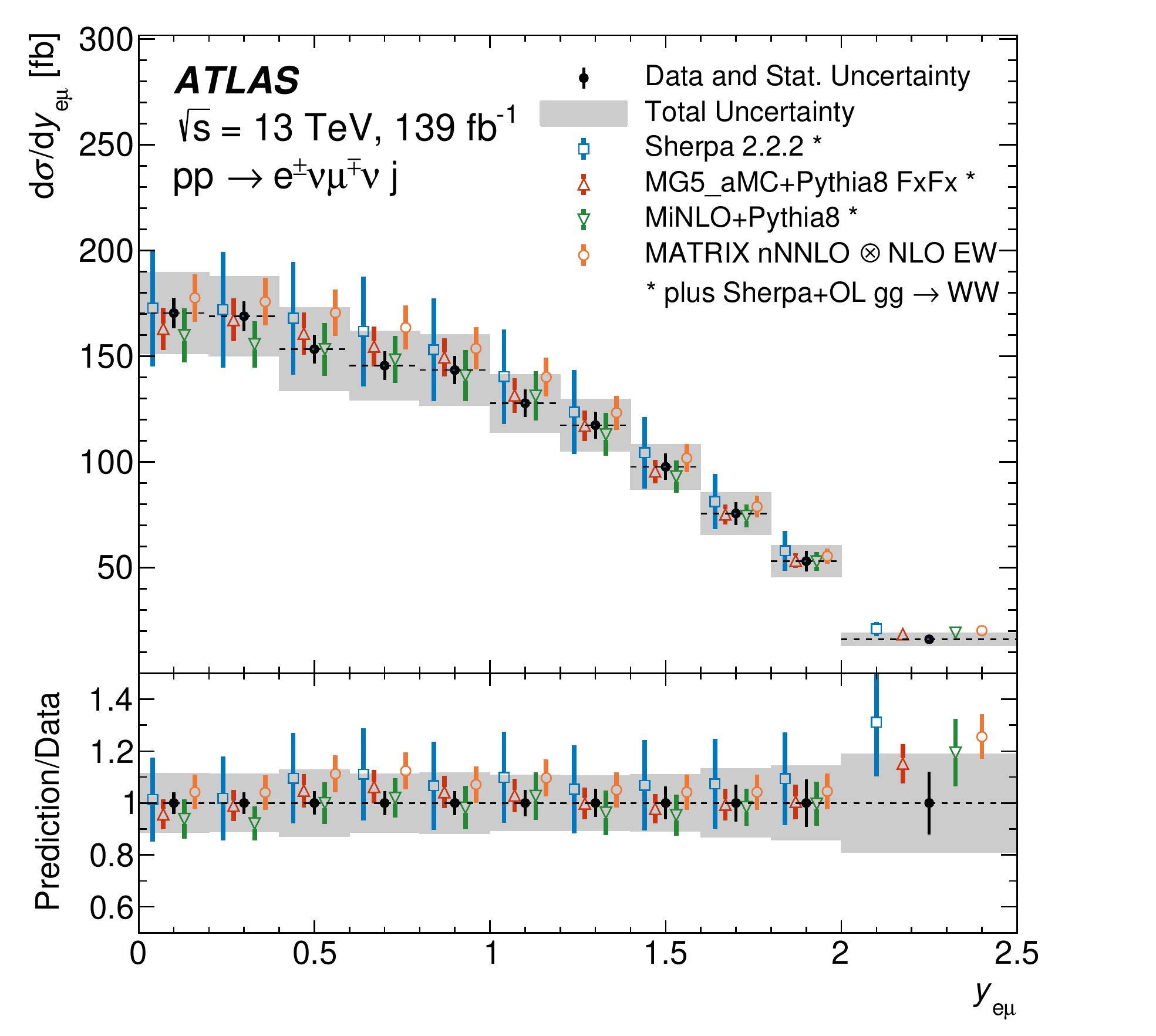}
\includegraphics[width=0.49\textwidth]{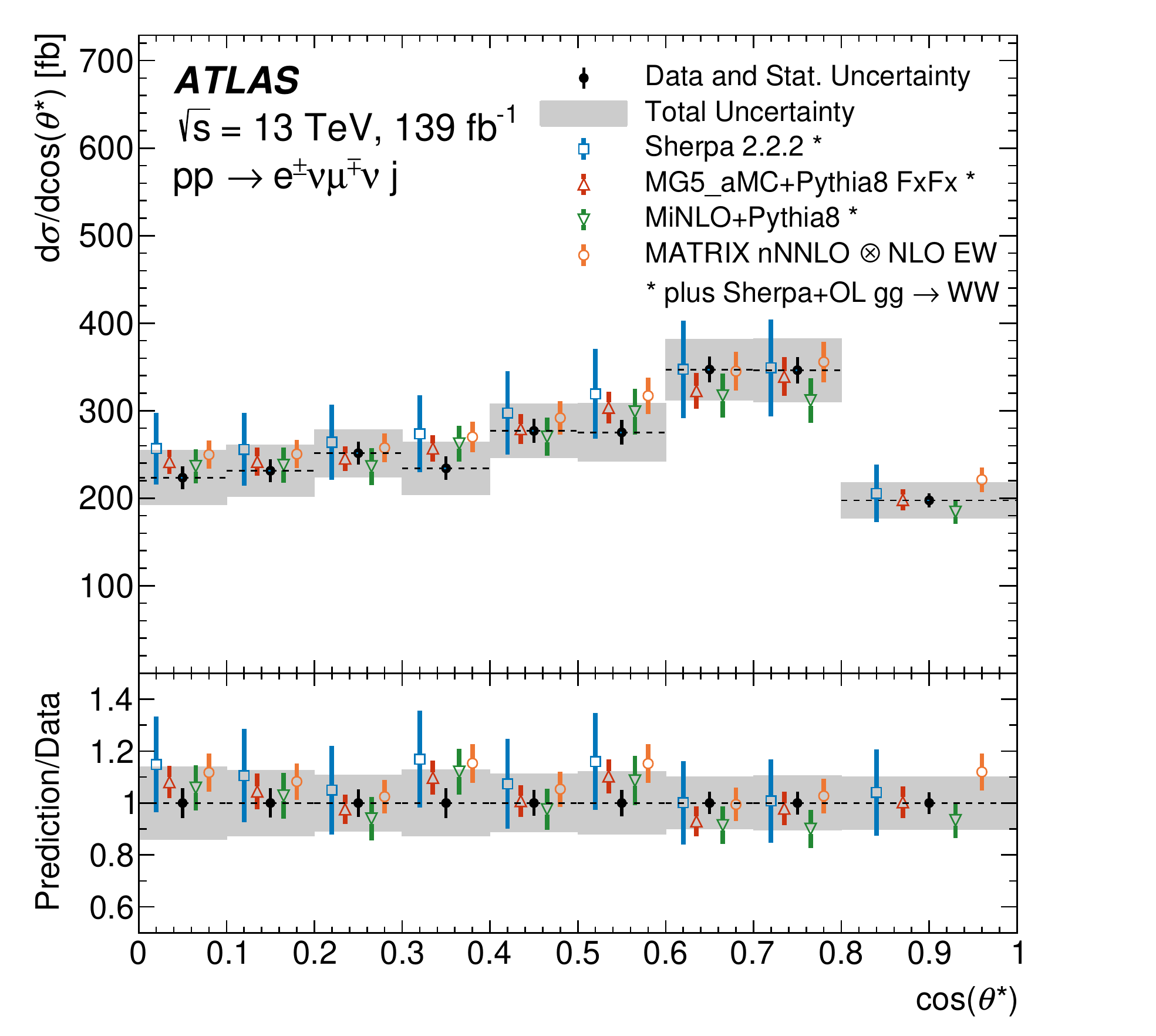}
\includegraphics[width=0.49\textwidth]{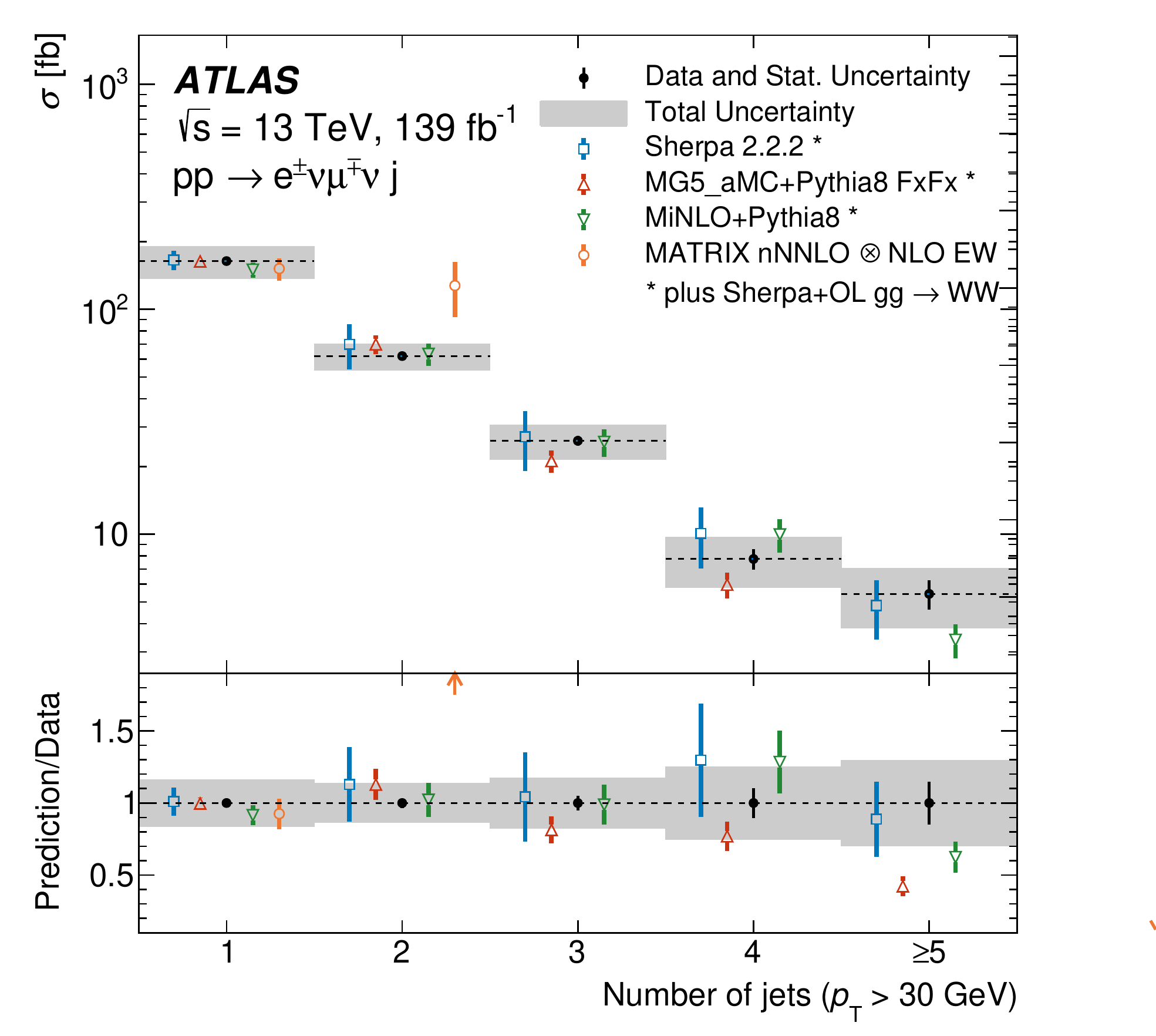}
\caption{Measured fiducial cross-sections of \WWj production for (from left to right and top to bottom): \dphill, \yll, \costhetastar, and the exclusive jet multiplicity.
\ResFigCaption
The MATRIX prediction is not defined for more than two jet emissions.
}
\label{fig:DiffResults-3}
\end{figure}
 
\begin{figure}
\centering
\includegraphics[width=0.49\textwidth]{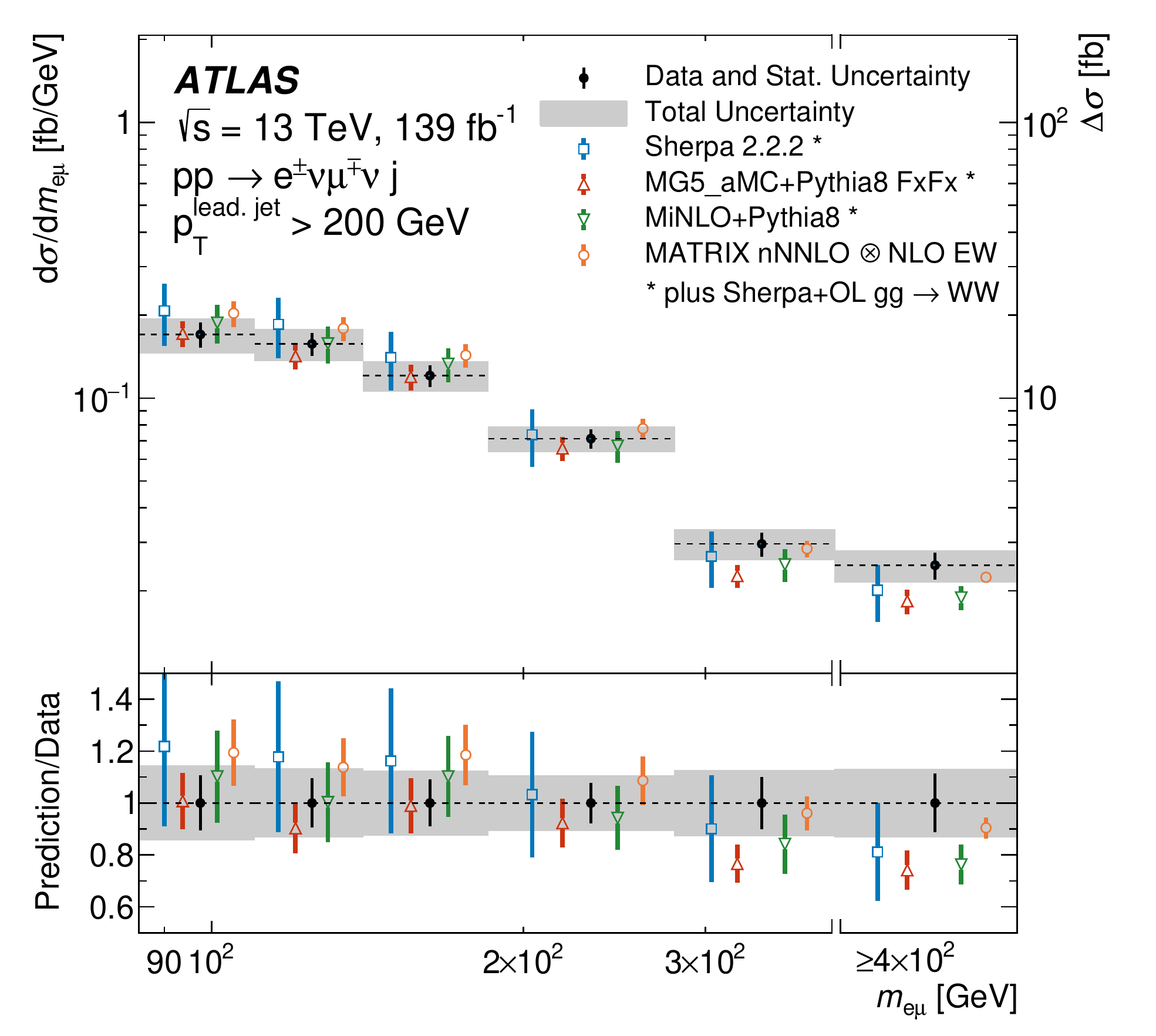}
\includegraphics[width=0.49\textwidth]{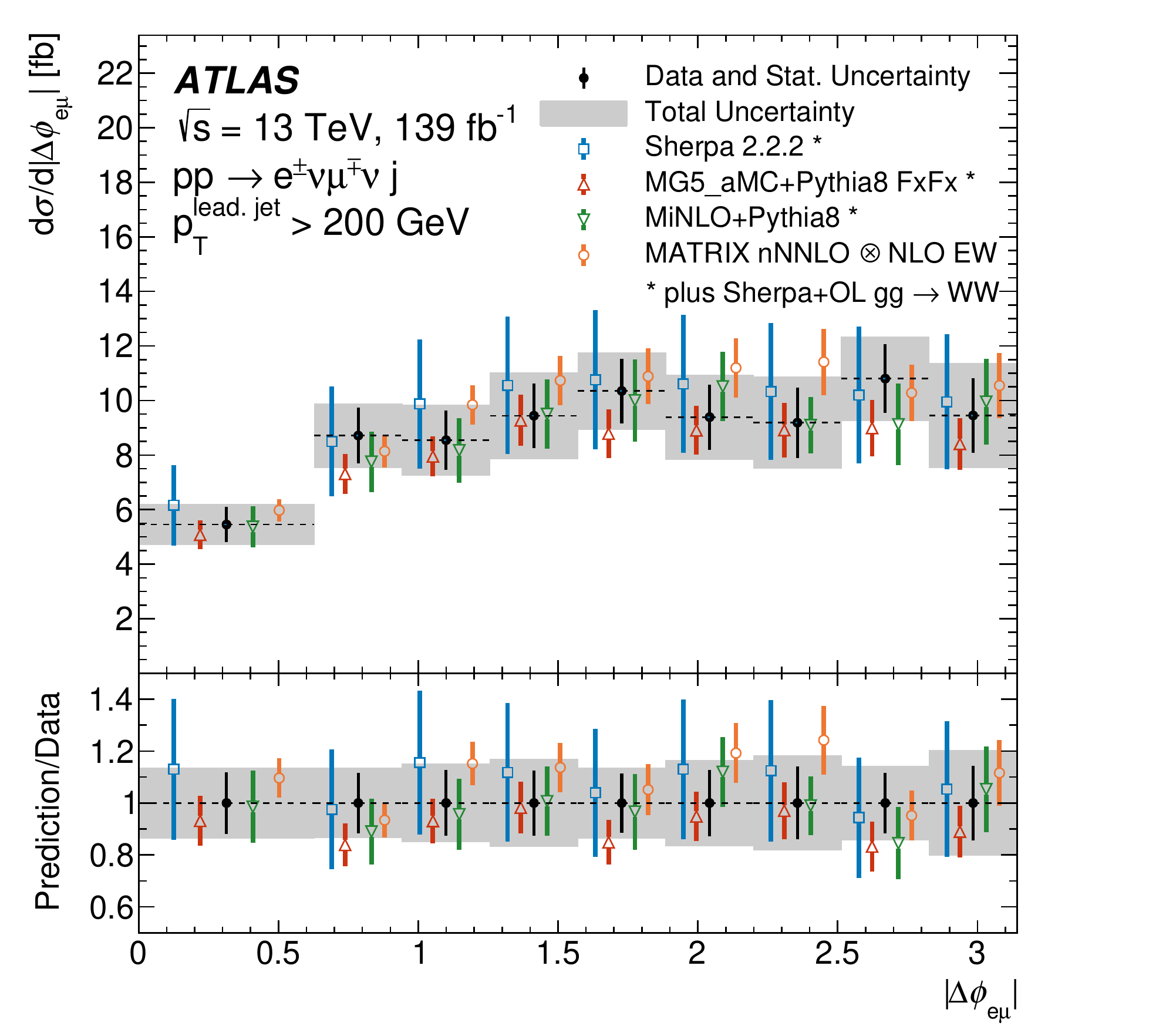}
\caption{Measured fiducial cross-sections of \WWj production for \mll (left) and \dphill (right)  in the fiducial phase space requiring $\ptleadjet>200\,\GeV$.
The last bin of the \mll distribution is inclusive and the corresponding integrated cross-section is indicated by the right-hand-side axis.
\ResFigCaption
}
\label{fig:DiffResults-4}
\end{figure}

\subsection{Effective field theory interpretation}
\label{sub:eft-results}
Many new-physics models that introduce new states at a high energy scale ($\Lambda$) can be described, at lower energy scales, by operators with mass dimensions larger than four in an effective field theory (EFT) framework. The higher-dimensional operators of the lowest order that can generate anomalous triple-gauge-boson couplings (aTGC) are of dimension six.
The $Q_{W}$ dimension-six operator, as defined in Ref.~\cite{Grzadkowski:2010es}, is of particular interest for an analysis of diboson production because it can only be measured in processes affected by modifications of the gauge-boson self-couplings.
Its effect increases rapidly with the centre-of-mass energy, making a measurement at the energies probed by the LHC important.
However, the interference of the SM and anomalous amplitudes, and, thus, the observable consequences of the operator, decrease with increasing energy due to the different helicities of the dominant contributions to the two amplitudes~\cite{Azatov:2016sqh,Falkowski:2016cxu}.
As a consequence, the square of the anomalous dimension-six amplitude, which is quadratic in the ratio of the Wilson coefficient, $c_W$, to $\Lambda^2$, dominates, while, in general, the interference of dimension-six operators with the SM is expected to be larger, as it is linear in $c_W/\Lambda^2$ and, thus, less suppressed by $\Lambda$.
The interference-suppression weakens the limits on $c_W$ that can be achieved by a measurement of diboson production and also poses a problem for the validity of an interpretation in a dimension-six model, since other terms of order $\Lambda^{-4}$, for example those due to dimension-eight operators, are neglected.
Requiring a hard jet in addition to the diboson pair alters the relative contributions of different helicity configurations and reduces the suppression of the interference of SM and anomalous amplitudes~\cite{Azatov:2017kzw}.

Constraints on the Wilson coefficient, $c_W$, are determined using the unfolded \mll cross-section, which is the measured distribution most sensitive to the interference of the $Q_W$ operator with the SM. The fit is performed both for jet $\pt > 30\,\GeV$ and for jet $\pt >200\,\GeV$.
The latter selection is used  to enhance the effect of the interference term per the above discussion.
\OnlySupportNote{Appendix~\ref{app:eft} shows a detailed documentation of the EFT interpretation.}
 
Templates of the distributions representing the pure SM contribution, the new-physics contribution, and the interference between the SM and the new-physics contributions at LO are prepared using \MGMCatNLOV{2.7.2}~\cite{Alwall:2011uj}, interfaced to \PYTHIAV{8.244}~\cite{Sjostrand:2014zea}, with the A14 tune~\cite{ATL-PHYS-PUB-2014-021}, for parton showering, and hadronization.
Events with zero or one jet are simulated in \MGMCatNLO and the overlap between matrix-element and parton-shower emissions is removed using the CKKW-L merging procedure~\cite{Lonnblad:2001iq,Lonnblad:2011xx}.
Agreement of the \MGMCatNLO prediction with the baseline \SHERPAV{2.2.2} generator is ensured by applying a bin-wise correction, determined as the ratio of the SM predictions from \SHERPA and \MGMCatNLO.
It is assumed that the relative scale-induced uncertainties of the \SHERPA prediction are also applicable, differentially in \mll, to the prediction that includes the effect of dimension-six operators.
The prediction and the measured cross-section are, then, used to construct a likelihood function. Measurement uncertainties are modelled using a multivariate Gaussian distribution, while QCD scale and PDF uncertainties affecting the theory prediction are considered as nuisance parameters, constrained with a Gaussian distribution. Two nuisance parameters are introduced to model the scale uncertainty affecting the predicted \mll distribution so that its effect is not fully correlated between bins.
The first (second) parameter models the full effect of the scale uncertainty in the first (last) bin of the distribution. The effect decreases linearly with $\log(\mll)$ such that the parameter has no effect in the last (first) bin.
The decorrelation of scale-uncertainty effects increases the width of confidence intervals by up to 40\% relative to a model in which the scale-uncertainty effects are assumed to be fully correlated between bins of \mll.
Confidence intervals for $c_W$ are derived using Wilk's theorem~\cite{Wilks:1938dza}, assuming that the profile likelihood test statistic is $\chi^2$ distributed~\cite{Cowan:2010js}.
 
Observed and expected 95\% confidence intervals for the EFT coefficients are summarized in Table~\ref{tab:eft-results}.
They are presented both for a fit that takes into account only linear terms in the cross-section parameterization
and for a fit that also takes into account quadratic terms due to the square of the dimension-six amplitude.
For jet $\pt > 200\,\GeV$, limits in the linearized EFT expansion are improved relative to a $\pt >30\,\GeV$ requirement, and the impact of the quadratic term is reduced.
As expected, the analysis of the phase space characterized by a high-\pt jet increases the experimental sensitivity to effects proportional to $c_W/\Lambda^2$ due to the reduced suppression of the interference between the SM amplitude and the dimension-six amplitude.
However, pure dimension-six contributions, which are $\mathcal{O}(\Lambda^{-4})$, are still dominant in this phase space, and the EFT expansion in $\Lambda^{-1}$ does not converge quickly. The limits are, thus, not valid in a general SM EFT scenario that includes additional $\Lambda^{-4}$ contributions due to dimension-eight operators.
 
The presented constraints on $c_W$, obtained accounting for quadratic terms, are weaker than those obtained by the ATLAS measurement of \WW events with no associated jets~\cite{STDM-2017-24}.
There, a dataset corresponding to only 36 fb$^{-1}$ was analysed and the results constrain $c_W/\Lambda^2$ to a 95\% confidence interval with a width of $0.5/\TeV^2$. Limits obtained from this measurement when only including linear terms are improved relative to the previous measurement, for which the corresponding confidence interval has a width of $11/\TeV^2$.
The limits from such a linear fit are, however, an order of magnitude weaker than those obtained by the ATLAS analysis of electroweak production of dijets in association with a $Z$ boson~\cite{STDM-2017-27}.

\begin{table}
\caption{Observed and expected confidence intervals (CI) for $c_W$ for a linearized and a quadratic EFT fit of \mll, when requiring either jet $\pt>30\,\GeV$ or jet $\pt>200\,\GeV$. The new-physics scale $\Lambda$ is set to 1\,\TeV.}
\label{tab:eft-results}
\centering
\begin{tabular}{cccccc}
\toprule
 
Jet \pt & Linear only & 68\% CI obs. & 95\% CI obs.& 68\% CI exp. & 95\% CI exp.\\
\midrule
~~$>30\,\GeV$  & yes & [--1.64, 2.86] & [--3.85, 4.97] & [--2.30, 2.27] & [--4.53, 4.41] \\
~~$>30\,\GeV$  & ~\,no  & [--0.20, 0.20] & [--0.33, 0.33] & [--0.28, 0.27] & [--0.39, 0.38] \\
$> 200\,\GeV$ & yes & [--0.29, 1.84] & [--1.37, 2.81] & [--1.12, 1.09] & [--2.24, 2.10] \\
$> 200\,\GeV$ & ~\,no  & [--0.43, 0.46] & [--0.60, 0.58] & [--0.38, 0.33] & [--0.53, 0.48] \\
\bottomrule
\end{tabular}
\end{table}
 
 
\section{Conclusion}
\label{sec:conclusion}
 
The cross-section for the production of $W$-boson pairs decaying into $e^\pm\nu_e\mu^\mp\nu_\mu$ final states in $pp$ collisions at $\sqrt{s}=13$\,\TeV
is measured in a fiducial phase space that requires the presence of at least one hadronic jet with transverse momentum of at least 30\,\gev, providing jet-inclusive measurements in \WW events.
The measurement is performed with data recorded by the ATLAS experiment at the LHC between 2015 and 2018 that correspond to an integrated luminosity of $139\,\ifb$.
The measured fiducial cross-section,
$\sigma_\text{fid}=258\pm4\text{ (stat)}\pm25\text{ (syst)}~\text{fb}$,
is found to be consistent with theoretical predictions.
With a total uncertainty of 10\%, this result represents a precise measurement of \WW production in association with jets at the LHC that probes a previously unexplored event topology.
Differential cross-sections for \WWj production are measured as a function of the kinematics of the final-state charged leptons, jets, and missing transverse momentum, and are compared with predictions from perturbative QCD calculations.
The data agree well with predictions in all differential distributions, up to the highest measured transverse momenta and for up to five jets.
Dimension six operators that produce anomalous triple-gauge-boson interactions are studied in a phase space that benefits from enhanced interference between the Standard Model amplitude and the anomalous amplitude.
 
\clearpage
\appendix
\part*{Appendix}
\addcontentsline{toc}{part}{Appendix}
 
\section{Measurement at high \ptleadlep}
\label{app:lep200}
 
At high vector-boson $\pt$, predictions for inclusive $WW$ events suffer from so-called `giant $K$-factors', which correspond to large higher-order corrections for QCD and electroweak effects~\cite{MATRIXGiantK}. These come in part from event topologies similar to those in $W$+jets production with the additional emission of a real $W$ boson from a hard jet.
 
In order to study kinematic configurations that are expected to be strongly affected by higher-order EW and QCD corrections, a sample of events is selected with the requirement that $\ptleadlep > 200\, \GeV$. This selects event topologies that generally have a high-\pt $W$ boson accompanied by a lower-\pt $W$ boson.
Here the cross-section is measured differentially in the azimuthal separation between the sub-leading lepton and the leading jet, \dphiljet, and their $\eta$--$\phi$ separation \dRljet, as well as in the ratio of the lepton transverse momenta, $\rlep$, and the ratio of the sub-leading lepton and leading jet transverse momenta, \rjet.
 
\OnlySupportNote{During the analysis it was found that with the requirement of at least one jet the effect of the `giant $K$-factors' is reduced. The measurements with $\ptleadlep > 200\, \GeV$ are thus only presented in an appendix for the publication.}
 
Table~\ref{tab:sr-yields-lep200} lists the selected \WW candidate events in this region, as well as the breakdown of the background estimates. Figure~\ref{fig:reco-level-dist-lep200} shows the measured distributions at detector level in the final analysis binning, comparing the observed data with the signal prediction and the background estimate.
 
The unfolded distributions are shown in Figure~\ref{fig:DiffResults-lep200}. In general, the predictions are in good agreement with the measurement.
 
\begin{table}[hb]
\caption{Selected \WW candidate events, together with the signal prediction and the background estimates for $\ptleadlep > 200\,\GeV$. The uncertainties include statistical and systematic contributions. The fractions in percent give the relative contribution to the total SM prediction.}
\label{tab:sr-yields-lep200}
\centering
\begin{tabular}{lr@{$\,\pm\,$}rr}
\toprule	 & \multicolumn{3}{r}{$\ptleadlep > 200\,\GeV$} \\
\midrule
Data & \multicolumn{2}{l}{3873} & \\
Total SM & 3960 & 120 & \\
\midrule
\WW & 1740 & 50 & 44\%\\
Total bkg. & 2210 & 110 & 56\%\\
\midrule
Top & 1920 & 90 & 49\%\\
Drell--Yan & 42 & 6 & 1\%\\
Fake leptons & 70 & 40 & 2\%\\
$WZ,ZZ,V\gamma$ & 180 & 40 & 4\%\\
\bottomrule
\end{tabular}
\end{table}
 
\begin{figure}
\centering
\includegraphics[width=.45\linewidth]{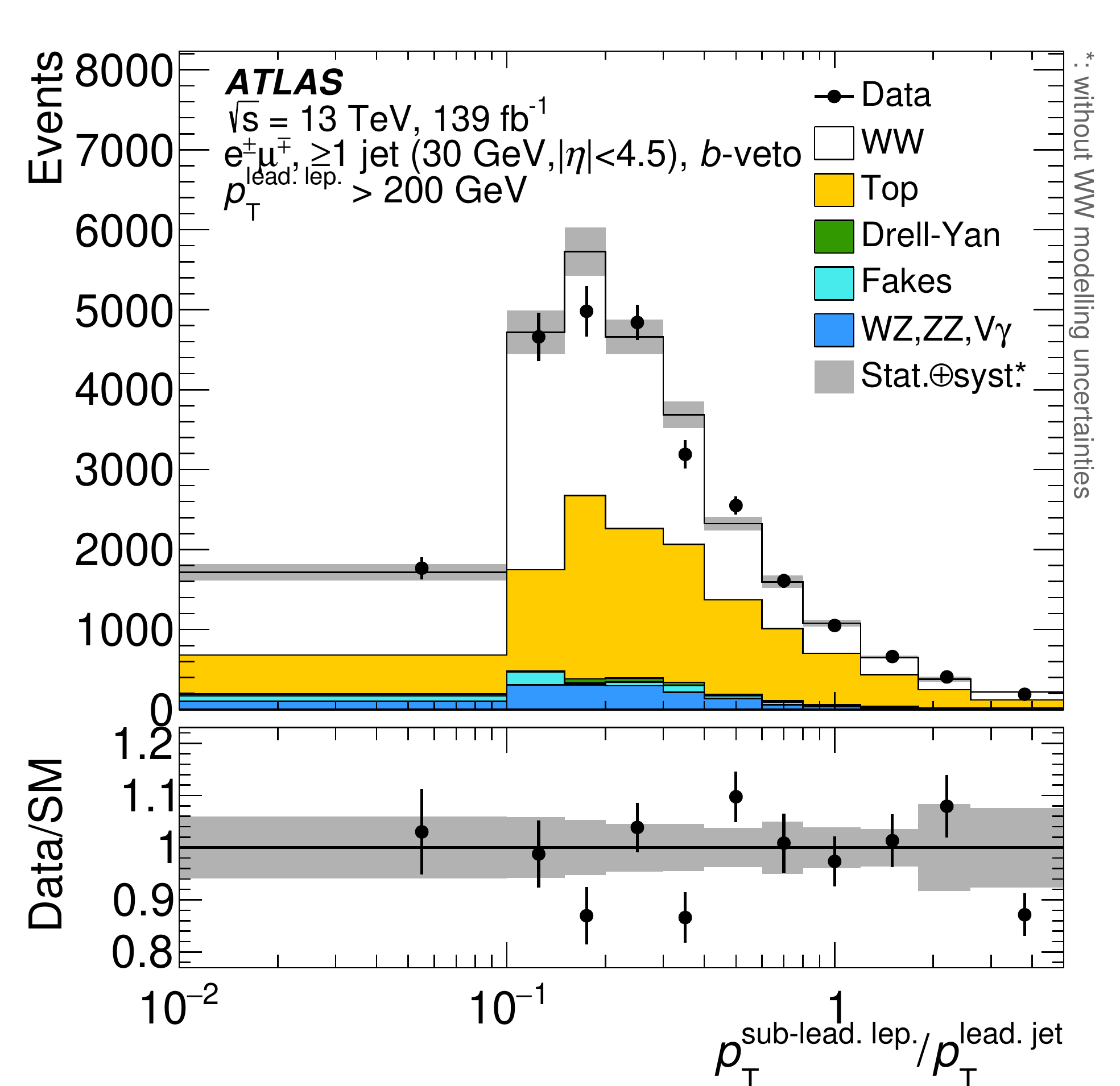}
\includegraphics[width=.45\linewidth]{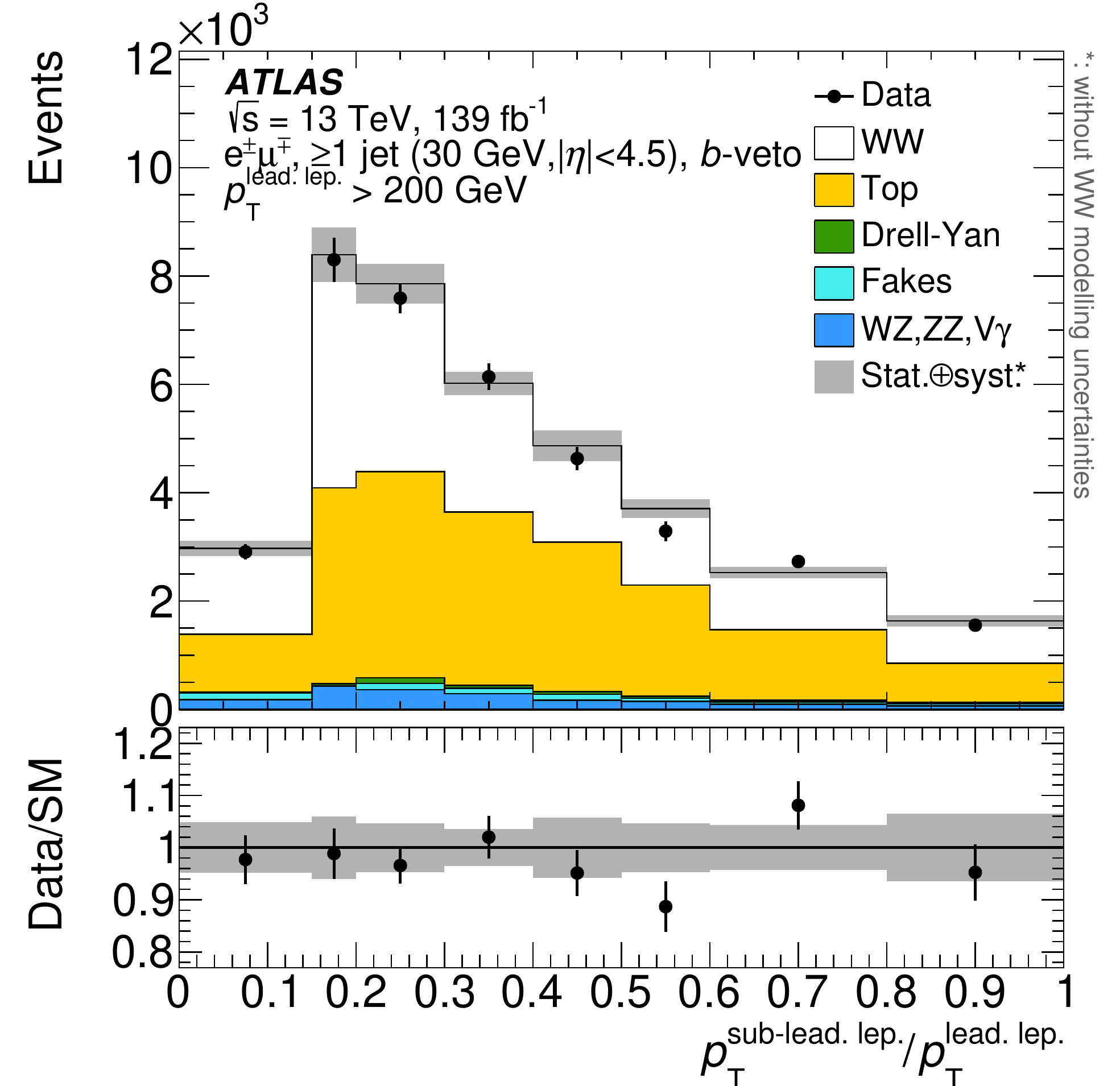}
 
\includegraphics[width=.45\linewidth]{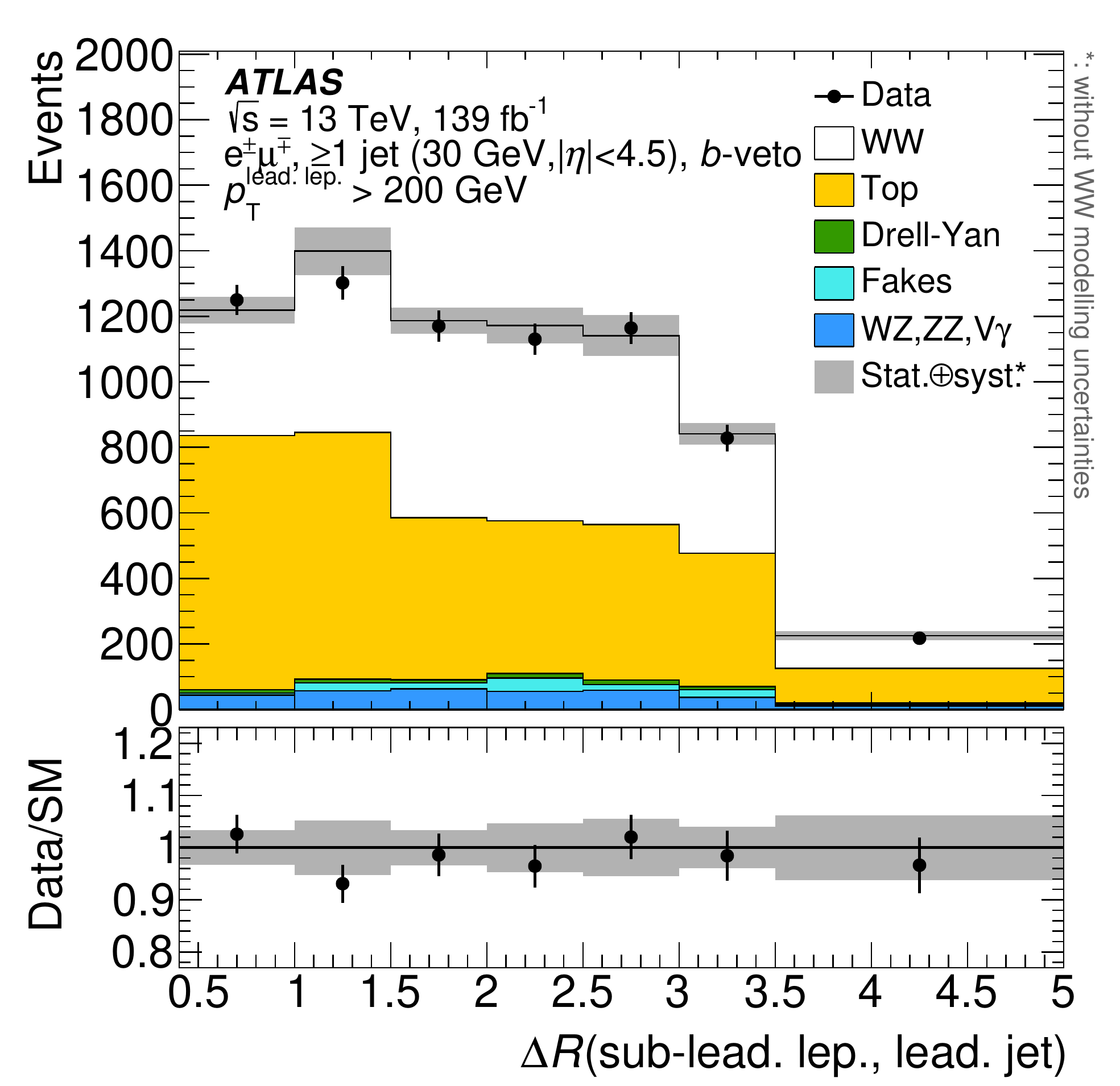}
\includegraphics[width=.45\linewidth]{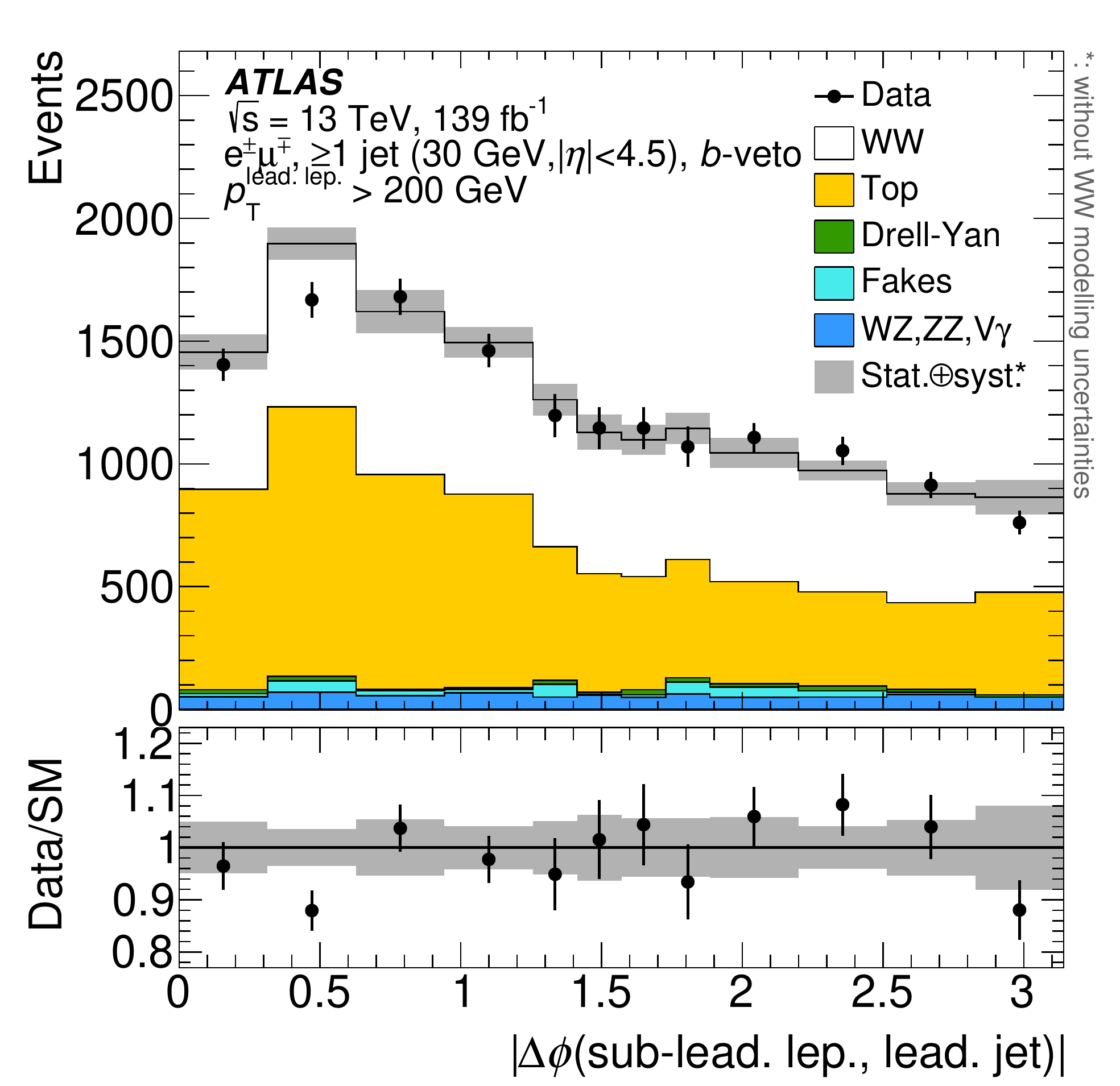}
 
\caption{Signal region detector-level distributions, requiring $\ptleadlep>200\,\GeV$, of \rjet (top left), \rlep (top right), \dRljet (bottom left) and \dphiljet (bottom right). Data are shown as black markers together with the predictions for the signal and background production processes. The last bin contains overflow events.
The lower panels show the ratio of the data to the total prediction.
The uncertainty bands shown include statistical and systematic uncertainties, excluding theory uncertainties on the signal.
}
\label{fig:reco-level-dist-lep200}
 
\end{figure}
 
\begin{figure}
\centering
\includegraphics[width=0.49\textwidth]{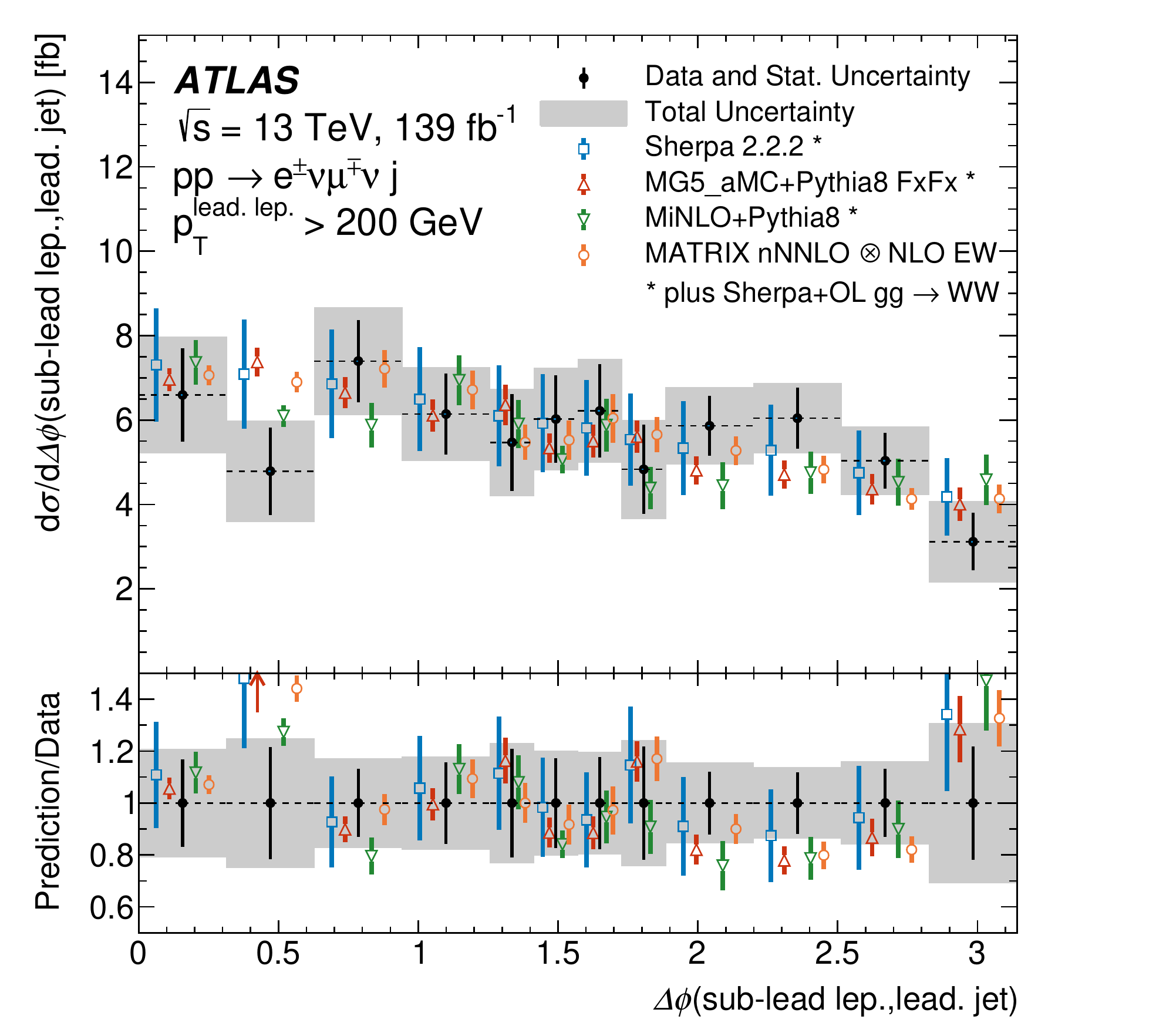}
\includegraphics[width=0.49\textwidth]{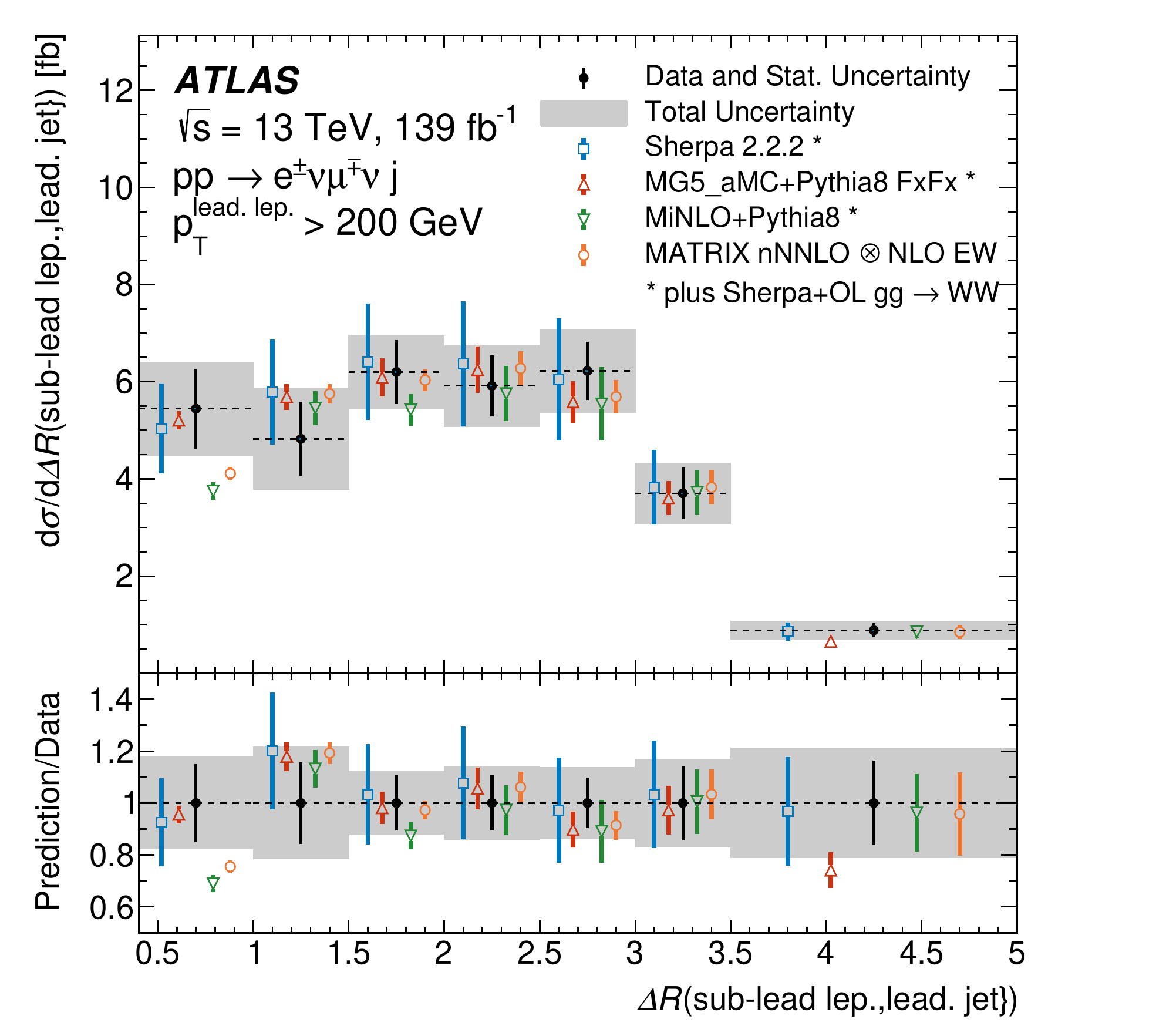}
\includegraphics[width=0.49\textwidth]{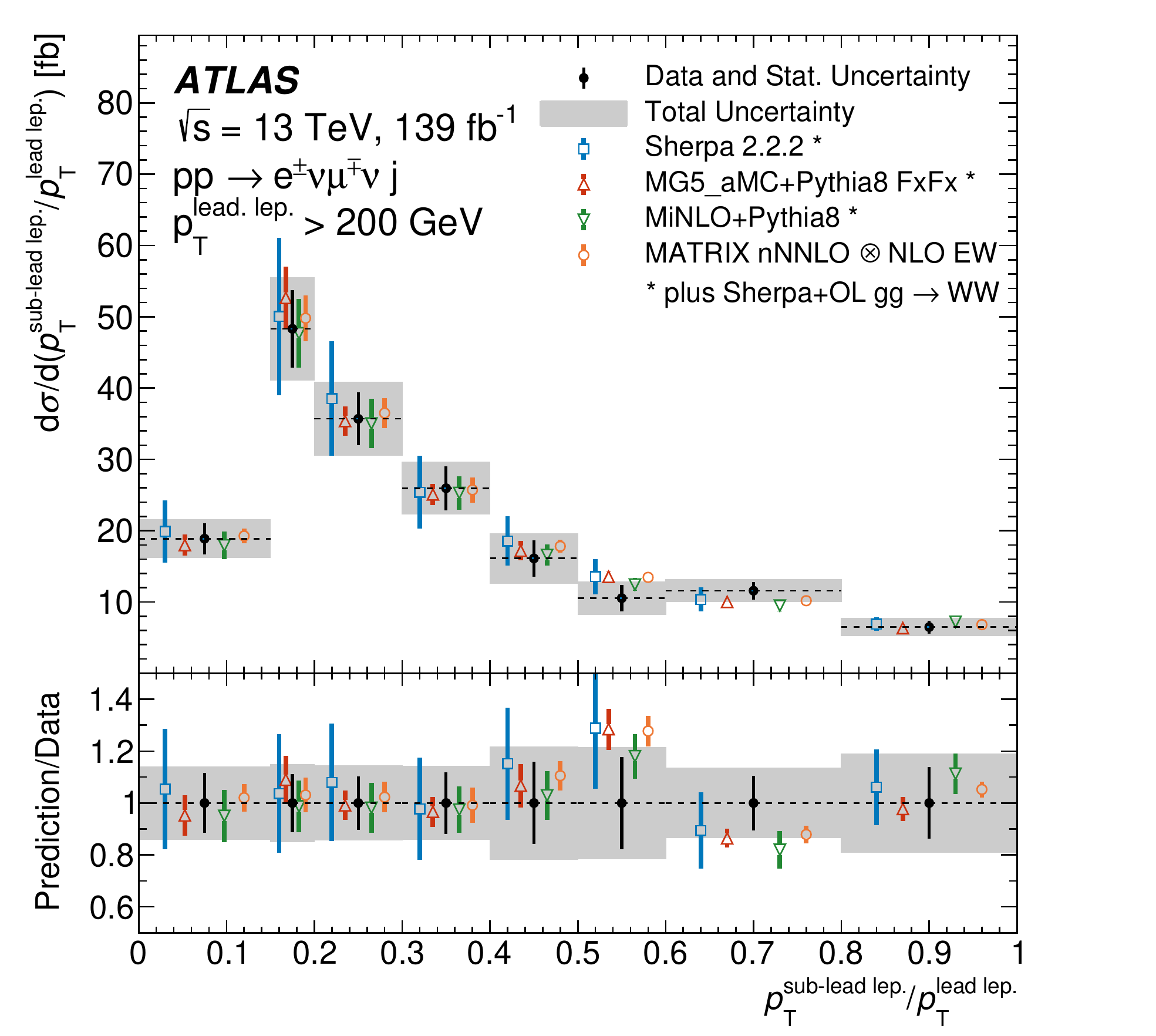}
\includegraphics[width=0.49\textwidth]{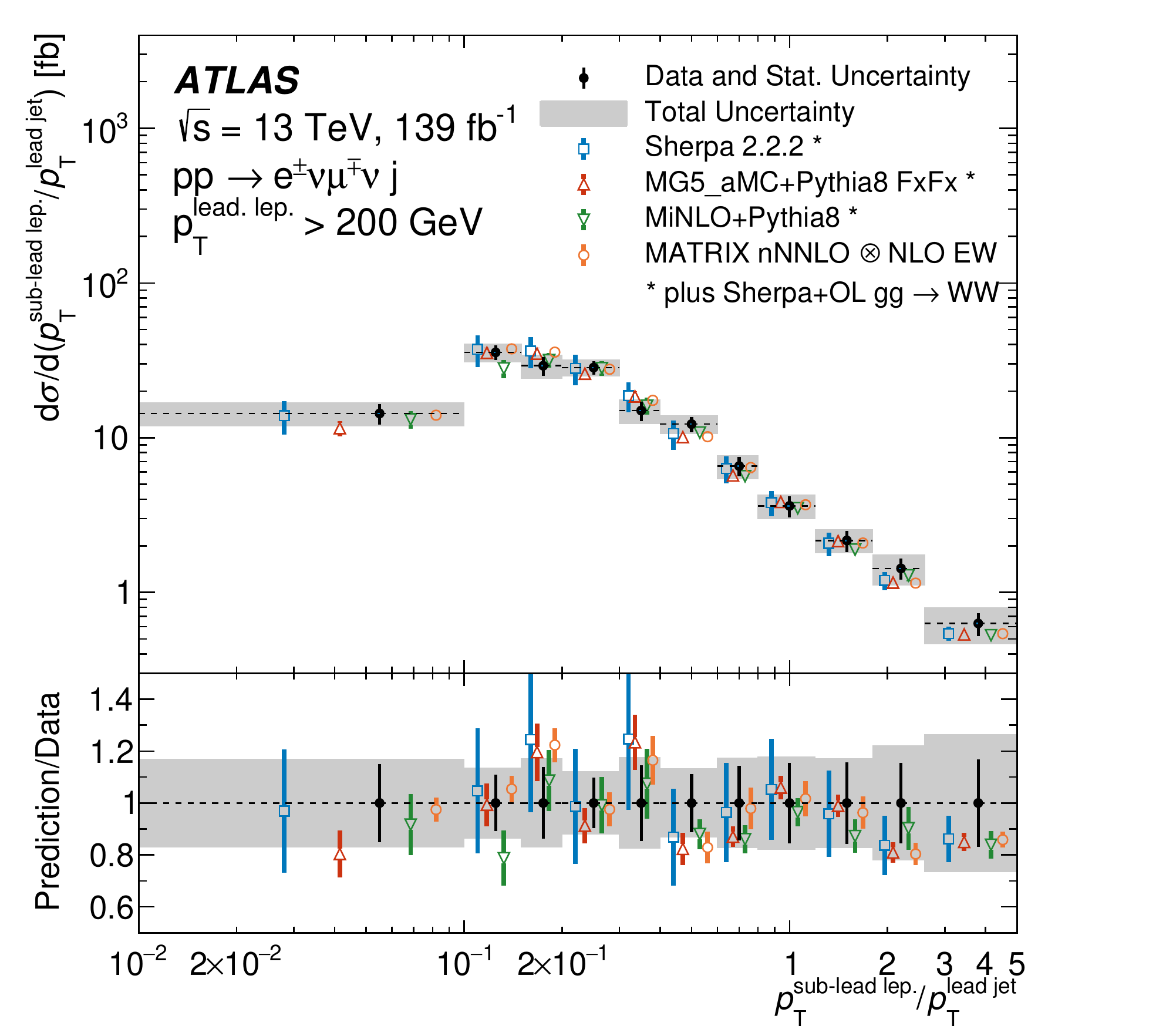}
\caption{Measured fiducial cross-sections of \WWj production for (from left to right and top to bottom): \dphiljet, \dRljet, \rlep, and \rjet in the fiducial phase space requiring $\ptleadlep>200\,\GeV$.
\ResFigCaption
}
\label{fig:DiffResults-lep200}
\end{figure}
 
\FloatBarrier
\section{\ttbar background estimate}
\label{app:topbkg}
 
Figure~\ref{fig:app:ttbar-cr} shows the \ptleadlep and jet multiplicity distributions in the two \ttbar control regions, which require exactly one and exactly two \bjets, respectively.
The \bjet correlation factor $C_b$ for the two distributions is shown in Figure~\ref{fig:app:ttbar-Cb}.
Figure~\ref{fig:app:ttbar-mll} shows the \mll distribution for $\ptleadjet > 200\,\GeV$ in the two control regions and for the top-enriched selection, together with the \bjet correlation factor $C_b$.
The excess of events predicted at high \ptleadlep, in comparison with data, is corrected for by the data-driven estimate, and no discrepancy is seen in the top-enriched selection, as shown in Figure~\ref{fig:vr-datamc} in the main body. The jet multiplicity is well modelled up to five selected jets.
 
\begin{figure}
\centering
\includegraphics[width=.45\linewidth]{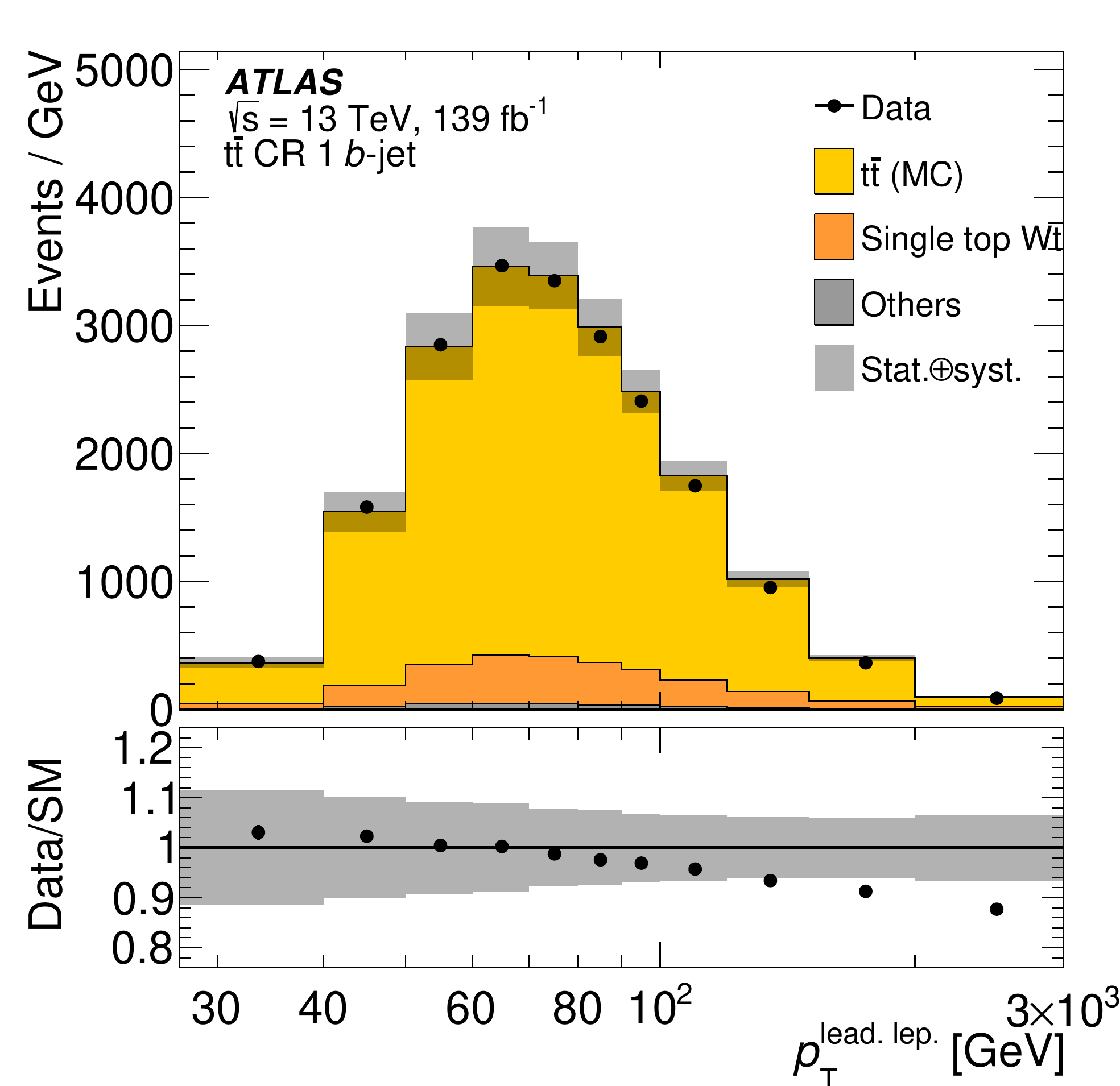}
\includegraphics[width=.45\linewidth]{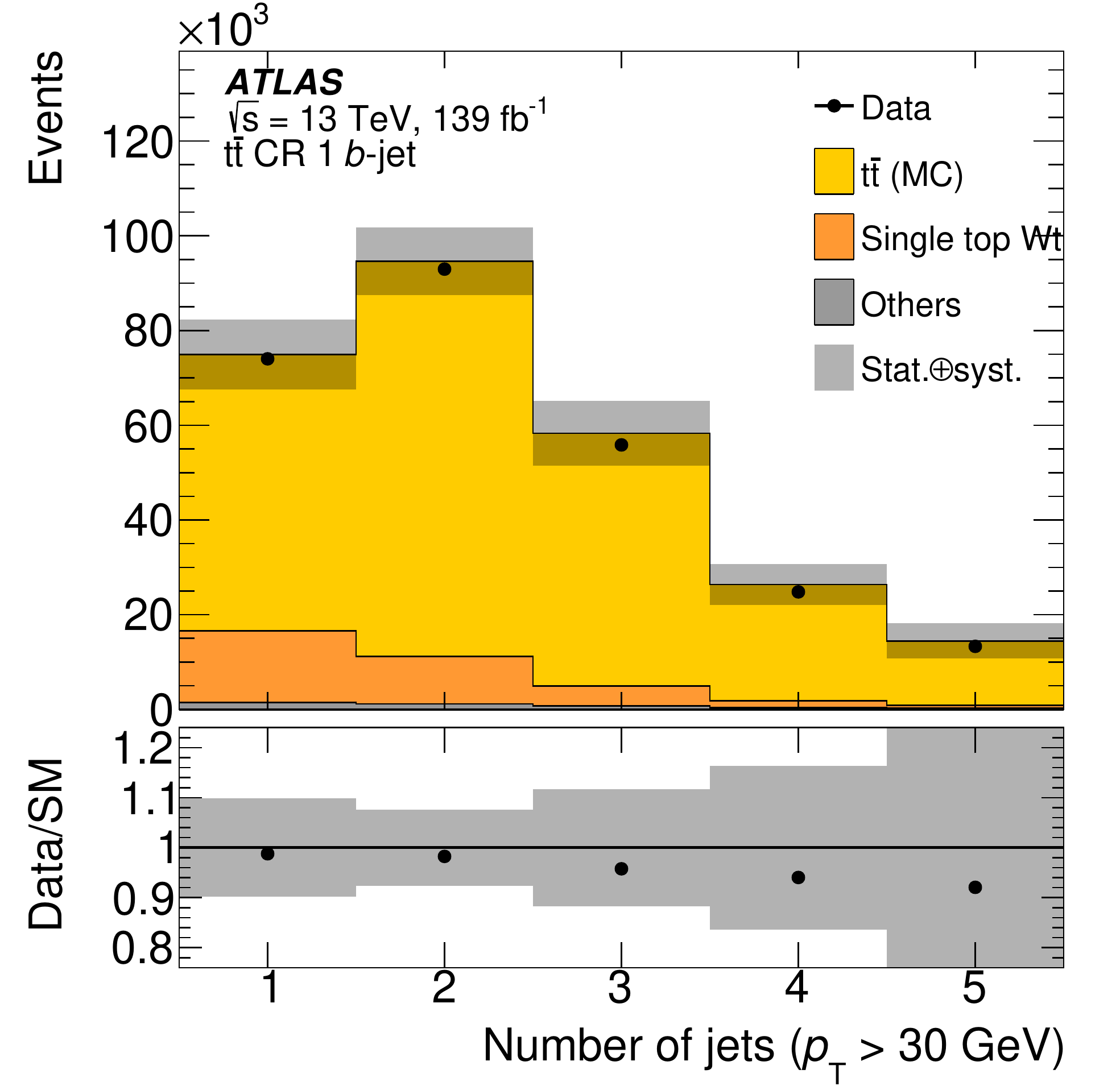}
 
\includegraphics[width=.45\linewidth]{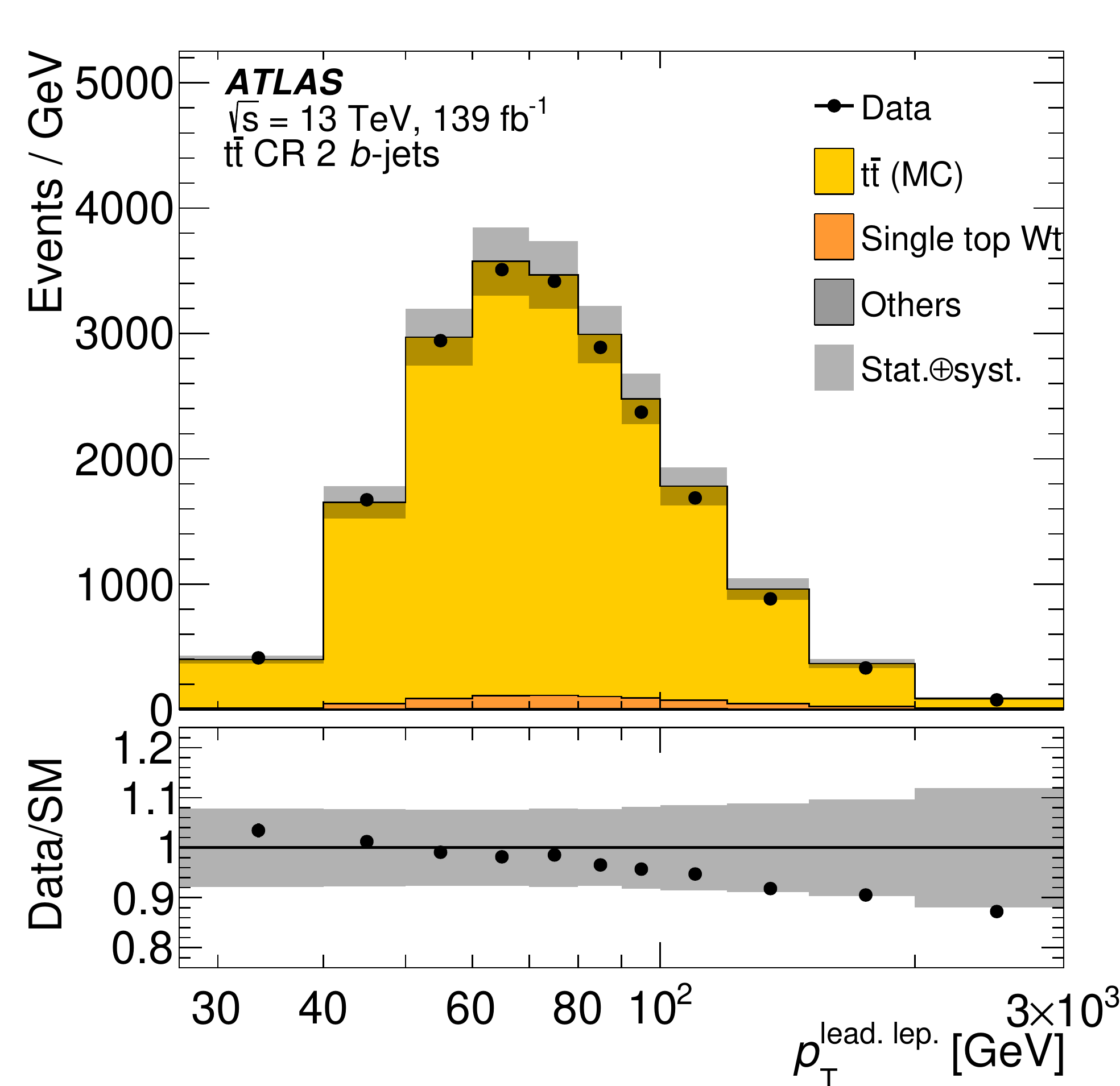}
\includegraphics[width=.45\linewidth]{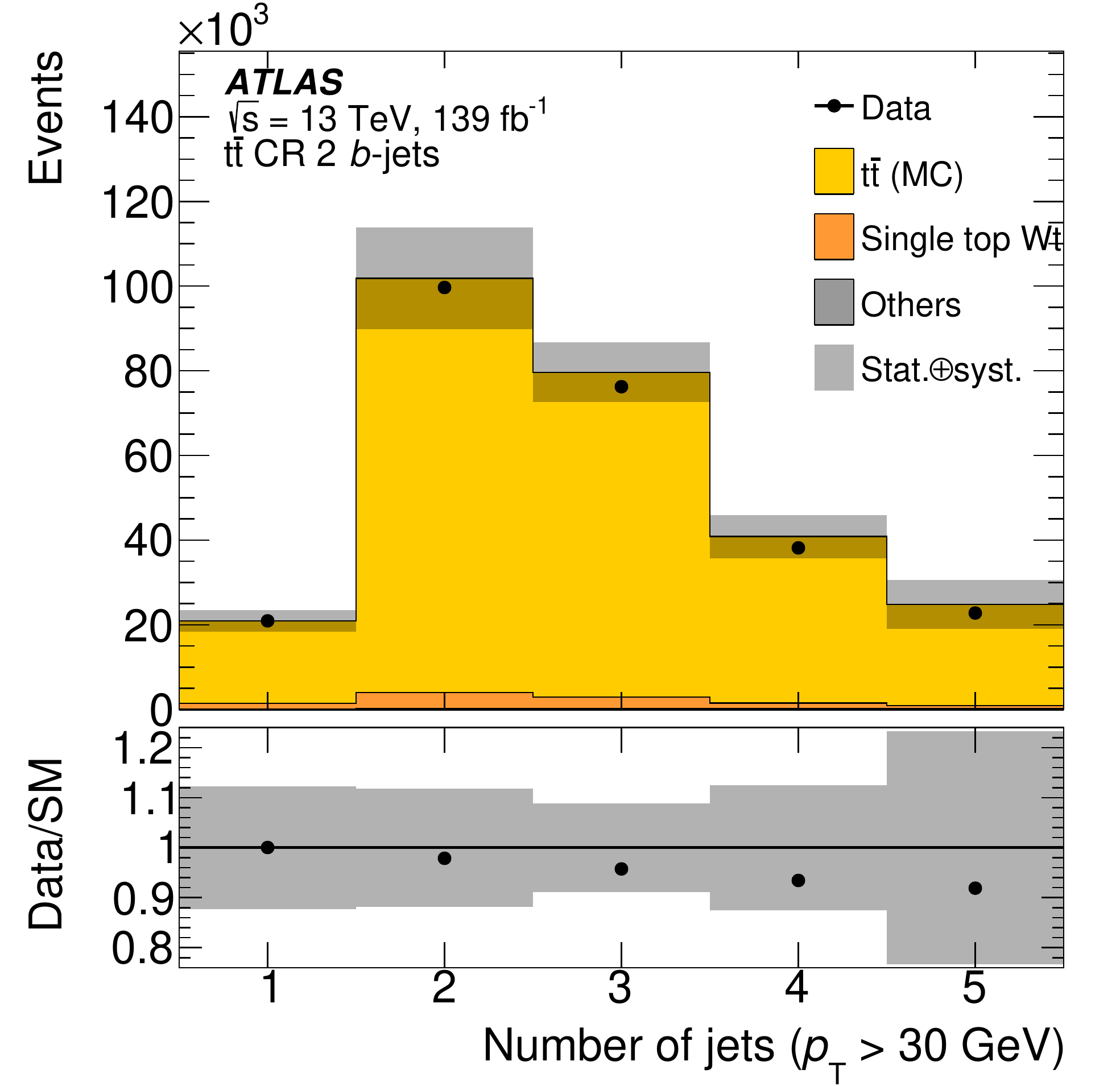}
 
\caption{Detector-level distributions of the \ptleadlep (left) and the jet multiplicity (right) in the \ttbar control regions with one \bjet (left) and two \bjets (right). Data are shown together with the predictions for \ttbar, single-top $Wt$ and other production processes from simulation. The last bin contains overflow events.
The lower panels show the ratio of the data to the total prediction.
The uncertainties shown include statistical and systematic uncertainties.
}
\label{fig:app:ttbar-cr}
\end{figure}
 
\begin{figure}
\centering
\includegraphics[width=.45\linewidth]{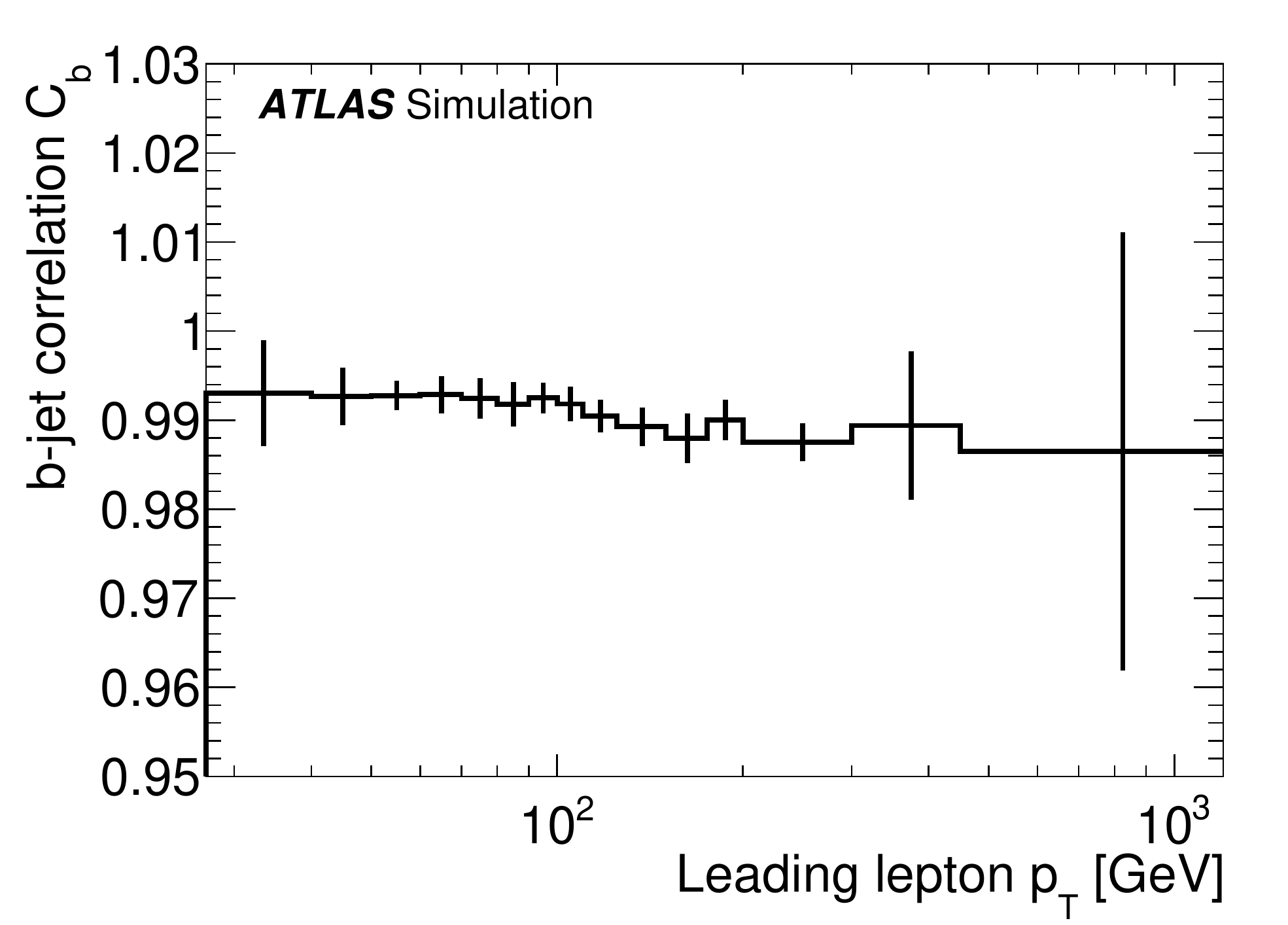}
\includegraphics[width=.45\linewidth]{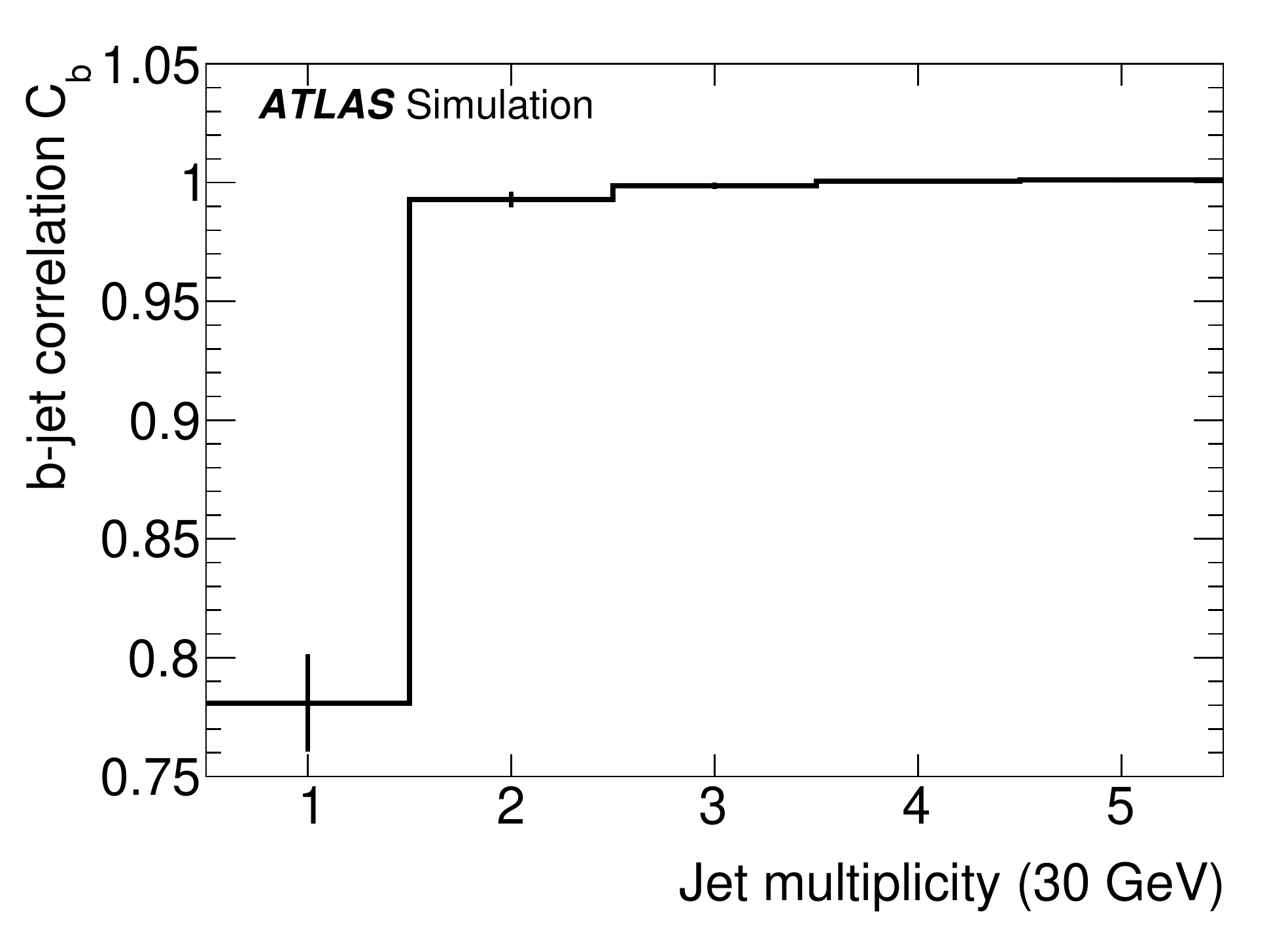}
 
\caption{Distribution of the \bjet correlation factor $C_b$ as a function of the \ptleadlep (left) and the jet multiplicity (right), as determined from the nominal \ttbar simulation. The uncertainties shown include MC statistical and systematic uncertainties. }
\label{fig:app:ttbar-Cb}
\end{figure}
 
\begin{figure}
\centering
\includegraphics[width=.45\linewidth]{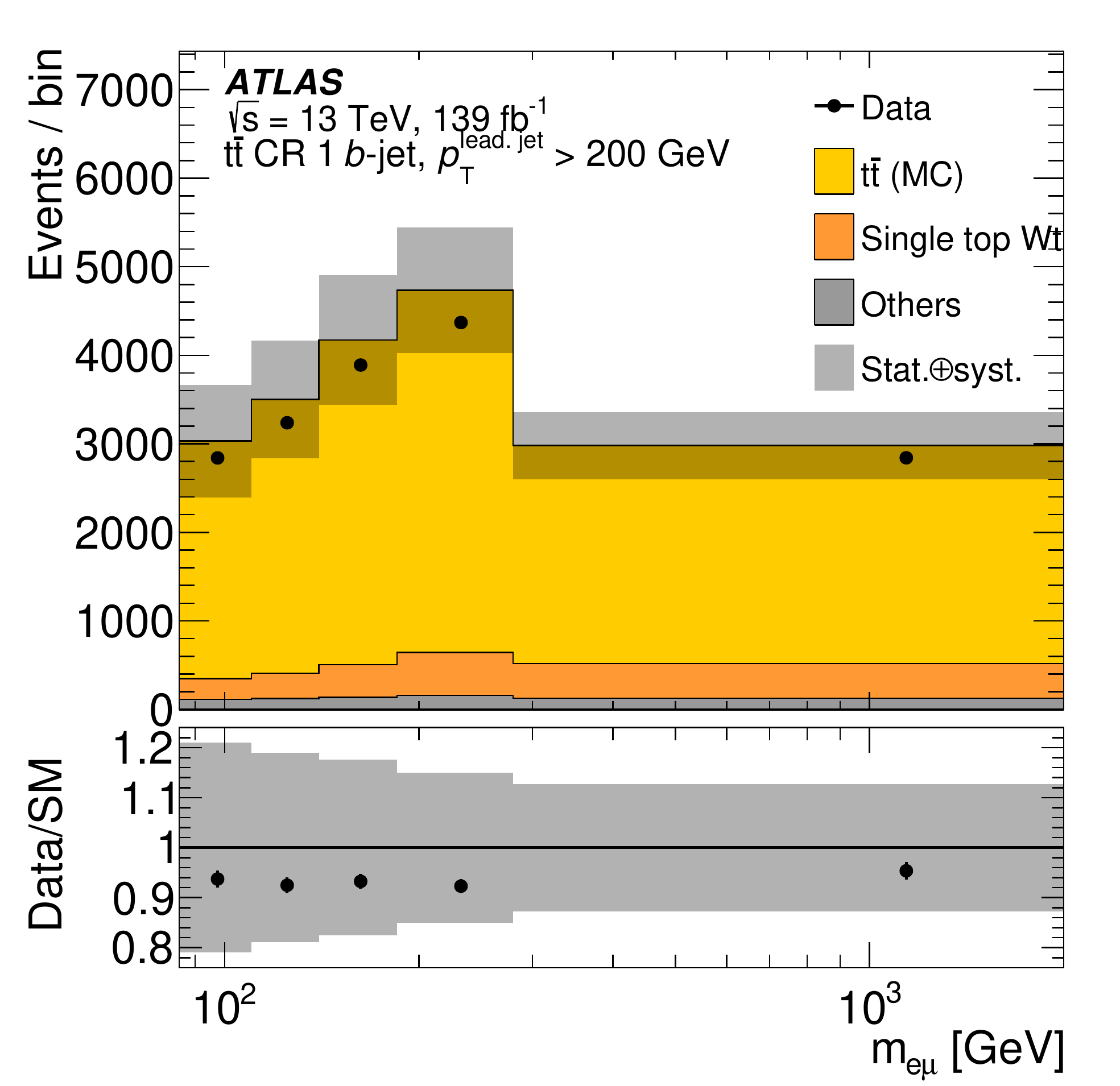}
\includegraphics[width=.45\linewidth]{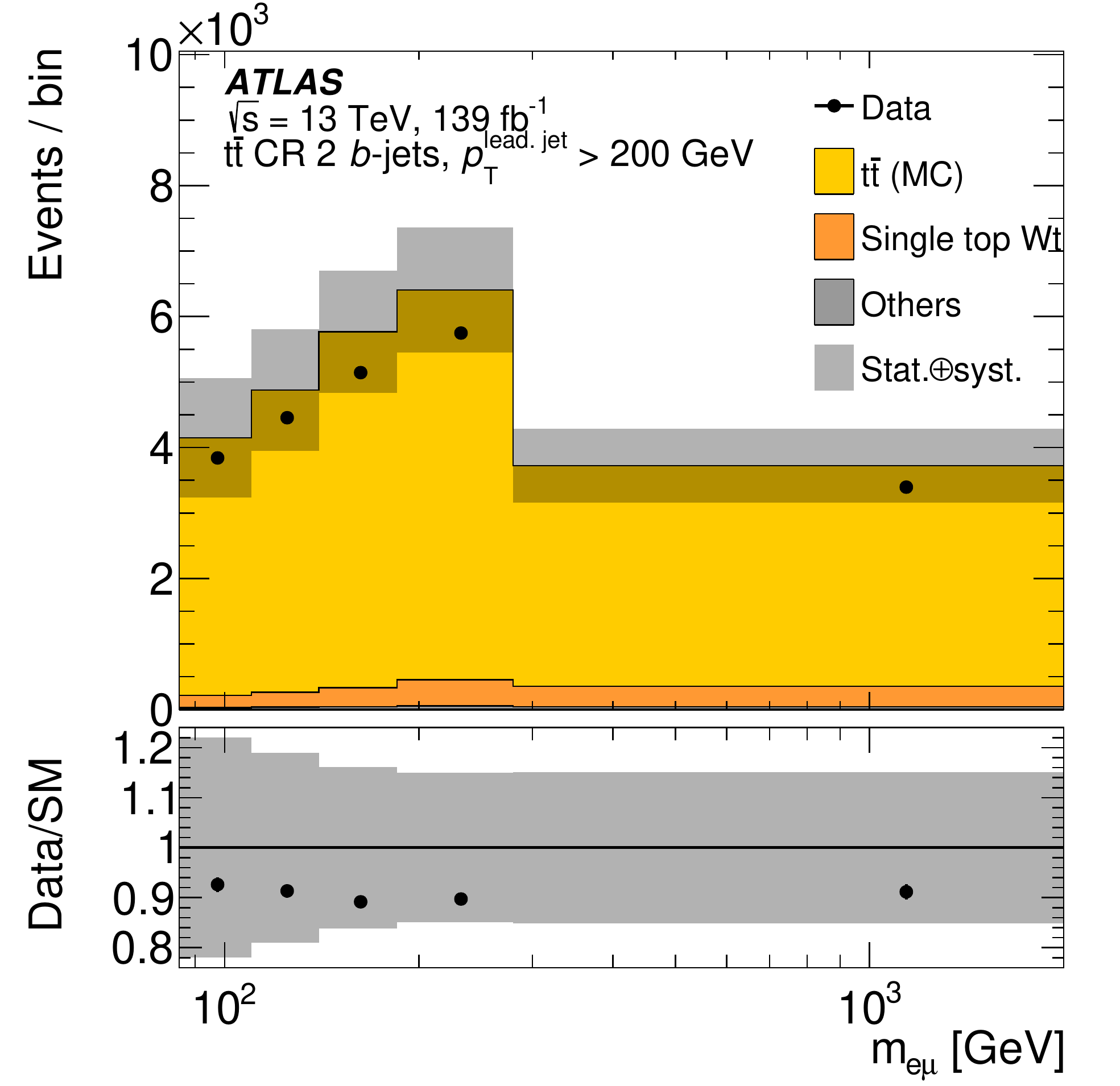}
\includegraphics[width=.45\linewidth]{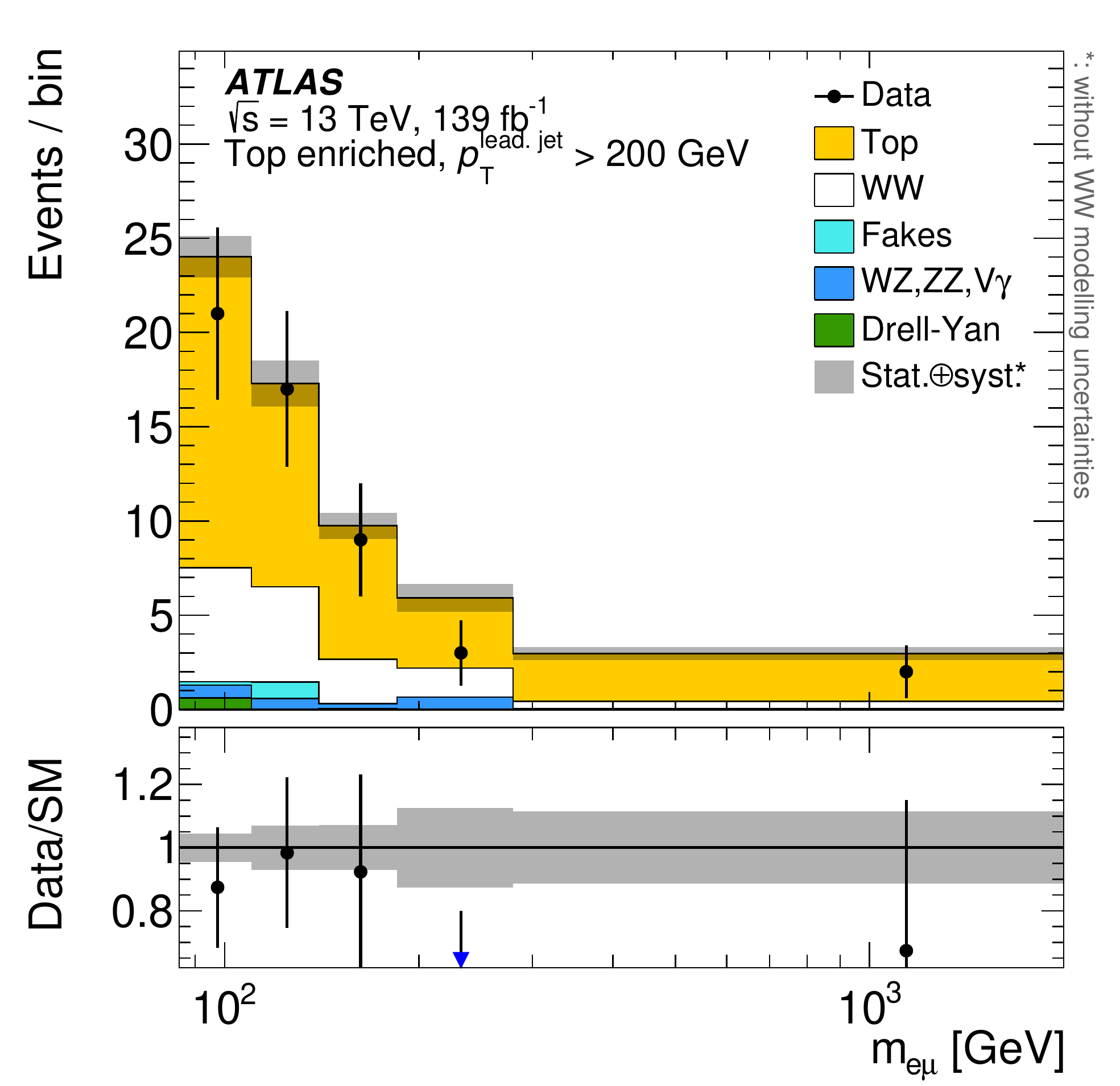}
\includegraphics[width=.45\linewidth]{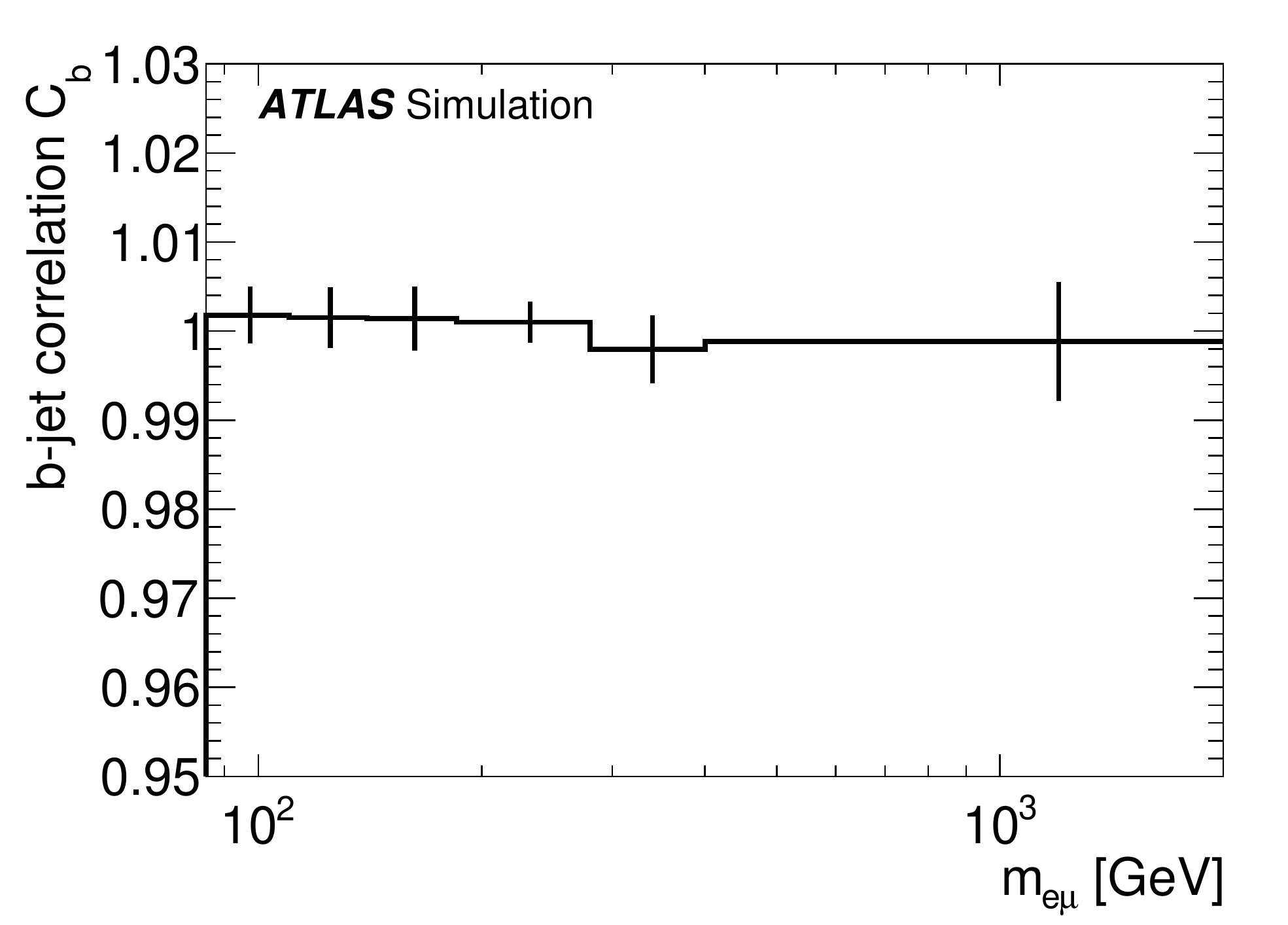}
\caption{
Detector-level distributions of the dilepton invariant mass \mll for $\ptleadjet > 200\,\GeV$ in the one \btag and two \btag control regions as well as the top-enriched region, together with the \bjet correlation factor $C_b$.
The lower panels in the plots of detector-level distributions show the ratio of the data to the total prediction.
The uncertainties shown include statistical and systematic uncertainties, for both the distributions and the correlation factor.
}
 
\label{fig:app:ttbar-mll}
\end{figure}

\FloatBarrier
 
\section*{Acknowledgements}
 

We thank CERN for the very successful operation of the LHC, as well as the
support staff from our institutions without whom ATLAS could not be
operated efficiently.
 
We acknowledge the support of ANPCyT, Argentina; YerPhI, Armenia; ARC, Australia; BMWFW and FWF, Austria; ANAS, Azerbaijan; SSTC, Belarus; CNPq and FAPESP, Brazil; NSERC, NRC and CFI, Canada; CERN; ANID, Chile; CAS, MOST and NSFC, China; COLCIENCIAS, Colombia; MSMT CR, MPO CR and VSC CR, Czech Republic; DNRF and DNSRC, Denmark; IN2P3-CNRS and CEA-DRF/IRFU, France; SRNSFG, Georgia; BMBF, HGF and MPG, Germany; GSRT, Greece; RGC and Hong Kong SAR, China; ISF and Benoziyo Center, Israel; INFN, Italy; MEXT and JSPS, Japan; CNRST, Morocco; NWO, Netherlands; RCN, Norway; MNiSW and NCN, Poland; FCT, Portugal; MNE/IFA, Romania; JINR; MES of Russia and NRC KI, Russian Federation; MESTD, Serbia; MSSR, Slovakia; ARRS and MIZ\v{S}, Slovenia; DST/NRF, South Africa; MICINN, Spain; SRC and Wallenberg Foundation, Sweden; SERI, SNSF and Cantons of Bern and Geneva, Switzerland; MOST, Taiwan; TAEK, Turkey; STFC, United Kingdom; DOE and NSF, United States of America. In addition, individual groups and members have received support from BCKDF, CANARIE, Compute Canada, CRC and IVADO, Canada; Beijing Municipal Science \& Technology Commission, China; COST, ERC, ERDF, Horizon 2020 and Marie Sk{\l}odowska-Curie Actions, European Union; Investissements d'Avenir Labex, Investissements d'Avenir Idex and ANR, France; DFG and AvH Foundation, Germany; Herakleitos, Thales and Aristeia programmes co-financed by EU-ESF and the Greek NSRF, Greece; BSF-NSF and GIF, Israel; La Caixa Banking Foundation, CERCA Programme Generalitat de Catalunya and PROMETEO and GenT Programmes Generalitat Valenciana, Spain; G\"{o}ran Gustafssons Stiftelse, Sweden; The Royal Society and Leverhulme Trust, United Kingdom.
 
The crucial computing support from all WLCG partners is acknowledged gratefully, in particular from CERN, the ATLAS Tier-1 facilities at TRIUMF (Canada), NDGF (Denmark, Norway, Sweden), CC-IN2P3 (France), KIT/GridKA (Germany), INFN-CNAF (Italy), NL-T1 (Netherlands), PIC (Spain), ASGC (Taiwan), RAL (UK) and BNL (USA), the Tier-2 facilities worldwide and large non-WLCG resource providers. Major contributors of computing resources are listed in Ref.~\cite{ATL-SOFT-PUB-2020-001}.
 

\printbibliography

@article{LUCID2,
	author={G. Avoni and others},
	title={The new LUCID-2 detector for luminosity measurement and monitoring in ATLAS},
	journal={JINST},
	volume={13},
	number={07},
	pages={P07017},
	doi="10.1088/1748-0221/13/07/P07017",
	year={2018},
}

@article{Gehrmann:2014fva,
	author         = "Gehrmann, T. and Grazzini, M. and Kallweit, S. and
	Maierhöfer, P. and von Manteuffel, A. and Pozzorini, S.
	and Rathlev, D. and Tancredi, L.",
	title          = "{$W^+W^-$ Production at Hadron Colliders in Next to Next
	to Leading Order QCD}",
	journal        = "Phys. Rev. Lett.",
	volume         = "113",
	year           = "2014",
	number         = "21",
	pages          = "212001",
	doi            = "10.1103/PhysRevLett.113.212001",
	eprint         = "1408.5243",
	archivePrefix  = "arXiv",
	primaryClass   = "hep-ph",
	reportNumber   = "ZU-TH-29-14, MITP-14-053",
	SLACcitation   = "%%CITATION = ARXIV:1408.5243;%%"
}

@article{TOPQ-2018-17,
	author         = "{ATLAS Collaboration}",
	title          = "{Measurement of the $t\bar{t}$ production cross-section
	and lepton differential distributions in $e\mu$ dilepton
	events from $pp$ collisions at $\sqrt{s}=13$ TeV with the
	ATLAS detector}",
	collaboration  = "ATLAS",
	year           = "2019",
	eprint         = "1910.08819",
	archivePrefix  = "arXiv",
	primaryClass   = "hep-ex",
	reportNumber   = "CERN-EP-2019-203",
	SLACcitation   = "%%CITATION = ARXIV:1910.08819;%%"
}

@techreport{TOPQ-2018-17-int,
	author        = "Hawkings, Richard",
	title         = "{Measurement of the $t\bar{t}$ production cross-section
	and lepton differential distributions in $pp$ collisions at
	$\sqrt{s}=13$ TeV with 2015-16 ATLAS data}",
	institution   = "CERN",
	address       = "Geneva",
	number        = "ATL-COM-PHYS-2018-1163",
	month         = "8",
	year          = "2018",
	reportNumber  = "ATL-COM-PHYS-2018-1163",
	url           = "https://cds.cern.ch/record/2634819",
}

@article{MATRIXGiantK,
	author = "Grazzini, Massimiliano and Kallweit, S. and Lindert, Jonas M. and Pozzorini, Stefano and Wiesemann, Marius",
	archivePrefix = "arXiv",
	doi = "10.1007/JHEP02(2020)087",
	eprint = "1912.00068",
	journal = "JHEP",
	pages = "087",
	primaryClass = "hep-ph",
	reportNumber = "MPP-2019-175, ZU-TH 49/19",
	title = "{NNLO QCD + NLO EW with Matrix+OpenLoops: precise predictions for vector-boson pair production}",
	volume = "02",
	year = "2020"
}

@article{Grazzini:2016swo,
	author = "Grazzini, Massimiliano and Kallweit, Stefan and Rathlev, Dirk and Wiesemann, Marius",
	title = "{$W^{\pm}Z$ production at hadron colliders in NNLO QCD}",
	eprint = "1604.08576",
	archivePrefix = "arXiv",
	primaryClass = "hep-ph",
	reportNumber = "ZU-TH-15-16, MITP-16-037, NSF-KITP-16-046, DESY-16-074",
	doi = "10.1016/j.physletb.2016.08.017",
	journal = "Phys. Lett. B",
	volume = "761",
	pages = "179--183",
	year = "2016"
}

@article{Grazzini:2017ckn,
	author = "Grazzini, Massimiliano and Kallweit, Stefan and Rathlev, Dirk and Wiesemann, Marius",
	title = "{$W^\pm Z$ production at the LHC: fiducial cross sections and distributions in NNLO QCD}",
	eprint = "1703.09065",
	archivePrefix = "arXiv",
	primaryClass = "hep-ph",
	reportNumber = "ZU-TH-06-17, CERN-TH-2017-065",
	doi = "10.1007/JHEP05(2017)139",
	journal = "JHEP",
	volume = "05",
	pages = "139",
	year = "2017"
}

@article{Grazzini:2015hta,
	author = "Grazzini, Massimiliano and Kallweit, Stefan and Rathlev, Dirk",
	title = "{ZZ production at the LHC: fiducial cross sections and distributions in NNLO QCD}",
	eprint = "1507.06257",
	archivePrefix = "arXiv",
	primaryClass = "hep-ph",
	doi = "10.1016/j.physletb.2015.09.055",
	journal = "Phys. Lett. B",
	volume = "750",
	pages = "407--410",
	year = "2015"
}

@article{Cascioli:2014yka,
	author = "Cascioli, F. and Gehrmann, T. and Grazzini, M. and Kallweit, S. and Maierhöfer, P. and von Manteuffel, A. and Pozzorini, S. and Rathlev, D. and Tancredi, L. and Weihs, E.",
	title = "{ZZ production at hadron colliders in NNLO QCD}",
	eprint = "1405.2219",
	archivePrefix = "arXiv",
	primaryClass = "hep-ph",
	reportNumber = "ZU-TH-19-14, MITP-14-033",
	doi = "10.1016/j.physletb.2014.06.056",
	journal = "Phys. Lett. B",
	volume = "735",
	pages = "311--313",
	year = "2014"
}

@article{BayesUnfold1,
	Author = {D'Agostini, G.},
	Doi = {10.1016/0168-9002(95)00274-X},
	Journal = {Nucl. Instr. Meth. A},
	Pages = {487-498},
	Title = {{A multidimensional unfolding method based on Bayes' theorem}},
	Volume = {362},
	Year = {1995}}

@article{BayesUnfold2,
	Archiveprefix = {arXiv},
	Author = {D'Agostini, G.},
	Eprint = {1010.0632},
	Primaryclass = {physics.data-an},
	Title = {{Improved iterative Bayesian unfolding}},
	Year = {2010}}

@article{Grazzini:2020stb,
	author = "Grazzini, Massimiliano and Kallweit, Stefan and Wiesemann, Marius and Yook, Jeong Yeon",
	title = "{$W^+W^-$ production at the LHC: NLO QCD corrections to the loop-induced gluon fusion channel}",
	eprint = "2002.01877",
	archivePrefix = "arXiv",
	primaryClass = "hep-ph",
	reportNumber = "MPP-2019-176, ZU-TH 04/20",
	doi = "10.1016/j.physletb.2020.135399",
	journal = "Phys. Lett. B",
	volume = "804",
	pages = "135399",
	year = "2020"
}

@article{Grazzini:2017mhc,
	author         = "Grazzini, Massimiliano and Kallweit, Stefan and
	Wiesemann, Marius",
	title          = "{Fully differential NNLO computations with MATRIX}",
	journal        = "Eur. Phys. J. C",
	volume         = "78",
	year           = "2018",
	number         = "7",
	pages          = "537",
	doi            = "10.1140/epjc/s10052-018-5771-7",
	eprint         = "1711.06631",
	archivePrefix  = "arXiv",
	primaryClass   = "hep-ph",
	reportNumber   = "ZU-TH-30-17, CERN-TH-2017-232, ZU-TH 30/17",
	SLACcitation   = "%%CITATION = ARXIV:1711.06631;%%"
}

@article{Cascioli:2011va,
	Archiveprefix = {arXiv},
	Author = {Cascioli, Fabio and Maierhofer, Philipp and Pozzorini, Stefano},
	Doi = {10.1103/PhysRevLett.108.111601},
	Eprint = {1111.5206},
	Journal = {Phys. Rev. Lett.},
	Pages = {111601},
	Primaryclass = {hep-ph},
	Reportnumber = {ZU-TH-23-11, LPN11-66},
	Slaccitation = {%%CITATION = ARXIV:1111.5206;%%},
	Title = {{Scattering Amplitudes with Open Loops}},
	Volume = {108},
	Year = {2012}}

@article{Azatov:2017kzw,
	author = "Azatov, A. and Elias-Miro, J. and Reyimuaji, Y. and Venturini, E.",
	title = "{Novel measurements of anomalous triple gauge couplings for the LHC}",
	eprint = "1707.08060",
	archivePrefix = "arXiv",
	primaryClass = "hep-ph",
	reportNumber = "SISSA-35-2017-FISI",
	doi = "10.1007/JHEP10(2017)027",
	journal = "JHEP",
	volume = "10",
	pages = "027",
	year = "2017"
}

@article{Gehrmann:2015ora,
	author         = "Gehrmann, Thomas and von Manteuffel, Andreas and
	Tancredi, Lorenzo",
	title          = "{The two-loop helicity amplitudes for $
	q\overline{q}^{\prime}\to {V}_1{V}_2\to 4 $ leptons}",
	journal        = "JHEP",
	volume         = "09",
	year           = "2015",
	pages          = "128",
	doi            = "10.1007/JHEP09(2015)128",
	eprint         = "1503.04812",
	archivePrefix  = "arXiv",
	primaryClass   = "hep-ph",
	reportNumber   = "ZU-TH-03-15, MITP-15-011, TTP15-011",
	SLACcitation   = "%%CITATION = ARXIV:1503.04812;%%"
}

@article{Grazzini:2015nwa,
	author = "Grazzini, Massimiliano and Kallweit, Stefan and Rathlev, Dirk",
	title = "{$W\gamma$ and $Z\gamma$ production at the LHC in NNLO QCD}",
	eprint = "1504.01330",
	archivePrefix = "arXiv",
	primaryClass = "hep-ph",
	reportNumber = "ZU-TH-04-15, MITP-15-021",
	doi = "10.1007/JHEP07(2015)085",
	journal = "JHEP",
	volume = "07",
	pages = "085",
	year = "2015"
}

@article{Barr:2005dz,
	author = "Barr, A.J.",
	title = "{Measuring slepton spin at the LHC}",
	eprint = "hep-ph/0511115",
	archivePrefix = "arXiv",
	reportNumber = "ATL-PHYS-PUB-2005-023",
	doi = "10.1088/1126-6708/2006/02/042",
	journal = "JHEP",
	volume = "02",
	pages = "042",
	year = "2006"
}

@article{Kallweit:2014xda,
    author = "Kallweit, Stefan and Lindert, Jonas M. and Maierhöfer, Philipp and Pozzorini, Stefano and Schönherr, Marek",
    title = "{NLO electroweak automation and precise predictions for W+multijet production at the LHC}",
    eprint = "1412.5157",
    archivePrefix = "arXiv",
    primaryClass = "hep-ph",
    reportNumber = "LPN14-127, IPPP-14-107, DCPT-14-214, MCNET-14-26, ZU-TH-42-14, MITP-14-102",
    doi = "10.1007/JHEP04(2015)012",
    journal = "JHEP",
    volume = "04",
    pages = "012",
    year = "2015"
}

@article{Schonherr:2017qcj,
    author = "Schönherr, Marek",
    title = "{An automated subtraction of NLO EW infrared divergences}",
    eprint = "1712.07975",
    archivePrefix = "arXiv",
    primaryClass = "hep-ph",
    reportNumber = "CERN-TH-17-281, CERN-TH-2017-281, MCNET-17-24",
    doi = "10.1140/epjc/s10052-018-5600-z",
    journal = "Eur. Phys. J. C",
    volume = "78",
    number = "2",
    pages = "119",
    year = "2018"
}

@article{Bothmann:2019yzt,
    author = "Bothmann, Enrico and others",
    collaboration = "Sherpa",
    title = "{Event Generation with Sherpa 2.2}",
    eprint = "1905.09127",
    archivePrefix = "arXiv",
    primaryClass = "hep-ph",
    reportNumber = "FERMILAB-PUB-19-218-T, SLAC-PUB-17433, IPPP/19/42, MCNET-19-11",
    doi = "10.21468/SciPostPhys.7.3.034",
    journal = "SciPost Phys.",
    volume = "7",
    number = "3",
    pages = "034",
    year = "2019"
}

@article{Buccioni:2019sur,
    author = "Buccioni, Federico and Lang, Jean-Nicolas and Lindert, Jonas M. and Maierhöfer, Philipp and Pozzorini, Stefano and Zhang, Hantian and Zoller, Max F.",
    title = "{OpenLoops 2}",
    eprint = "1907.13071",
    archivePrefix = "arXiv",
    primaryClass = "hep-ph",
    reportNumber = "IPPP/19/62, FR-PHENO-2019-12, PSI-PR-19-15, ZU-TH 37/19",
    doi = "10.1140/epjc/s10052-019-7306-2",
    journal = "Eur. Phys. J. C",
    volume = "79",
    number = "10",
    pages = "866",
    year = "2019"
}

@article{Grzadkowski:2010es,
    author = "Grzadkowski, B. and Iskrzynski, M. and Misiak, M. and Rosiek, J.",
    title = "{Dimension-Six Terms in the Standard Model Lagrangian}",
    eprint = "1008.4884",
    archivePrefix = "arXiv",
    primaryClass = "hep-ph",
    reportNumber = "IFT-9-2010, TTP10-35",
    doi = "10.1007/JHEP10(2010)085",
    journal = "JHEP",
    volume = "10",
    pages = "085",
    year = "2010"
}

@article{Azatov:2016sqh,
    author = "Azatov, Aleksandr and Contino, Roberto and Machado, Camila S. and Riva, Francesco",
    title = "{Helicity selection rules and noninterference for BSM amplitudes}",
    eprint = "1607.05236",
    archivePrefix = "arXiv",
    primaryClass = "hep-ph",
    reportNumber = "CERN-TH-2016-165",
    doi = "10.1103/PhysRevD.95.065014",
    journal = "Phys. Rev. D",
    volume = "95",
    number = "6",
    pages = "065014",
    year = "2017"
}

@article{Falkowski:2016cxu,
    author = "Falkowski, Adam and Gonzalez-Alonso, Martin and Greljo, Admir and Marzocca, David and Son, Minho",
    title = "{Anomalous Triple Gauge Couplings in the Effective Field Theory Approach at the LHC}",
    eprint = "1609.06312",
    archivePrefix = "arXiv",
    primaryClass = "hep-ph",
    reportNumber = "ZU-TH-34-16",
    doi = "10.1007/JHEP02(2017)115",
    journal = "JHEP",
    volume = "02",
    pages = "115",
    year = "2017"
}

@Article{STDM-2017-27,
    author         = "{ATLAS Collaboration}",
    title          = "{Differential cross-section measurements for the electroweak production of dijets in association with a \(Z\) boson in proton--proton collisions at ATLAS}",
    year           = "2020",
    doi            = "10.3204/PUBDB-2020-02663",
    reportNumber   = "CERN-EP-2020-045",
    eprint         = "2006.15458",
    archivePrefix  = "arXiv",
    primaryClass   = "hep-ex",
}

@article{Hamilton:2016bfu,
    author = "Hamilton, Keith and Melia, Tom and Monni, Pier Francesco and Re, Emanuele and Zanderighi, Giulia",
    title = "{Merging WW and WW+jet with MINLO}",
    eprint = "1606.07062",
    archivePrefix = "arXiv",
    primaryClass = "hep-ph",
    reportNumber = "CERN-TH-2016-146, LAPTH-031-16, OUTP-16-17P",
    doi = "10.1007/JHEP09(2016)057",
    journal = "JHEP",
    volume = "09",
    pages = "057",
    year = "2016"
}

@article{schael:2013ita,
	Archiveprefix = {arXiv},
	Author = {{ALEPH, DELPHI, L3 and OPAL Collaborations and the LEP Electroweak Working Group}},
	Collaboration = {DELPHI, OPAL, LEP Electroweak, ALEPH, L3},
	Doi = {10.1016/j.physrep.2013.07.004},
	Eprint = {1302.3415},
	Journal = {Phys. Rept.},
	Pages = {119-244},
	Primaryclass = {hep-ex},
	Reportnumber = {CERN-PH-EP-2013-022},
	Slaccitation = {%%CITATION = ARXIV:1302.3415;%%},
	Title = {{Electroweak measurements in electron--positron collisions at W-boson-pair energies at LEP}},
	Volume = {532},
	Year = {2013}
}

@article{Abe:1996dw,
	Author = {{CDF Collaboration}},
	Collaboration = {CDF},
	Doi = {10.1103/PhysRevLett.78.4536},
	Journal = {Phys. Rev. Lett.},
	Pages = {4536-4540},
	Title = {{Observation of $W^+ W^-$ Production in $p\bar{p}$ collisions at $\sqrt{s}$ = 1.8 \TeV}},
	Volume = {78},
	Year = {1997}}

@article{Aaltonen:2009aa,
	Archiveprefix = {arXiv},
	Author = {{CDF Collaboration}},
	Collaboration = {CDF},
	Doi = {10.1103/PhysRevLett.104.201801},
	Eprint = {0912.4500},
	Journal = {Phys. Rev. Lett.},
	Note = {[Erratum: Phys. Rev. Lett. 105 (2010) 019905]},
	Pages = {201801},
	Primaryclass = {hep-ex},
	Reportnumber = {FERMILAB-PUB-09-636},
	Slaccitation = {%%CITATION = ARXIV:0912.4500;%%},
	Title = {{Measurement of the $W^+ W^-$ Production Cross Section and Search for Anomalous $WW \gamma$ and $WWZ$ Couplings in $p \bar{p}$ Collisions at $\sqrt{s}=1.96$ \TeV}},
	Volume = {104},
	Year = {2010}
}

@article{Abazov:2009ys,
	Archiveprefix = {arXiv},
	Author = {{D\O{} Collaboration}},
	Doi = {10.1103/PhysRevLett.103.191801},
	Eprint = {0904.0673},
	Journal = {Phys. Rev. Lett.},
	Pages = {191801},
	Primaryclass = {hep-ex},
	Reportnumber = {FERMILAB-PUB-09-094-E},
	Slaccitation = {%%CITATION = ARXIV:0904.0673;%%},
	Title = {{Measurement of the WW Production Cross Section with Dilepton Final States in $p\bar{p}$ Collisions at $\sqrt{s}$ = 1.96 \TeV\ and Limits on Anomalous Trilinear Gauge Couplings}},
	Volume = {103},
	Year = {2009}
}

@article{Wilks:1938dza,
	author = "Wilks, S.S.",
	title = "{The Large-Sample Distribution of the Likelihood Ratio for Testing Composite Hypotheses}",
	doi = "10.1214/aoms/1177732360",
	journal = "Annals Math. Statist.",
	volume = "9",
	number = "1",
	pages = "60--62",
	year = "1938"
}

@article{MUON-2018-03,
	author         = "{ATLAS Collaboration}",
	collaboration = "ATLAS",
	title = "{Muon reconstruction and identification efficiency in ATLAS using the full Run 2 $pp$ collision data set at $\sqrt{s}=13$ TeV}",
	eprint = "2012.00578",
	archivePrefix = "arXiv",
	primaryClass = "hep-ex",
	reportNumber = "CERN Preprint ID: CERN-EP-2020-199",
	month = "12",
	year = "2020"
}

@article{Cacciari:2007fd,
    author = "Cacciari, Matteo and Salam, Gavin P.",
    title = "{Pileup subtraction using jet areas}",
    eprint = "0707.1378",
    archivePrefix = "arXiv",
    primaryClass = "hep-ph",
    reportNumber = "LPTHE-07-01",
    doi = "10.1016/j.physletb.2007.09.077",
    journal = "Phys. Lett. B",
    volume = "659",
    pages = "119--126",
    year = "2008"
}

@article{Grazzini:2016ctr,
    author = "Grazzini, Massimiliano and Kallweit, Stefan and Pozzorini, Stefano and Rathlev, Dirk and Wiesemann, Marius",
    title = "{$W^{+}W^{-}$ production at the LHC: fiducial cross sections and distributions in NNLO QCD}",
    eprint = "1605.02716",
    archivePrefix = "arXiv",
    primaryClass = "hep-ph",
    reportNumber = "ZU-TH-16-16, MITP-16-038, NSF-KITP-16-047, DESY-16-075",
    doi = "10.1007/JHEP08(2016)140",
    journal = "JHEP",
    volume = "08",
    pages = "140",
    year = "2016"
}

@article{Catani:2012qa,
    author = "Catani, Stefano and Cieri, Leandro and de Florian, Daniel and Ferrera, Giancarlo and Grazzini, Massimiliano",
    title = "{Vector boson production at hadron colliders: hard-collinear coefficients at the NNLO}",
    eprint = "1209.0158",
    archivePrefix = "arXiv",
    primaryClass = "hep-ph",
    reportNumber = "ZU-TH-16-12",
    doi = "10.1140/epjc/s10052-012-2195-7",
    journal = "Eur. Phys. J. C",
    volume = "72",
    pages = "2195",
    year = "2012"
}

@Article{Evans:2008zzb,
      author         = "Evans, Lyndon and Bryant, Philip",
      title          = "{LHC Machine}",
      journal        = "JINST",
      volume         = "3",
      pages          = "S08001",
      doi            = "10.1088/1748-0221/3/08/S08001",
      year           = "2008",
      SLACcitation   = "%%CITATION = JINST,3,S08001;%%",
}

@Article{Cacciari:2008gp,
     author    = "Cacciari, Matteo and Salam, Gavin P. and Soyez, Gregory",
     title     = "{The anti-\(k_{t}\) jet clustering algorithm}",
     journal   = "JHEP",
     volume    = "04",
     year      = "2008",
     pages     = "063",
     eprint    = "0802.1189",
     archivePrefix = "arXiv",
     primaryClass  =  "hep-ph",
     doi       = "10.1088/1126-6708/2008/04/063",
     SLACcitation  = "%%CITATION = 0802.1189;%%"
}

@Article{Pumplin:2002vw,
     author    = "Pumplin, J. and others",
     title     = "{New Generation of Parton Distributions with Uncertainties from Global QCD Analysis}",
     journal   = "JHEP",
     volume    = "07",
     year      = "2002",
     pages     = "012",
     doi       = "10.1088/1126-6708/2002/07/012",
     eprint    = "hep-ph/0201195",
     archivePrefix = "arXiv",
     SLACcitation  = "%%CITATION = HEP-PH/0201195;%%"
}

@article{Lai:2010vv,
      author         = "Lai, H.-L. and others",
      title          = "{New parton distributions for collider physics}",
      journal        = "Phys. Rev. D",
      volume         = "82",
      pages          = "074024",
      doi            = "10.1103/PhysRevD.82.074024",
      year           = "2010",
      eprint         = "1007.2241",
      archivePrefix  = "arXiv",
      primaryClass   = "hep-ph",
      reportNumber   = "MSUHEP-100707, SMU-HEP-10-10",
      SLACcitation   = "%%CITATION = ARXIV:1007.2241;%%",
}

@article{Harland-Lang:2014zoa,
      author         = "Harland-Lang, L.A. and Martin, A.D. and Motylinski, P.
                        and Thorne, R.S.",
      title          = "{Parton distributions in the LHC era: MMHT 2014 PDFs}",
      journal        = "Eur. Phys. J. C",
      number         = "5",
      volume         = "75",
      pages          = "204",
      doi            = "10.1140/epjc/s10052-015-3397-6",
      year           = "2015",
      eprint         = "1412.3989",
      archivePrefix  = "arXiv",
      primaryClass   = "hep-ph",
      reportNumber   = "LCTS-2014-47, IPPP-14-97, DCPT-14-194",
      SLACcitation   = "%%CITATION = ARXIV:1412.3989;%%",
}

@article{Ball:2012cx,
      author         = "Ball, Richard D. and others",
      title          = "{Parton distributions with LHC data}",
      journal        = "Nucl. Phys. B",
      volume         = "867",
      year           = "2013",
      pages          = "244",
      doi            = "10.1016/j.nuclphysb.2012.10.003",
      eprint         = "1207.1303",
      archivePrefix  = "arXiv",
      primaryClass   = "hep-ph",
      reportNumber   = "EDINBURGH-2012-08, IFUM-FT-997, FR-PHENO-2012-014,
                        RWTH-TTK-12-25, CERN-PH-TH-2012-037, SFB-CPP-12-47\,
                        --CERN-PH-TH-2012-037",
      SLACcitation   = "%%CITATION = ARXIV:1207.1303;%%"
}

@article{Ball:2014uwa,
      author         = "Ball, Richard D. and others",
      title          = "{Parton distributions for the LHC run II}",
      collaboration  = "NNPDF",
      journal        = "JHEP",
      volume         = "04",
      year           = "2015",
      pages          = "040",
      doi            = "10.1007/JHEP04(2015)040",
      eprint         = "1410.8849",
      archivePrefix  = "arXiv",
      primaryClass   = "hep-ph",
      reportNumber   = "EDINBURGH-2014-15, IFUM-1034-FT, CERN-PH-TH-2013-253,
                        OUTP-14-11P, CAVENDISH-HEP-14-11",
      SLACcitation   = "%%CITATION = ARXIV:1410.8849;%%"
}

@article{Czakon:2012pz,
      author         = "Czakon, Michal and Mitov, Alexander",
      title          = "{NNLO corrections to top pair production at hadron
                        colliders: the quark-gluon reaction}",
      journal        = "JHEP",
      volume         = "01",
      pages          = "080",
      doi            = "10.1007/JHEP01(2013)080",
      year           = "2013",
      eprint         = "1210.6832",
      archivePrefix  = "arXiv",
      primaryClass   = "hep-ph",
      SLACcitation   = "%%CITATION = ARXIV:1210.6832;%%",
}

@Article{Sjostrand:2007gs,
	author    = "Sj{\"o}strand, T. and Mrenna, S. and Skands, P.",
	title     = "{A brief introduction to PYTHIA 8.1}",
	journal   = "Comput. Phys. Commun.",
	volume    = "178",
	year      = "2008",
	pages     = "852-867",
	eprint    = "0710.3820",
	archivePrefix = "arXiv",
	primaryClass  =  "hep-ph",
	doi       = "10.1016/j.cpc.2008.01.036",
	SLACcitation  = "%%CITATION = 0710.3820;%%"
}

@article{Sjostrand:2014zea,
      author         = "Sj{\"o}strand, Torbj{\"o}rn and Ask, Stefan and Christiansen,
                        Jesper R. and Corke, Richard and Desai, Nishita and Ilten,
                        Philip and Mrenna, Stephen and Prestel, Stefan and
                        Rasmussen, Christine O. and Skands, Peter Z.",
      title          = "{An introduction to PYTHIA 8.2}",
      journal        = "Comput. Phys. Commun.",
      volume         = "191",
      year           = "2015",
      pages          = "159",
      doi            = "10.1016/j.cpc.2015.01.024",
      eprint         = "1410.3012",
      archivePrefix  = "arXiv",
      primaryClass   = "hep-ph",
      reportNumber   = "LU-TP-14-36, MCNET-14-22, CERN-PH-TH-2014-190,
                        FERMILAB-PUB-14-316-CD, DESY-14-178, SLAC-PUB-16122,
                        --FERMILAB-PUB-14-316-CD",
      SLACcitation   = "%%CITATION = ARXIV:1410.3012;%%"
}

@article{Lonnblad:2001iq,
      author         = "L{\"o}nnblad, Leif",
      title          = "{Correcting the Colour-Dipole Cascade Model
                  with Fixed Order Matrix Elements}",
      journal        = "JHEP",
      volume         = "05",
      year           = "2002",
      pages          = "046",
      doi            = "10.1088/1126-6708/2002/05/046",
      eprint         = "hep-ph/0112284",
      archivePrefix  = "arXiv",
      XprimaryClass   = "hep-ph",
      reportNumber   = "LU-TP-01-38",
      SLACcitation   = "%%CITATION = HEP-PH/0112284;%%"
}

@article{Lonnblad:2011xx,
      author         = "L{\"o}nnblad, Leif and Prestel, Stefan",
      title          = "{Matching tree-level matrix elements with interleaved
                        showers}",
      journal        = "JHEP",
      volume         = "03",
      year           = "2012",
      pages          = "019",
      doi            = "10.1007/JHEP03(2012)019",
      eprint         = "1109.4829",
      archivePrefix  = "arXiv",
      primaryClass   = "hep-ph",
      SLACcitation   = "%%CITATION = ARXIV:1109.4829;%%"
}

@Article{Frixione:2008yi,
     author    = "Frixione, Stefano and Laenen, Eric and Motylinski, Patrick and White, Chris and Webber, Bryan R.",
     title     = "{Single-top hadroproduction in association with a \(W\) boson}",
     journal   = "JHEP",
     volume    = "07",
     year      = "2008",
     pages     = "029",
     eprint    = "0805.3067",
     archivePrefix = "arXiv",
     primaryClass  =  "hep-ph",
     doi       = "10.1088/1126-6708/2008/07/029"
}

@Article{Alwall:2011uj,
      author         = "Alwall, Johan and Herquet, Michel and Maltoni, Fabio and
                        Mattelaer, Olivier and Stelzer, Tim",
      title          = "{MadGraph 5 : Going Beyond}",
      journal        = "JHEP",
      volume         = "06",
      pages          = "128",
      doi            = "10.1007/JHEP06(2011)128",
      year           = "2011",
      eprint         = "1106.0522",
      archivePrefix  = "arXiv",
      primaryClass   = "hep-ph"
}

@Article{Alwall:2014hca,
      author         = "Alwall, J. and Frederix, R. and Frixione, S. and Hirschi,
                        V. and Maltoni, F. and Mattelaer, O. and Shao, H. -S. and
                        Stelzer, T. and Torrielli, P. and Zaro, M.",
      title          = "{The automated computation of tree-level and
                        next-to-leading order differential cross sections, and
                        their matching to parton shower simulations}",
      journal        = "JHEP",
      volume         = "07",
      year           = "2014",
      pages          = "079",
      doi            = "10.1007/JHEP07(2014)079",
      eprint         = "1405.0301",
      archivePrefix  = "arXiv",
      primaryClass   = "hep-ph",
      reportNumber   = "CERN-PH-TH-2014-064, CP3-14-18, LPN14-066, MCNET-14-09,
                        ZU-TH-14-14",
      SLACcitation   = "%%CITATION = ARXIV:1405.0301;%%"
}

@article{Frederix:2012ps,
      author         = "Frederix, Rikkert and Frixione, Stefano",
      title          = "{Merging meets matching in MC@NLO}",
      journal        = "JHEP",
      volume         = "12",
      year           = "2012",
      pages          = "061",
      doi            = "10.1007/JHEP12(2012)061",
      eprint         = "1209.6215",
      archivePrefix  = "arXiv",
      primaryClass   = "hep-ph",
      reportNumber   = "CERN-PH-TH-2012-247, ZU-TH-21-12",
      SLACcitation   = "%%CITATION = ARXIV:1209.6215;%%"
}

@Article{Bahr:2008pv,
      author         = "B{\"a}hr, M. and others",
      title          = "{Herwig++ physics and manual}",
      journal        = "Eur. Phys. J. C",
      volume         = "58",
      year           = "2008",
      pages          = "639",
      doi            = "10.1140/epjc/s10052-008-0798-9",
      eprint         = "0803.0883",
      archivePrefix  = "arXiv",
      primaryClass   = "hep-ph",
      reportNumber   = "CERN-PH-TH-2008-038, CAVENDISH-HEP-08-03, KA-TP-05-2008,
                        DCPT-08-22, IPPP-08-11, CP3-08-05",
      SLACcitation   = "%%CITATION = ARXIV:0803.0883;%%"
}

@Article{Bellm:2015jjp,
      author         = "Bellm, Johannes and others",
      title          = "{Herwig 7.0/Herwig++ 3.0 release note}",
      journal        = "Eur. Phys. J. C",
      volume         = "76",
      year           = "2016",
      number         = "4",
      pages          = "196",
      doi            = "10.1140/epjc/s10052-016-4018-8",
      eprint         = "1512.01178",
      archivePrefix  = "arXiv",
      primaryClass   = "hep-ph",
      reportNumber   = "CERN-PH-TH-2015-289, MAN-HEP-2015-15, IFJPAN-IV-2015-13,
                        HERWIG-2015-01, KA-TP-18-2015, DCPT-15-142, MCNET-15-28,
                        IPPP-15-71, --HERWIG-2015-01",
      SLACcitation   = "%%CITATION = ARXIV:1512.01178;%%"
}

@Article{Nason:2004rx,
      author         = "Nason, Paolo",
      title          = "{A new method for combining NLO QCD with shower Monte Carlo algorithms}",
      journal        = "JHEP",
      volume         = "11",
      pages          = "040",
      doi            = "10.1088/1126-6708/2004/11/040",
      year           = "2004",
      eprint         = "hep-ph/0409146",
      archivePrefix  = "arXiv",
}

@Article{Frixione:2007vw,
      author         = "Frixione, Stefano and Nason, Paolo and Oleari, Carlo",
      title          = "{Matching NLO QCD computations with parton shower
                        simulations: the POWHEG method}",
      journal        = "JHEP",
      volume         = "11",
      pages          = "070",
      doi            = "10.1088/1126-6708/2007/11/070",
      year           = "2007",
      eprint         = "0709.2092",
      archivePrefix  = "arXiv",
      primaryClass   = "hep-ph",
}

@Article{Alioli:2010xd,
      author         = "Alioli, Simone and Nason, Paolo and Oleari, Carlo and Re,
                        Emanuele",
      title          = "{A general framework for implementing NLO calculations in
                        shower Monte Carlo programs: the POWHEG BOX}",
      journal        = "JHEP",
      volume         = "06",
      pages          = "043",
      doi            = "10.1007/JHEP06(2010)043",
      year           = "2010",
      eprint         = "1002.2581",
      archivePrefix  = "arXiv",
      primaryClass   = "hep-ph",
}

@article{Nason:2013ydw,
      author         = "Nason, Paolo and Zanderighi, Giulia",
      title          = "{\(W^+ W^-\), \(W Z\) and \(Z Z\) production in the
                        POWHEG-BOX-V2}",
      journal        = "Eur. Phys. J. C",
      volume         = "74",
      year           = "2014",
      number         = "1",
      pages          = "2702",
      doi            = "10.1140/epjc/s10052-013-2702-5",
      eprint         = "1311.1365",
      archivePrefix  = "arXiv",
      primaryClass   = "hep-ph",
      SLACcitation   = "%%CITATION = ARXIV:1311.1365;%%"
}

@article{Kidonakis:2010ux,
      author         = "Kidonakis, Nikolaos",
      title          = "{Two-loop soft anomalous dimensions for single top quark
                        associated production with a \(W^{-}\) or \(H^{-}\)}",
      journal        = "Phys. Rev. D",
      volume         = "82",
      year           = "2010",
      pages          = "054018",
      doi            = "10.1103/PhysRevD.82.054018",
      eprint         = "1005.4451",
      archivePrefix  = "arXiv",
      primaryClass   = "hep-ph",
      SLACcitation   = "%%CITATION = ARXIV:1005.4451;%%"
}

@inproceedings{Kidonakis:2013zqa,
      author         = "Kidonakis, Nikolaos",
      title          = "{Top Quark Production}",
      booktitle      = "{Proceedings, Helmholtz International Summer School on
                        Physics of Heavy Quarks and Hadrons (HQ 2013)}",
      eventdate      = "2013-07-15/2013-07-28",
      venue          = "JINR, Dubna, Russia",
      pages          = "139-168",
      doi            = "10.3204/DESY-PROC-2013-03/Kidonakis",
      eprint         = "1311.0283",
      archivePrefix  = "arXiv",
      primaryClass   = "hep-ph",
      reportNumber   = "KSU-HEP-110113",
      SLACcitation   = "%%CITATION = ARXIV:1311.0283;%%"
}

@article{Beneke:2011mq,
      author         = "Beneke, M. and Falgari, P. and Klein, S. and Schwinn, C.",
      title          = "{Hadronic top-quark pair production with NNLL threshold
                        resummation}",
      journal        = "Nucl. Phys. B",
      volume         = "855",
      year           = "2012",
      pages          = "695-741",
      doi            = "10.1016/j.nuclphysb.2011.10.021",
      eprint         = "1109.1536",
      archivePrefix  = "arXiv",
      primaryClass   = "hep-ph",
      reportNumber   = "TTK-11-38, ITP-UU-11-26, SPIN-11-19, FR-PHENO-2011-015,
                        SFB-CPP-11-49",
      SLACcitation   = "%%CITATION = ARXIV:1109.1536;%%"
}

@article{Anastasiou:2003ds,
      author         = "Anastasiou, Charalampos and Dixon, Lance J. and Melnikov,
                        Kirill and Petriello, Frank",
      title          = "{High precision QCD at hadron colliders: Electroweak
                        gauge boson rapidity distributions at next-to-next-to leading order}",
      journal        = "Phys. Rev. D",
      volume         = "69",
      year           = "2004",
      pages          = "094008",
      doi            = "10.1103/PhysRevD.69.094008",
      eprint         = "hep-ph/0312266",
      archivePrefix  = "arXiv",
      reportNumber   = "SLAC-PUB-10288, UH-511-1042-03",
      SLACcitation   = "%%CITATION = HEP-PH/0312266;%%"
}

@article{Cacciari:2011hy,
      author         = "Cacciari, Matteo and Czakon, Michal and Mangano,
                        Michelangelo and Mitov, Alexander and Nason, Paolo",
      title          = "{Top-pair production at hadron colliders with
                        next-to-next-to-leading logarithmic soft-gluon
                        resummation}",
      journal        = "Phys. Lett. B",
      volume         = "710",
      pages          = "612-622",
      doi            = "10.1016/j.physletb.2012.03.013",
      year           = "2012",
      eprint         = "1111.5869",
      primaryClass   = "hep-ph",
      archivePrefix  = "arXiv",
      reportNumber   = "CERN-PH-TH-2011-277, TTK-11-54",
      SLACcitation   = "%%CITATION = ARXIV:1111.5869;%%",
}

@article{Czakon:2012zr,
      author         = "Czakon, Michal and Mitov, Alexander",
      title          = "{NNLO corrections to top-pair production at hadron
                        colliders: the all-fermionic scattering channels}",
      journal        = "JHEP",
      volume         = "12",
      pages          = "054",
      doi            = "10.1007/JHEP12(2012)054",
      year           = "2012",
      eprint         = "1207.0236",
      archivePrefix  = "arXiv",
      primaryClass   = "hep-ph",
      SLACcitation   = "%%CITATION = ARXIV:1207.0236;%%",
}

@article{Czakon:2013goa,
      author         = "Czakon, Michal and Fiedler, Paul and Mitov, Alexander",
      title          = "{Total Top-Quark Pair-Production Cross Section at
                        Hadron Colliders Through \(O(\alpha_S^4)\)}",
      journal        = "Phys. Rev. Lett.",
      volume         = "110",
      pages          = "252004",
      doi            = "10.1103/PhysRevLett.110.252004",
      year           = "2013",
      eprint         = "1303.6254",
      archivePrefix  = "arXiv",
      primaryClass   = "hep-ph",
      reportNumber   = "CERN-PH-TH-2013-056, TTK-13-08",
      SLACcitation   = "%%CITATION = ARXIV:1303.6254;%%",
}

@article{Baernreuther:2012ws,
      author         = "B{\"a}rnreuther, Peter and Czakon, Michal and Mitov,
                        Alexander",
      title          = "{Percent-Level-Precision Physics at the
                  Tevatron: Next-to-Next-to-Leading Order QCD
                  Corrections to \(q \bar{q} \to t \bar{t} + X\)}",
      journal        = "Phys. Rev. Lett.",
      volume         = "109",
      pages          = "132001",
      doi            = "10.1103/PhysRevLett.109.132001",
      year           = "2012",
      eprint         = "1204.5201",
      archivePrefix  = "arXiv",
      primaryClass   = "hep-ph",
      SLACcitation   = "%%CITATION = ARXIV:1204.5201;%%",
}

@article{Frixione:2007nw,
      author         = "Frixione, Stefano and Nason, Paolo and Ridolfi, Giovanni",
      title          = "{A positive-weight next-to-leading-order Monte Carlo for
                        heavy flavour hadroproduction}",
      journal        = "JHEP",
      volume         = "09",
      pages          = "126",
      doi            = "10.1088/1126-6708/2007/09/126",
      year           = "2007",
      eprint         = "0707.3088",
      archivePrefix  = "arXiv",
      primaryClass   = "hep-ph",
      SLACcitation   = "%%CITATION = ARXIV:0707.3088;%%",
}

@Article{Hoeche:2009rj,
      author         = "H{\"o}che, Stefan and Krauss, Frank and Schumann, Steffen and Siegert, Frank",
      title          = "{QCD matrix elements and truncated showers}",
      journal        = "JHEP",
      volume         = "05",
      pages          = "053",
      doi            = "10.1088/1126-6708/2009/05/053",
      year           = "2009",
      eprint         = "0903.1219",
      archivePrefix  = "arXiv",
      primaryClass   = "hep-ph",
}

@Article{Gleisberg:2008fv,
      author         = "Gleisberg, Tanju and H{\"o}che, Stefan",
      title          = "{Comix, a new matrix element generator}",
      journal        = "JHEP",
      volume         = "12",
      pages          = "039",
      doi            = "10.1088/1126-6708/2008/12/039",
      year           = "2008",
      eprint         = "0808.3674",
      archivePrefix  = "arXiv",
      primaryClass   = "hep-ph",
}

@Article{Schumann:2007mg,
      author         = "Schumann, Steffen and Krauss, Frank",
      title          = "{A parton shower algorithm based on Catani--Seymour dipole factorisation}",
      journal        = "JHEP",
      volume         = "03",
      pages          = "038",
      doi            = "10.1088/1126-6708/2008/03/038",
      year           = "2008",
      eprint         = "0709.1027",
      archivePrefix  = "arXiv",
      primaryClass   = "hep-ph",
}

@article{Hoeche:2012yf,
      author         = "H{\"o}che, Stefan and Krauss, Frank and Sch{\"o}nherr, Marek and
                        Siegert, Frank",
      title          = "{QCD matrix elements + parton showers. The NLO case}",
      journal        = "JHEP",
      volume         = "04",
      year           = "2013",
      pages          = "027",
      doi            = "10.1007/JHEP04(2013)027",
      eprint         = "1207.5030",
      archivePrefix  = "arXiv",
      primaryClass   = "hep-ph",
      reportNumber   = "SLAC-PUB-15191, IPPP-12-52, DCPT-12-104, LPN12-081,
                        FR-PHENO-2012-017, MCNET-12-09, --FR-PHENO-2012-017",
      SLACcitation   = "%%CITATION = ARXIV:1207.5030;%%"
}

@article{Hoeche:2011fd,
      author         = "H{\"o}che, Stefan and Krauss, Frank and Sch{\"o}nherr, Marek and
                        Siegert, Frank",
      title          = "{A critical appraisal of NLO+PS matching methods}",
      journal        = "JHEP",
      volume         = "09",
      year           = "2012",
      pages          = "049",
      doi            = "10.1007/JHEP09(2012)049",
      eprint         = "1111.1220",
      archivePrefix  = "arXiv",
      primaryClass   = "hep-ph",
      reportNumber   = "SLAC-PUB-14661, IPPP-11-67, DCPT-11-134, LPN11-58,
                        FR-PHENO-2011-019, MCNET-11-24",
      SLACcitation   = "%%CITATION = ARXIV:1111.1220;%%"
}

@article{Catani:2001cc,
      author         = "Catani, S. and Krauss, F. and Kuhn, R. and Webber, B. R.",
      title          = "{QCD Matrix Elements + Parton Showers}",
      journal        = "JHEP",
      volume         = "11",
      year           = "2001",
      pages          = "063",
      doi            = "10.1088/1126-6708/2001/11/063",
      eprint         = "hep-ph/0109231",
      archivePrefix  = "arXiv",
      reportNumber   = "CERN-TH-2000-367, CAVENDISH-HEP-00-03",
      SLACcitation   = "%%CITATION = HEP-PH/0109231;%%"
}

@article{Denner:2016kdg,
      author         = "Denner, Ansgar and Dittmaier, Stefan and Hofer, Lars",
      title          = "{\textsc{Collier}: A fortran-based complex one-loop library in
                        extended regularizations}",
      journal        = "Comput. Phys. Commun.",
      volume         = "212",
      year           = "2017",
      pages          = "220-238",
      doi            = "10.1016/j.cpc.2016.10.013",
      eprint         = "1604.06792",
      archivePrefix  = "arXiv",
      primaryClass   = "hep-ph",
      reportNumber   = "FR-PHENO-2016-003, ICCUB-16-016",
      SLACcitation   = "%%CITATION = ARXIV:1604.06792;%%"
}

@article{Czakon:2011xx,
  author         = "Czakon, Michal and Mitov, Alexander",
  title          = "{Top++: A program for the calculation of the top-pair cross-section at hadron colliders}",
  journal        = "Comput. Phys. Commun.",
  volume         = "185",
  year           = "2014",
  pages          = "2930",
  doi            = "10.1016/j.cpc.2014.06.021",
  eprint         = "1112.5675",
  archivePrefix  = "arXiv",
  primaryClass   = "hep-ph",
  reportNumber   = "CERN-PH-TH-2011-303, TTK-11-58",
}

@article{Catani:2007vq,
      author         = "Catani, Stefano and Grazzini, Massimiliano",
      title          = "{An NNLO subtraction formalism in hadron collisions and
                        its application to Higgs boson production at the LHC}",
      journal        = "Phys. Rev. Lett.",
      volume         = "98",
      year           = "2007",
      pages          = "222002",
      doi            = "10.1103/PhysRevLett.98.222002",
      eprint         = "hep-ph/0703012",
      archivePrefix  = "arXiv",
      primaryClass   = "hep-ph",
      SLACcitation   = "%%CITATION = HEP-PH/0703012;%%"
}

@article{Agostinelli:2002hh,
      author         = "Agostinelli, S. and others",
      title          = "{\textsc{Geant4} -- a simulation toolkit}",
      journal        = "Nucl. Instrum. Meth. A",
      volume         = "506",
      year           = "2003",
      pages          = "250",
      doi            = "10.1016/S0168-9002(03)01368-8",
      reportNumber   = "SLAC-PUB-9350, FERMILAB-PUB-03-339",
      SLACcitation   = "%%CITATION = NUIMA,A506,250;%%"
}

@Article{Cowan:2010js,
   Author = {Cowan, Glen and Cranmer, Kyle and Gross, Eilam and Vitells, Ofer},
   Title = {{Asymptotic formulae for likelihood-based tests of new physics}},
   Journal = {Eur. Phys. J. C},
   Volume = {71},
   Year = {2011},
   Pages = {1554},
   doi = "10.1140/epjc/s10052-011-1554-0",
   Eprint = "1007.1727",
   Archiveprefix = {arXiv},
   Primaryclass = {physics.data-an},
   related = "Cowan:2010js-err",
   relatedstring  = "Erratum:",
}

@Article{PERF-2007-01,
    author         = "{ATLAS Collaboration}",
    title          = "{The ATLAS Experiment at the CERN Large Hadron Collider}",
    journal        = "JINST",
    volume         = "3",
    year           = "2008",
    pages          = "S08003",
    doi            = "10.1088/1748-0221/3/08/S08003",
    primaryClass   = "hep-ex",
}

@Article{SOFT-2010-01,
    author         = "{ATLAS Collaboration}",
    title          = "{The ATLAS Simulation Infrastructure}",
    journal        = "Eur. Phys. J. C",
    volume         = "70",
    year           = "2010",
    pages          = "823",
    doi            = "10.1140/epjc/s10052-010-1429-9",
    eprint         = "1005.4568",
    archivePrefix  = "arXiv",
    primaryClass   = "physics.ins-det",
}

@Article{STDM-2012-01,
    author         = "{ATLAS Collaboration}",
    title          = "{Measurement of \(W^+W^-\) production in \(pp\) collisions at \(\sqrt{s} = 7\,\text{TeV}\) with the ATLAS detector and limits on anomalous \(WWZ\) and \(WW\gamma\) couplings}",
    journal        = "Phys. Rev. D",
    volume         = "87",
    year           = "2013",
    pages          = "112001",
    doi            = "10.1103/PhysRevD.87.112001",
    reportNumber   = "CERN-PH-EP-2012-242",
    eprint         = "1210.2979",
    archivePrefix  = "arXiv",
    primaryClass   = "hep-ex",
    related        = "STDM-2012-01-err",
    relatedstring  = "Erratum:",
}

@Article{STDM-2012-23,
    author         = "{ATLAS Collaboration}",
    title          = "{Measurement of the \(Z/\gamma^*\) boson transverse momentum distribution in \(pp\) collisions at \(\sqrt{s} = 7\,\text{TeV}\) with the ATLAS detector}",
    journal        = "JHEP",
    volume         = "09",
    year           = "2014",
    pages          = "145",
    doi            = "10.1007/JHEP09(2014)145",
    reportNumber   = "CERN-PH-EP-2014-075",
    eprint         = "1406.3660",
    archivePrefix  = "arXiv",
    primaryClass   = "hep-ex",
}

@Article{STDM-2013-07,
    author         = "{ATLAS Collaboration}",
    title          = "{Measurement of total and differential \(W^+W^-\) production cross sections in proton--proton collisions at \(\sqrt{s} = 8\,\text{TeV}\) with the ATLAS detector and limits on anomalous triple-gauge-boson couplings}",
    journal        = "JHEP",
    volume         = "09",
    year           = "2016",
    pages          = "029",
    doi            = "10.1007/JHEP09(2016)029",
    reportNumber   = "CERN-PH-EP-2015-323",
    eprint         = "1603.01702",
    archivePrefix  = "arXiv",
    primaryClass   = "hep-ex",
}

@Article{PERF-2014-03,
    author         = "{ATLAS Collaboration}",
    title          = "{Performance of pile-up mitigation techniques for jets in \(pp\) collisions at \(\sqrt{s} = 8\,\text{TeV}\) using the ATLAS detector}",
    journal        = "Eur. Phys. J. C",
    volume         = "76",
    year           = "2016",
    pages          = "581",
    doi            = "10.1140/epjc/s10052-016-4395-z",
    reportNumber   = "CERN-PH-EP-2015-206",
    eprint         = "1510.03823",
    archivePrefix  = "arXiv",
    primaryClass   = "hep-ex",
}

@Article{PERF-2015-09,
    author         = "{ATLAS Collaboration}",
    title          = "{Jet reconstruction and performance using particle flow with the ATLAS Detector}",
    journal        = "Eur. Phys. J. C",
    volume         = "77",
    year           = "2017",
    pages          = "466",
    doi            = "10.1140/epjc/s10052-017-5031-2",
    reportNumber   = "CERN-EP-2017-024",
    eprint         = "1703.10485",
    archivePrefix  = "arXiv",
    primaryClass   = "hep-ex",
}

@Article{PERF-2015-10,
    author         = "{ATLAS Collaboration}",
    title          = "{Muon reconstruction performance of the ATLAS detector in proton--proton collision data at \(\sqrt{s} = 13\,\text{TeV}\)}",
    journal        = "Eur. Phys. J. C",
    volume         = "76",
    year           = "2016",
    pages          = "292",
    doi            = "10.1140/epjc/s10052-016-4120-y",
    reportNumber   = "CERN-EP-2016-033",
    eprint         = "1603.05598",
    archivePrefix  = "arXiv",
    primaryClass   = "hep-ex",
}

@Article{STDM-2015-03,
    author         = "{ATLAS Collaboration}",
    title          = "{Measurement of \(W^{\pm}\) and \(Z\)-boson production cross sections in \(pp\) collisions at \(\sqrt{s} = 13\,\text{TeV}\) with the ATLAS detector}",
    journal        = "Phys. Lett. B",
    volume         = "759",
    year           = "2016",
    pages          = "601",
    doi            = "10.1016/j.physletb.2016.06.023",
    reportNumber   = "CERN-EP-2016-069",
    eprint         = "1603.09222",
    archivePrefix  = "arXiv",
    primaryClass   = "hep-ex",
}

@Article{STDM-2015-20,
    author         = "{ATLAS Collaboration}",
    title          = "{Measurement of the \(W^+W^-\) production cross section in \(pp\) collisions at a centre-of-mass energy of \(\sqrt{s} = 13\,\text{TeV}\) with the ATLAS experiment}",
    journal        = "Phys. Lett. B",
    volume         = "773",
    year           = "2017",
    pages          = "354",
    doi            = "10.1016/j.physletb.2017.08.047",
    reportNumber   = "CERN-EP-2016-267",
    eprint         = "1702.04519",
    archivePrefix  = "arXiv",
    primaryClass   = "hep-ex",
}

@Article{STDM-2015-24,
    author         = "{ATLAS Collaboration}",
    title          = "{Measurement of \(W^+W^-\) production in association with one jet in proton--proton collisions at \(\sqrt{s} = 8\,\text{TeV}\) with the ATLAS detector}",
    journal        = "Phys. Lett. B",
    volume         = "763",
    year           = "2016",
    pages          = "114",
    doi            = "10.1016/j.physletb.2016.10.014",
    reportNumber   = "CERN-EP-2016-186",
    eprint         = "1608.03086",
    archivePrefix  = "arXiv",
    primaryClass   = "hep-ex",
}

@Article{PERF-2016-07,
    author         = "{ATLAS Collaboration}",
    title          = "{Performance of missing transverse momentum reconstruction with the ATLAS detector using proton--proton collisions at \(\sqrt{s} = 13\,\text{TeV}\)}",
    journal        = "Eur. Phys. J. C",
    volume         = "78",
    year           = "2018",
    pages          = "903",
    doi            = "10.1140/epjc/s10052-018-6288-9",
    reportNumber   = "CERN-EP-2017-274",
    eprint         = "1802.08168",
    archivePrefix  = "arXiv",
    primaryClass   = "hep-ex",
}

@Article{STDM-2017-06,
    author         = "{ATLAS Collaboration}",
    title          = "{Observation of Electroweak Production of a Same-Sign \(W\) Boson Pair in Association with Two Jets in \(pp\) Collisions at \(\sqrt{s} = 13\,\text{TeV}\) with the ATLAS Detector}",
    journal        = "Phys. Rev. Lett.",
    volume         = "123",
    year           = "2019",
    pages          = "161801",
    doi            = "10.1103/PhysRevLett.123.161801",
    reportNumber   = "CERN-EP-2019-008",
    eprint         = "1906.03203",
    archivePrefix  = "arXiv",
    primaryClass   = "hep-ex",
}

@Article{STDM-2017-24,
    author         = "{ATLAS Collaboration}",
    title          = "{Measurement of fiducial and differential \(W^+W^-\) production cross-sections at \(\sqrt{s} = 13\,\text{TeV}\) with the ATLAS detector}",
    journal        = "Eur. Phys. J. C",
    volume         = "79",
    year           = "2019",
    pages          = "884",
    doi            = "10.1140/epjc/s10052-019-7371-6",
    reportNumber   = "CERN-EP-2019-055",
    eprint         = "1905.04242",
    archivePrefix  = "arXiv",
    primaryClass   = "hep-ex",
}

@Article{DAPR-2018-01,
    author         = "{ATLAS Collaboration}",
    title          = "{ATLAS data quality operations and performance for 2015--2018 data-taking}",
    journal        = "JINST",
    volume         = "15",
    year           = "2020",
    pages          = "P04003",
    doi            = "10.1088/1748-0221/15/04/P04003",
    reportNumber   = "CERN-EP-2019-207",
    eprint         = "1911.04632",
    archivePrefix  = "arXiv",
    primaryClass   = "physics.ins-det",
}

@Article{EGAM-2018-01,
    author         = "{ATLAS Collaboration}",
    title          = "{Electron and photon performance measurements with the ATLAS detector using the 2015--2017 LHC proton--proton collision data}",
    journal        = "JINST",
    volume         = "14",
    year           = "2019",
    pages          = "P12006",
    doi            = "10.1088/1748-0221/14/12/P12006",
    reportNumber   = "CERN-EP-2019-145",
    eprint         = "1908.00005",
    archivePrefix  = "arXiv",
    primaryClass   = "hep-ex",
}

@Article{FTAG-2018-01,
    author         = "{ATLAS Collaboration}",
    title          = "{ATLAS \(b\)-jet identification performance and efficiency measurement with \(t\bar{t}\) events in \(pp\) collisions at \(\sqrt{s} = 13\,\text{TeV}\)}",
    journal        = "Eur. Phys. J. C",
    volume         = "79",
    year           = "2019",
    pages          = "970",
    doi            = "10.1140/epjc/s10052-019-7450-8",
    reportNumber   = "CERN-EP-2019-132",
    eprint         = "1907.05120",
    archivePrefix  = "arXiv",
    primaryClass   = "hep-ex",
}

@Article{JETM-2018-05,
    author         = "{ATLAS Collaboration}",
    title          = "{Jet energy scale and resolution measured in proton--proton collisions at \(\sqrt{s} = 13\,\text{TeV}\) with the ATLAS detector}",
    year           = "2020",
    reportNumber   = "CERN-EP-2020-083",
    eprint         = "2007.02645",
    archivePrefix  = "arXiv",
    primaryClass   = "hep-ex",
}

@Article{PIX-2018-001,
    author         = "Abbott, B. and others",
    title          = "{Production and integration of the ATLAS Insertable B-Layer}",
    journal        = "JINST",
    volume         = "13",
    year           = "2018",
    pages          = "T05008",
    doi            = "10.1088/1748-0221/13/05/T05008",
    eprint         = "1803.00844",
    archivePrefix  = "arXiv",
    primaryClass   = "physics.ins-det",
}

@Article{STDM-2018-03,
    author         = "{ATLAS Collaboration}",
    title          = "{Measurement of \(W^{\pm}Z\) production cross sections and gauge boson polarisation in \(pp\) collisions at \(\sqrt{s} = 13\,\text{TeV}\) with the ATLAS detector}",
    journal        = "Eur. Phys. J. C",
    volume         = "79",
    year           = "2019",
    pages          = "535",
    doi            = "10.1140/epjc/s10052-019-7027-6",
    reportNumber   = "CERN-EP-2018-327",
    eprint         = "1902.05759",
    archivePrefix  = "arXiv",
    primaryClass   = "hep-ex",
}

@Article{TRIG-2018-01,
    author         = "{ATLAS Collaboration}",
    title          = "{Performance of the ATLAS muon triggers in Run~2}",
    journal        = "JINST",
    volume         = "15",
    year           = "2020",
    pages          = "P09015",
    doi            = "10.1088/1748-0221/15/09/p09015",
    reportNumber   = "CERN-EP-2020-031",
    eprint         = "2004.13447",
    archivePrefix  = "arXiv",
    primaryClass   = "hep-ex",
}

@Article{TRIG-2018-05,
    author         = "{ATLAS Collaboration}",
    title          = "{Performance of electron and photon triggers in ATLAS during LHC Run~2}",
    journal        = "Eur. Phys. J. C",
    volume         = "80",
    year           = "2020",
    pages          = "47",
    doi            = "10.1140/epjc/s10052-019-7500-2",
    reportNumber   = "CERN-EP-2019-169",
    eprint         = "1909.00761",
    archivePrefix  = "arXiv",
    primaryClass   = "hep-ex",
}

@Booklet{ATL-SOFT-PUB-2020-001,
    author         = "{ATLAS Collaboration}",
    title          = "{ATLAS Computing Acknowledgements}",
    howpublished   = "{ATL-SOFT-PUB-2020-001}",
    url            = "https://cds.cern.ch/record/2717821",
}

@Report{ATLAS-TDR-2010-19,
    author         = "{ATLAS Collaboration}",
    title          = "{ATLAS Insertable B-Layer Technical Design Report}",
    type           = "ATLAS-TDR-19; CERN-LHCC-2010-013",
    year           = "2010",
    url            = "https://cds.cern.ch/record/1291633",
}

@Article{CMS-SMP-12-005,
    author         = "{CMS Collaboration}",
    title          = "{Measurement of the \(W^+W^-\) cross section in \(pp\) collisions at \(\sqrt{s} = 7\,\text{TeV}\) and limits on anomalous \(WW\gamma\) and \(WWZ\) couplings}",
    journal        = "Eur. Phys. J. C",
    volume         = "73",
    year           = "2013",
    pages          = "2610",
    doi            = "10.1140/epjc/s10052-013-2610-8",
    reportNumber   = "CERN-PH-EP-2013-075",
    eprint         = "1306.1126",
    archivePrefix  = "arXiv",
    primaryClass   = "hep-ex",
}

@Article{CMS-SMP-14-016,
    author         = "{CMS Collaboration}",
    title          = "{Measurement of the \(W^+W^-\) cross section in \(pp\) collisions at \(\sqrt{s} = 8\,\text{TeV}\) and limits on anomalous gauge couplings}",
    journal        = "Eur. Phys. J. C",
    volume         = "76",
    year           = "2016",
    pages          = "401",
    doi            = "10.1140/epjc/s10052-016-4219-1",
    reportNumber   = "CERN-PH-EP-2015-122",
    eprint         = "1507.03268",
    archivePrefix  = "arXiv",
    primaryClass   = "hep-ex",
}

@Article{CMS-SMP-18-004,
    author         = "{CMS Collaboration}",
    title          = "{\(W^+W^-\) boson pair production in proton--proton collisions at \(\sqrt{s} = 13\,\text{TeV}\)}",
    journal        = "Phys. Rev. D",
    volume         = "102",
    year           = "2020",
    pages          = "092001",
    doi            = "10.1103/PhysRevD.102.092001",
    reportNumber   = "CERN-EP-2020-144",
    eprint         = "2009.00119",
    archivePrefix  = "arXiv",
    primaryClass   = "hep-ex",
}

@Booklet{ATLAS-CONF-2019-021,
    author         = "{ATLAS Collaboration}",
    title          = "{Luminosity determination in \(pp\) collisions at \(\sqrt{s} = 13\,\text{TeV}\) using the ATLAS detector at the LHC}",
    howpublished   = "{ATLAS-CONF-2019-021}",
    url            = "https://cds.cern.ch/record/2677054",
    year           = "2019",
}

@Article{EvtGen,
    author         = "{D. J. Lange}",
    title          = "{The EvtGen particle decay simulation package}",
    journal        = "Nucl. Instrum. Meth. A",
    volume         = "462",
    year           = "2001",
    pages          = "152",
    doi            = "10.1016/S0168-9002(01)00089-4",
    primaryClass   = "hep-ex",
}

@Booklet{ATL-PHYS-PUB-2014-021,
    author         = "{ATLAS Collaboration}",
    title          = "{ATLAS Pythia~8 tunes to \(7~\text{TeV}\) data}",
    howpublished   = "{ATL-PHYS-PUB-2014-021}",
    url            = "https://cds.cern.ch/record/1966419",
    year           = "2014",
}

@Booklet{ATL-PHYS-PUB-2016-002,
    author         = "{ATLAS Collaboration}",
    title          = "{Multi-boson simulation for \(13~\text{TeV}\) ATLAS analyses}",
    howpublished   = "{ATL-PHYS-PUB-2016-002}",
    url            = "https://cds.cern.ch/record/2119986",
    year           = "2016",
}

@Booklet{ATL-PHYS-PUB-2016-017,
    author         = "{ATLAS Collaboration}",
    title          = "{The Pythia~8 A3 tune description of ATLAS minimum bias and inelastic measurements incorporating the Donnachie--Landshoff diffractive model}",
    howpublished   = "{ATL-PHYS-PUB-2016-017}",
    url            = "https://cds.cern.ch/record/2206965",
    year           = "2016",
}

@Booklet{ATL-PHYS-PUB-2016-020,
    author         = "{ATLAS Collaboration}",
    title          = "{Studies on top-quark Monte Carlo modelling for Top2016}",
    howpublished   = "{ATL-PHYS-PUB-2016-020}",
    url            = "https://cds.cern.ch/record/2216168",
    year           = "2016",
}

@Booklet{ATL-PHYS-PUB-2017-007,
    author         = "{ATLAS Collaboration}",
    title          = "{Studies on top-quark Monte Carlo modelling with Sherpa and MG5\_aMC@NLO}",
    howpublished   = "{ATL-PHYS-PUB-2017-007}",
    url            = "https://cds.cern.ch/record/2261938",
    year           = "2017",
}

@Booklet{ATL-PHYS-PUB-2017-013,
    author         = "{ATLAS Collaboration}",
    title          = "{Optimisation and performance studies of the ATLAS \(b\)-tagging algorithms for the 2017-18 LHC run}",
    howpublished   = "{ATL-PHYS-PUB-2017-013}",
    url            = "https://cds.cern.ch/record/2273281",
    year           = "2017",
}

\clearpage 
 
\begin{flushleft}
\hypersetup{urlcolor=black}
{\Large The ATLAS Collaboration}

\bigskip

\AtlasOrcid[0000-0002-6665-4934]{G.~Aad}$^\textrm{\scriptsize 101}$,    
\AtlasOrcid[0000-0002-5888-2734]{B.~Abbott}$^\textrm{\scriptsize 127}$,    
\AtlasOrcid[0000-0002-7248-3203]{D.C.~Abbott}$^\textrm{\scriptsize 102}$,    
\AtlasOrcid[0000-0002-2788-3822]{A.~Abed~Abud}$^\textrm{\scriptsize 36}$,    
\AtlasOrcid[0000-0002-1002-1652]{K.~Abeling}$^\textrm{\scriptsize 53}$,    
\AtlasOrcid[0000-0002-2987-4006]{D.K.~Abhayasinghe}$^\textrm{\scriptsize 93}$,    
\AtlasOrcid[0000-0002-8496-9294]{S.H.~Abidi}$^\textrm{\scriptsize 29}$,    
\AtlasOrcid[0000-0002-8279-9324]{O.S.~AbouZeid}$^\textrm{\scriptsize 40}$,    
\AtlasOrcid{N.L.~Abraham}$^\textrm{\scriptsize 155}$,    
\AtlasOrcid[0000-0001-5329-6640]{H.~Abramowicz}$^\textrm{\scriptsize 160}$
% ...
\bigskip
\\

% ...
$^{29}$Physics Department, Brookhaven National Laboratory, Upton NY; United States of America.\\
% ...
$^{36}$CERN, Geneva; Switzerland.\\
% ...
$^{40}$Niels Bohr Institute, University of Copenhagen, Copenhagen; Denmark.\\
% ...
$^{53}$II. Physikalisches Institut, Georg-August-Universit\"{a}t G\"ottingen, G\"ottingen; Germany.\\
% ...
$^{93}$Department of Physics, Royal Holloway University of London, Egham; United Kingdom.\\
% ...
$^{101}$CPPM, Aix-Marseille Universit\'e, CNRS/IN2P3, Marseille; France.\\
$^{102}$Department of Physics, University of Massachusetts, Amherst MA; United States of America.\\
% ...
$^{127}$Homer L. Dodge Department of Physics and Astronomy, University of Oklahoma, Norman OK; United States of America.\\
% ...
$^{155}$Department of Physics and Astronomy, University of Sussex, Brighton; United Kingdom.\\
% ...
$^{160}$Raymond and Beverly Sackler School of Physics and Astronomy, Tel Aviv University, Tel Aviv; Israel.\\
% ...

\end{flushleft}


\end{document}